\documentclass[notitlepage,twocolumn,prd,showpacs,preprintnumbers,nofootinbib,superscriptaddress,floatfix,tightenlines]{revtex4-1}
\usepackage{mathrsfs}
\usepackage{amssymb}
\usepackage{amsmath}
\usepackage{graphicx}
\usepackage{hyperref}
\usepackage{amsfonts,amsbsy}
\usepackage{animate}
\usepackage{color}

\newcommand{\p}{{\bf p}}
\newcommand{\Qt}{t_Q}

\newcommand{\xvect}{{\bf x}_{\bot}}

\newcommand{\pvect}{{\bf p}_{\bot}}
\newcommand{\kvect}{{\bf k}_{\bot}}
\newcommand{\qvect}{{\bf q}_{\bot}}

\newcommand{\ihat}{\hat{\imath}}

\newcommand{\pt}{\text{p}_\text{T}}
\newcommand{\pz}{\text{p}_\text{z}}
\newcommand{\psqr}{\text{p}^2}

\newcommand{\dInt}{\text{d}}

\newcommand{\bx}{\text{x}}

\newcommand{\Nt}{N_{\bot}}

\newcommand{\OneOverQ}{Q^{-1}}
\newcommand{\lnSqrOneOverAlpha}{\ln^{2}(\alpha_s^{-1})}

\newcommand{\nHard}{n_{\text{Hard}}}

\usepackage{nicefrac}

\def\dblone{\hbox{$1\hskip -1.2pt\vrule depth 0pt height 1.6ex width 0.7pt
                  \vrule depth 0pt height 0.3pt width 0.12em$}}

\makeatletter
  \def\l@subsubsection#1#2{}
\makeatother

\begin{document}

\title{Universal attractor in a highly occupied non-Abelian plasma}

\author{J.~Berges}

\affiliation{Institut f\"{u}r Theoretische Physik, Universit\"{a}t Heidelberg, Philosophenweg 16, 69120 Heidelberg, Germany}
\affiliation{ExtreMe Matter Institute (EMMI), GSI Helmholtzzentrum f\"ur Schwerionenforschung GmbH, 
Planckstra\ss e~1, 64291~Darmstadt, Germany}

\author{K.~Boguslavski}
\affiliation{Institut f\"{u}r Theoretische Physik, Universit\"{a}t Heidelberg, Philosophenweg 16, 69120 Heidelberg, Germany}

\author{S.~Schlichting}
\email{soeren@kaiden.de}
\affiliation{Institut f\"{u}r Theoretische Physik, Universit\"{a}t Heidelberg, Philosophenweg 16, 69120 Heidelberg, Germany}
\affiliation{Brookhaven National Laboratory, Physics Department, Bldg. 510A, Upton, NY 11973, USA}

\author{R.~Venugopalan}
\affiliation{Brookhaven National Laboratory, Physics Department, Bldg. 510A, Upton, NY 11973, USA}


\begin{abstract}
We study the thermalization process in highly occupied non-Abelian plasmas at weak coupling. The non-equilibrium dynamics of such systems is classical in nature and can be simulated with real-time lattice gauge theory techniques. We provide a detailed discussion of this framework and elaborate on the results reported in~\cite{Berges:2013eia} along with novel findings. We demonstrate the emergence of universal attractor solutions, which govern the non-equilibrium evolution on large time scales both for non-expanding and expanding non-Abelian plasmas. The turbulent attractor for a non-expanding plasma drives the system close to thermal equilibrium on a time scale $t\sim \OneOverQ \alpha_s^{-7/4}$. The attractor solution for an expanding non-Abelian plasma leads to a strongly interacting albeit highly anisotropic system at the transition to the low-occupancy or quantum regime. This evolution in the classical regime is, within the uncertainties of our simulations, consistent with the ``bottom up'' thermalization scenario~\cite{Baier:2000sb}. While the focus of this paper is to understand the non-equilibrium dynamics in weak coupling asymptotics, we also discuss the relevance of our results for larger couplings in the early time dynamics of heavy ion collision experiments.
\end{abstract}

\maketitle

\tableofcontents

\section{Introduction}
An ab initio understanding of how a non-Abelian plasma thermalizes in heavy ion collisions even at asymptotically high energies remains elusive and is an outstanding problem in theoretical physics~\cite{Berges:2012ks}. 
In recent years, significant progress in a first principles understanding of non-Abelian plasmas out-of-equilibrium has been achieved in two limiting cases. One of these is the study of the strong-coupling limit using gauge-string dualities in supersymmetric Yang-Mills theories. The other case that is amenable to ab initio calculations is Quantum Chromodynamics (QCD) in the weak-coupling limit $\alpha_s \ll 1$.\\

In the case of strongly coupled supersymmetric Yang-Mills theory, the gauge-string dualities provide a valuable tool to study non-equilibrium phenomena. This holographic thermalization process has been studied extensively in the literature~\cite{Chesler:2008hg,Chesler:2009cy,Chesler:2010bi,Grumiller:2008va,Balasubramanian:2010ce,Balasubramanian:2011ur,Heller:2011ju,Heller:2012je} and generally leads to fast thermalization of the plasma. Here the results for the longitudinally expanding system -- relevant to heavy ion collisions -- indicate the important role of anisotropies even at the transition to the hydrodynamic regime~\cite{Heller:2011ju,Heller:2012je}.\\ 

In the weak coupling limit, the colliding nuclei may be described as Color Glass Condensates (CGC) in an effective field theory description of high energy QCD~\cite{Iancu:2002xk,Iancu:2003xm,Gelis:2007kn,Gelis:2010nm}. Models constructed within this framework describe rather well the bulk properties of phenomena observed in high energy heavy ion collisions at RHIC and the LHC~\cite{Schenke:2012wb,Schenke:2012hg,Gale:2012rq,Bzdak:2013zma}. The dynamics of the non-equilibrium ``Glasma'' created in such a collision is that of highly occupied gluon fields with typical momentum $Q$~\cite{Kovner:1995ja,Kovner:1995ts,Krasnitz:1998ns,Krasnitz:1999wc,Krasnitz:2000gz,Krasnitz:2001qu,Krasnitz:2002mn,Lappi:2003bi,Lappi:2006hq,Lappi:2006fp,Lappi:2009xa,Lappi:2011ju}. Since the characteristic occupancies $\sim 1/\alpha_s (Q)$ are large, the gauge fields are strongly correlated even for small gauge coupling.\\ 

The dynamics of highly occupied gauge fields is classical in nature and can be studied from first principles using real-time lattice techniques. For a non-expanding ensemble of such initial conditions, classical-statistical lattice simulations of weakly coupled non-Abelian plasmas are well understood. The results obtained in independent computations are unambiguous~\cite{Berges:2008mr,Berges:2012ev,Schlichting:2012es,Kurkela:2012hp} and in line with analytical expectations. In contrast, the dynamics of anisotropically expanding non-Abelian plasmas is much less understood. Even in the theoretically ``clean'' limit of asymptotically high collision energies, where the characteristic coupling is very weak, no consensus has been reached concerning the dynamics of the thermalization process~\cite{Berges:2012ks}. A key objective of this paper is to address the thermalization process, and its sensitivity to the initial conditions, using classical-statistical lattice simulations.\\

In the weak coupling limit, the CGC framework provides the initial conditions of the evolution at time scales $\tau \lesssim \OneOverQ$ after the collisions. They are expressed in terms of strong classical fields and small quantum fluctuations around them. Both of these can be computed from first principles~\cite{Dusling:2011rz,Epelbaum:2013waa}. At these early times, the expectation value of the classical field, averaged over a statistical ensemble of quantum fluctuations, is large $A_{cl}\sim\mathcal{O}(1/\sqrt{\alpha}_s)$. In contrast, the quantum fluctuations are initially of order unity.\\

While classical-statistical methods can already be applied at the earliest times after the collision, our interest is primarily in later times $\tau \gtrsim \OneOverQ \lnSqrOneOverAlpha$ where plasma instabilities have caused quantum fluctuations in the initial conditions to become of the order of the classical fields~\cite{Romatschke:2005pm,Romatschke:2006nk,Romatschke:2005ag,Fukushima:2011nq,Berges:2012cj}. At these time scales, the system may be expected to experience phase decoherence leading to fluctuation dominated dynamics. In such a regime, the ensemble average of the classical field is zero. However the statistical fluctuations in this ensemble are large; field amplitudes of each element of the ensemble can be as large as $\mathcal{O}(1/\sqrt{\alpha}_s)$. This situation can also be described in the framework of kinetic theory \cite{Mueller:2002gd,Jeon:2004dh} as an overoccupied plasma with occupation numbers $1 \ll f \ll 1/\alpha_s$ and is a common starting point for weak coupling thermalization scenarios.\\ 

Depending on the value of the gauge coupling for the expanding non-Abelian theory, it is an open question whether the classical field dominated initial conditions at $\tau \lesssim \OneOverQ$~\cite{Epelbaum:2013waa,Gelis:2013rba} evolve into fluctuation dominated initial conditions at $\tau \gtrsim \OneOverQ \lnSqrOneOverAlpha$. This issue will not be addressed in the present work. Our focus here will be to study a wide class of fluctuation dominated initial conditions to differentiate between several thermalization scenarios in the weak coupling limit~\cite{Baier:2000sb,Bodeker:2005nv,MSW,Kurkela:2011ti,Kurkela:2011ub,Blaizot:2011xf}.\\

The available parametric estimates in different thermalization scenarios allow for different types of solutions of the underlying kinetic equations. Since classical-statistical lattice gauge theory and kinetic theory have an overlapping range of validity when the occupation numbers are in the range $1 \ll f \ll 1/\alpha_s$~\cite{Mueller:2002gd,Jeon:2004dh,Berges:2004yj} the former can be used to determine the correct kinetic equations that describe the complex non-Abelian dynamics in the classical regime.\\

A principal result of classical-statistical simulations of expanding non-Abelian plasmas was the discovery of a non-thermal attractor solution~\cite{Berges:2013eia}. For a wide class of fluctuation dominated initial conditions, the system exhibits the same self-similar scaling behavior at later times before thermal equilibrium is achieved. In particular, the single particle gluon distribution function satisfies the self-similarity relation 
\begin{eqnarray}
\label{eq:scalingf}
f(\pt,\pz,\tau)=(Q\tau)^{\alpha}f_S\Big((Q\tau)^\beta \pt,(Q\tau)^\gamma \pz \Big) ,
\end{eqnarray}
where $f_S$ denotes a {\em stationary} distribution function independent of time. The scaling exponent $\alpha$ describes the temporal evolution of the amplitude of the stationary scaling function, while $\beta$ and $\gamma$ respectively describe the temporal evolution of the hard transverse and longitudinal momentum scales. These exponents are universal numbers: as we shall show, their extraction only relies on conservation laws and the dominance of small angle scattering processes. They do not depend on any other features of the underlying theory. Interestingly, such universal behavior far from equilibrium has also been predicted for systems ranging from early-universe inflaton dynamics~\cite{Micha:2004bv,Berges:2008wm,Berges:2008sr,Berges:2010ez,Berges:2012us} to table-top experiments with cold atoms~\cite{Semikoz:1994zp,Semikoz:1995rd,Scheppach:2009wu,Nowak:2010tm,Nowak:2011sk,Nowak:2013juc,Karl:2013kua}. In these examples, attractor solutions are characterized by a self-similar scaling behavior associated with the phenomenon of wave turbulence.\\

The scaling exponents we obtained in~\cite{Berges:2013eia} are, within the numerical accuracy of the lattice simulations, consistent with those obtained in the ``bottom up'' kinetic thermalization scenario of Baier, Mueller, Schiff and Son (BMSS)~\cite{Baier:2000sb}. In the BMSS implementation of the kinetic equations, the dynamics of the classical regime is dominated by small angle elastic scattering of hard modes. Since each of the collisions is soft, the transverse momenta $\pt$ of hard modes remain unchanged in the classical regime. The longitudinal momenta $\pz$ however are redshifted due to the rapid expansion of the system. After early free streaming where the longitudinal momenta decrease as $1/\tau$, the classical scattering dynamics slows the redshift of the longitudinal momenta to fall as $1/\tau^{1/3}$ instead. The relative anisotropy of longitudinal to transverse momenta increases at this rate for the rest of the classical regime where $f \gtrsim 1$. As we shall discuss further later, the outlined dynamics is a consequence of the universal attractor solution leading to the phenomenon of wave turbulence.\\

In the quantum regime of $f \lesssim 1$, the classical-statistical approach is no longer applicable. In the bottom up scenario, for $f \lesssim 1$, the onset of inelastic number changing processes becomes significant. It is this dynamics that eventually leads to isotropization and thermalization of the system. While this is a plausible scenario, we are unable to follow the evolution of the system in the quantum regime within our framework.\\ 

The paper is organized as follows. In sec.~\ref{sec:CGC}, we begin with a brief introduction to the classical-statistical framework for an expanding non-Abelian plasma. We discuss our choice of initial conditions as well as the range of validity of the classical-statistical framework. We provide details of the numerical solution of the $SU(2)$ Yang-Mills equations in 3+1 dimensions in sec.~\ref{sec:lattice} along with suitable gauge invariant and gauge-dependent observables measured in our simulations. In sec.~\ref{sec:StaticBox}, we briefly review the thermalization process in a non-expanding plasma. Numerical results for the classical-statistical evolution are presented and the turbulent thermalization mechanism is discussed in kinetic theory. In sec.~\ref{sec:ExpandingBox}, we study the dynamics of an expanding non-Abelian plasma. We elaborate significantly on the results presented in~\cite{Berges:2013eia}. Several new results are presented that corroborate and strengthen the conclusions in this earlier work. We discuss the discretization dependence and show that our results interpolate smoothly between results for non-expanding and expanding non-Abelian plasmas. We conclude in sec.~\ref{sec:Conclusion}, where we discuss the implications of our results and outline future work. The appendices contain additional information essential for a detailed understanding of the quantitative results in this paper.

\section{Heavy ion collisions and classical-statistical dynamics}
\label{sec:CGC}
While Quantum Chromodynamics (QCD) provides a fundamental framework to treat nucleus-nucleus collisions, an {\it ab initio} theoretical description remains a challenge. At high energies, the nuclear wavefunction represents a complex many body system of quark and gluon fields. One might therefore expect collisions of nuclei to be significantly more complicated than elementary proton-proton collisions. Nevertheless, effective descriptions of the dynamics of nucleus-nucleus collisions have been developed in different contexts.\\

One such effective description is based on the holographic principle relating classical gravity in higher dimensions to strongly coupled (super-symmetric) gauge theories. Holography has been used recently to understand thermalization in nucleus-nucleus collisions~\cite{Chesler:2008hg,Chesler:2009cy,Chesler:2010bi,Grumiller:2008va,Balasubramanian:2010ce,Balasubramanian:2011ur,Heller:2011ju,Heller:2012je}. The study of thermalization is accomplished either by following the collision of gravitational shock waves which carry the energy content of the colliding nuclei~\cite{Grumiller:2008va,Chesler:2010bi} or by a matching of boundary conditions immediately after the collision~\cite{Balasubramanian:2010ce,Balasubramanian:2011ur,Heller:2011ju,Heller:2012je}. While the holographic approach provides unique insights into the dynamics of strongly coupled (super-symmetric) gauge theories, the fundamental degrees of freedom of the underlying field theory can be very different from actual QCD. The results of such studies are therefore difficult to incorporate in quantitative frameworks. \\

The other effective description of the dynamics of nucleus-nucleus collisions is the Color Glass Condensate (CGC) framework. In this weak coupling framework, gluon states of very high occupancy are generated in the collision of large nuclei. For the rest of this paper, we will discuss the subsequent time evolution of such highly overoccupied but weakly coupled states in nucleus-nucleus collisions at very high energies.\\ 

Before we discuss the weak coupling approach in more detail, we will briefly introduce the coordinates best suited to describe the geometry of nucleus-nucleus collisions. These are the lightcone coordinates $x^{\pm}=(t\pm z)/\sqrt{2}$. At sufficiently high collider energies, the incoming nuclei travel close to the lightcone (denoted by $x^{\pm}=0)$. The collision takes place around the time when $x^{+}=x^{-}=0$, where an approximately boost invariant plasma is created at mid-rapidity. The dynamics of the plasma in the forward lightcone ($x^{\pm}>0$) is subject to longitudinal expansion and is conveniently described in terms of the co-moving coordinates
\begin{eqnarray}
\tau=\sqrt{t^2-z^2} \;, \qquad \eta=\text{atanh}(z/t) \;.
\end{eqnarray}
Here $\tau$ is the proper time in the longitudinal direction, $\eta$ is the longitudinal rapidity and the transverse coordinates, to be denoted by $\xvect$, are unaffected by this transformation.\\

\begin{figure}[t!]
\centering
\includegraphics[width=0.45\textwidth]{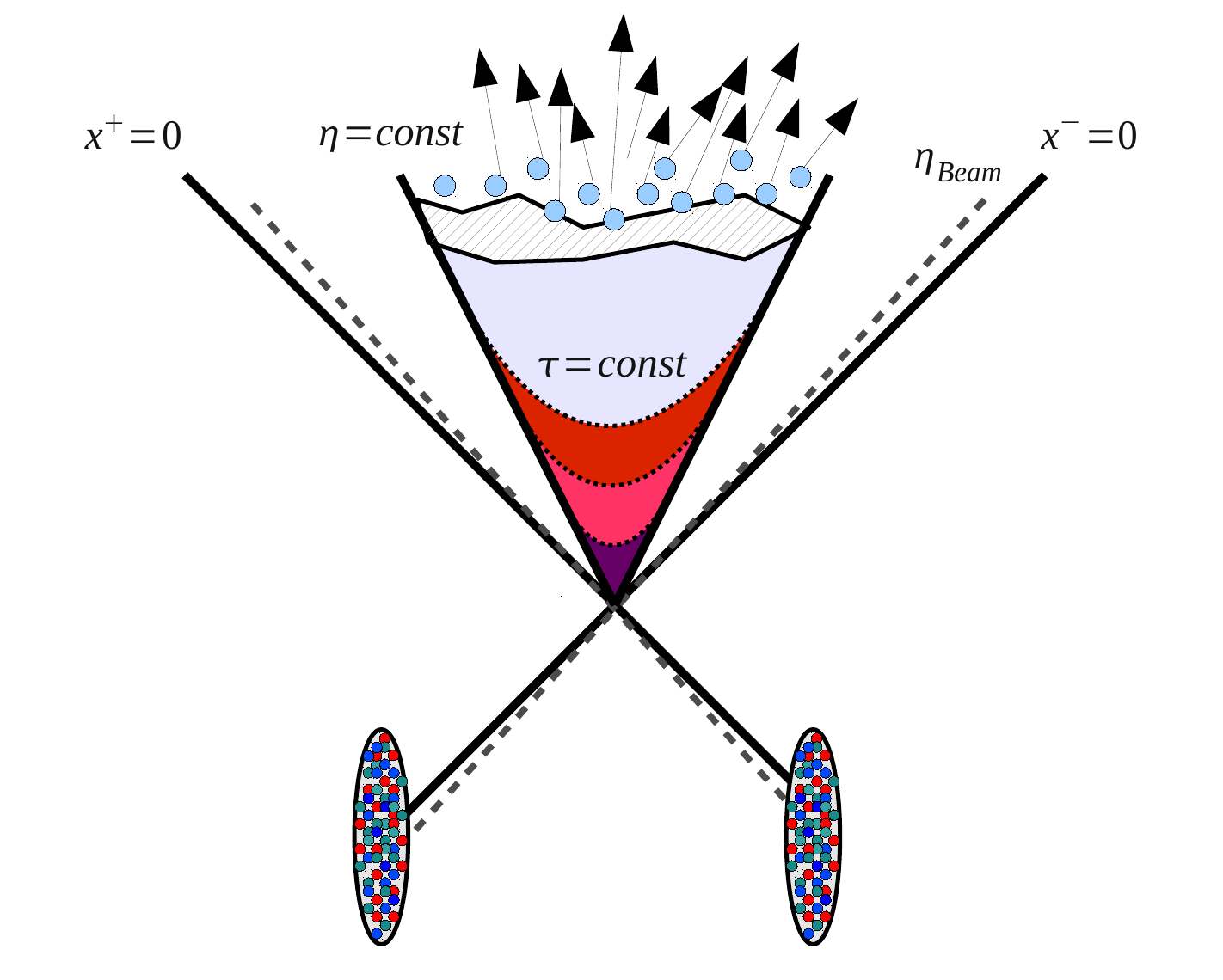}
\caption{\label{fig:cartoon} (color online) Illustration of the spacetime evolution of a high energy heavy ion collision. The (approximately) boost invariant plasma formed after the collision is subject to a longitudinal expansion in the beam direction. The non-equilibrium dynamics of the plasma can be described in terms of $(\tau,\xvect,\eta)$ coordinates.}
\end{figure}

The metric in $x^{\mu}=(\tau,\xvect,\eta)$ coordinates takes the form $g_{\mu\nu}(x)=\text{diag}(1,-1,-1,-\tau^2)$ and we will denote the metric determinant as $g(x)=\text{det}g_{\mu\nu}(x)=-\tau^2$. The explicit dependence on the proper time coordinate characterizes the longitudinal expansion of the system with an expansion rate of $1/\tau$. The corresponding spacetime evolution of the collision as well as the geometry of the coordinates are illustrated in fig.~\ref{fig:cartoon}. The different colors in the forward lightcone represent the different stages of the non-equilibrium evolution after the collision. We will discuss these further below.
 
\subsection{The Color Glass Condensate}
In the CGC framework, the colliding nuclei are represented in terms of fast moving hard ``valence'' partons (with large parton momentum fraction $\bx\sim 1$) and an abundance of soft ``wee" partons (with $\bx\ll 1$). A consistent dynamical separation of the hard and soft degrees of freedom is described by a renormalization group procedure~\cite{Iancu:2002xk,Iancu:2003xm,Gelis:2007kn,Gelis:2010nm}. The wee parton dynamics is governed by the classical Yang-Mills action
\begin{eqnarray}
S[A]=-\frac{1}{4}\int \dInt^4x~\sqrt{-g(x)}~ \mathcal{F}_{\mu\nu}^a(x)~g^{\mu\alpha}(x)g^{\nu\beta}(x)~\mathcal{F}_{\alpha\beta}^a(x)\;, \nonumber \\
\end{eqnarray}
where 
\begin{eqnarray}
\mathcal{F}_{\mu\nu}^a(x)=\partial_{\mu} A_{\nu}^{a}(x)-\partial_{\nu}A_{\mu}^{a}(x)+gf^{abc}A_{\mu}^b(x)A_{\nu}^c(x)\;,
\end{eqnarray}
denotes the non-Abelian field strength tensor and $g$ denotes the gauge coupling with $g^2 = 4 \pi \alpha_s$.\\

These wee gluon fields are coupled to the large $\bx$ hard partons via an eikonal current $J^{\mu}_a(x)$ -- they ``see" static sources of color charge. In the high energy limit, the hard partons inside the nuclei move along the $x^+$ and $x^-$ directions respectively at the speed of light. The associated current takes the generic form~\cite{Iancu:2003xm,Gelis:2007kn,Gelis:2010nm}
\begin{eqnarray}
\label{C3eq:EikCurr}
J^{\mu}_a(t,\xvect,z)=\delta^{\mu+}\varrho^{(1)}_a(\xvect)\delta(x^-)+\delta^{\mu-}\varrho^{(2)}_a(\xvect)\delta(x^+)\;, \nonumber \\
\end{eqnarray}
where $\delta^{\mu\pm}$ is the Kronecker delta in lightcone coordinates and the superscript labels the two different nuclei. The color charge density of hard partons $\varrho^{(1/2)}_a(\xvect)$ varies from event to event according to a statistical distribution which contains detailed non-perturbative information about the nuclear species and impact parameter dependence of the collision. As we shall discuss in a little more detail shortly, the color charge densities $\varrho^{(1/2)}_a(\xvect)$ evolve with the value of $\bx$ separating the hard valence partons from the soft wee partons. They become of order $\mathcal{O}(1/g)$ in the high energy limit. This non-perturbatively large color charge density is also realized in heavy nuclei because these naturally contain a large number of $(\bx\sim1)$ color sources.

\subsection{Classical dynamics and power counting}
The dynamics of nucleus-nucleus collisions within the CGC framework is described as the time evolution of the initial vacuum state in the presence of the eikonal currents in eq.~(\ref{C3eq:EikCurr}). Self-consistency also requires a non-trivial constraint imposed by the covariant conservation of the current. Since the initial conditions of the time evolution are formulated in the remote past $t\to-\infty$, this constitutes a (retarded) boundary value problem, which can be studied in non-equilibrium quantum field theory.\footnote{This formulation as an initial value problem is only valid for ``inclusive'' observables -- ones that have no kinematic restrictions of the final state.}\\ 

The formalism to compute inclusive observables in a series expansion in powers of the coupling constant was developed in~\cite{Gelis:2006yv,Gelis:2006cr,Gelis:2008rw,Gelis:2008ad,Gelis:2008sz}. In this formalism, the leading order contribution is computed from the solution of the classical Yang-Mills equations with vanishing boundary conditions in the remote past. The classical field configurations $A_{\mu}^{cl}$, which describe the state of the system immediately after the collision, are known analytically at leading order~\cite{Kovner:1995ja,Kovner:1995ts}. Their contribution to the energy momentum tensor $T^{\mu\nu}$ is $T^{\mu\nu}_{LO}\sim1/g^2$. The (LO) classical dynamics after the collision can be studied by numerically solving the classical Yang-Mills equations~\cite{Krasnitz:1998ns,Krasnitz:1999wc,Krasnitz:2000gz,Krasnitz:2001qu,Krasnitz:2002mn,Lappi:2003bi,Lappi:2006hq,Lappi:2006fp,Lappi:2009xa,Lappi:2011ju}. Though the leading order description is robust, it proves essential to consider higher order contributions. This is because at leading order thermalization and isotropization of the system are not accessible since the underlying approximations lead to an effectively 2+1-dimensional dynamics.\\

At next-to-leading (NLO) order, one needs to consider how quantum fluctuations of the initial vacuum state propagate immediately after the nuclear collision. The strategy to properly include quantum fluctuations was developed in~\cite{Gelis:2008rw,Gelis:2008ad,Gelis:2008sz} and can be summarized as follows. In the remote past ($t\rightarrow -\infty$), the initial state is the perturbative vacuum. The classical field $A_{\rm cl}$ vanishes and quantum fluctuations can be represented as plane waves. As the fluctuations develop in the nuclei, they are time dilated due to the rapid motion of the nucleus. A coherent cascade of such fluctuations develops, with the longer lived (large $\bx$) gluons acting as sources of color charge for the shorter lived (small $\bx$) fields. 
This is a dynamical process which depends on the $\bx$ value (or equivalently, the rapidity) of interest. Consequently, the corresponding source density $\varrho^{(1/2)}_a(\xvect)$ is implicitly dependent on $\bx$.\footnote{The very large density of color sources that arise from this process, and couple coherently to the small $\bx$ modes, is parametrically of order $\mathcal{O}(1/g)$ -- the largest value possible in QCD. Such large values for the source densities are also generated for large nuclei, where valence currents from different nucleons couple coherently with the small $\bx$ fields. The large density of sources thus explains why $A_{\rm cl}$ in the effective theory is also of order $1/g$.}\\

The requirement that physical observables do not depend on the separation between static sources and dynamical fields inside the incoming nuclei leads to a renormalization group description of these quantum fluctuations before the collision. In practice, all quantum fluctuations (from either nucleus) with $\bx$ values above those that populate the central rapidity region of the collision, can be factorized into distributions of the color source densities of each nucleus. The renormalization group equation that describes how the color source distributions in each of the nuclei evolves with $\bx$ (to leading logarithmic accuracy) is called the JIMWLK equation~\cite{Balitsky:1995ub,JalilianMarian:1997gr,JalilianMarian:1997dw,Iancu:2000hn,Ferreiro:2001qy}.\\

This factorization of the quantum fluctuations at larger $\bx$ is essential to treat the subsequent evolution of the smaller $x$ quantum fluctuations at central rapidity in the forward lightcone of the collision. The linearized Yang-Mills equations (around the classical background induced by the color sources) that propagate the quantum fluctuations into the forward lightcone at central rapidities have been solved and the spectrum of fluctuations has been computed analytically~\cite{Epelbaum:2013waa}. However, the quantum fluctuations that are propagated into the forward lightcone are unstable and will grow rapidly~\cite{Romatschke:2005pm,Romatschke:2006nk,Romatschke:2005ag,Fukushima:2011nq,Berges:2012cj} . 
The linearized evolution equations thus no longer provide a good approximation of their dynamics and shows pathological behavior at late times~\cite{Dusling:2010rm,Epelbaum:2011pc,Dusling:2012ig}. One therefore needs to partially resum higher order corrections to obtain an improved description. This can be efficiently achieved within a classical-statistical field theory. Here one keeps track of non-linearities by considering the classical evolution of a composite field $A=A_{cl}+a$, where different realizations of the NLO quantum fluctuations $a$ are added to the LO classical field $A_{cl}$. Averaging over the statistical distribution of quantum fluctuations\footnote{The interpretation in terms of a classical-statistical ensemble of course requires the statistical distribution of vacuum fluctuations to be positive semi-definite. This property holds for the Wigner function of the perturbative vacuum.} then allows one to compute observables in a single event to NLO accuracy, while partially including higher order (classical) corrections to ensure convergence~\cite{Dusling:2011rz,Epelbaum:2013waa}.\\ 

In summary, the non-equilibrium dynamics after the nuclear collision is classical in nature at leading (LO) and next-to-leading order (NLO). This is convenient because classical dynamics can be studied numerically with well-established real-time lattice simulation techniques (c.f. sec.~\ref{sec:lattice}). Genuine quantum evolution effects only contribute starting at next-to-next-to-leading order (N$^2$LO)~\cite{Jeon:2013zga}. Unfortunately, there is no satisfactory framework at present to implement such higher order corrections in non-Abelian gauge theories.\\ 

Based on the classical-statistical approach outlined above, we shall now discuss the initial conditions immediately after the collision as well as the LO and NLO dynamics at early times. We will then formulate initial conditions for the evolution at later times. These are employed in sec.~\ref{sec:ExpandingBox} to study the thermalization process at weak coupling.

\subsection{\texorpdfstring{Initial conditions at early times $(\tau \lesssim \OneOverQ)$: The ``Glasma''}{Initial conditions at early times: The ``Glasma''}}
The leading order classical fields $A_{cl}$ as well as the spectrum of next-to-leading order vacuum fluctuations $a$ are known analytically at $\tau=0^+$ immediately after the collision of heavy-nuclei. Adopting the Fock-Schwinger gauge condition ($A_{\tau}=0$), the leading order classical fields are given by~\cite{Kovner:1995ja,Kovner:1995ts}
\begin{align}
\label{C3eq:IC1}
&A_i^{cl}(\xvect)=\alpha_i^{(1)}(\xvect)+\alpha_i^{(2)}(\xvect) \;, \qquad A_{\eta}^{cl}=0\;, \nonumber \\
&E_i^{cl}=0 \;, \quad E_{\eta}^{cl}(\xvect)=ig[\alpha_i^{(1)}(\xvect)\;,\alpha_i^{(2)}(\xvect)]\;,
\end{align}
where $i=1,2$ denotes Lorentz indices of transverse coordinates. The pure gauge configurations $\alpha_i^{(1/2)}(\xvect)$ in eq.~(\ref{C3eq:IC1}) describe the Yang-Mills field outside the lightcone. They are related to the nuclear color charge densities by~\cite{Kovner:1995ja,Kovner:1995ts}
\begin{eqnarray}
\label{C3eq:alpha}
\alpha^{(N)}_i(\xvect)&=&\frac{-i}{g}e^{ig\Lambda^{(N)}(\xvect)}\partial_ie^{-ig\Lambda^{(N)}(\xvect)} \;, \nonumber \\
\partial_i\partial^{i}\Lambda^{(N)}(\xvect)&=&\varrho^{(N)}(\xvect)\;,
\end{eqnarray}
and depend on the transverse coordinates $\xvect$ only.\\

While the collision is instantaneous in the high-energy limit (at $\tau=0$), the boost invariant chromo-electric $(E_\eta^{cl})$ and chromo-magnetic $(B_\eta^{cl})$ color fields in eq.~(\ref{C3eq:IC1}) mediate the color flux between the two nuclei as soon as they establish causal contact with each other ($\tau=0^{+}$). Subsequently the two nuclei separate from each other at nearly the speed of light, and the large $\bx$ partons quickly escape from the mid-rapidity region. The initial small $\bx$ color fields that interact at mid-rapidity decay on a time scale $\tau \sim \OneOverQ$, when a description in terms of on-shell gluons becomes applicable.\\

The (LO) classical dynamics of the decaying Glasma fields has been studied extensively in the literature~\cite{Krasnitz:1998ns,Krasnitz:1999wc,Krasnitz:2000gz,Krasnitz:2001qu,Krasnitz:2002mn,Lappi:2003bi,Lappi:2006hq,Lappi:2006fp,Lappi:2009xa,Lappi:2011ju}. Initially, the longitudinal chromo-electric and chromo-magnetic fields give rise to a negative pressure. This relaxes towards zero rapidly on a time scale $\tau\sim \OneOverQ$ when the transverse pressure begins to decrease according to a $\propto \tau^{-1}$ free streaming behavior. The basic properties of the system at this time $(\tau\sim \OneOverQ)$ are reminiscent of a highly occupied and highly anisotropic quasi-particle system. The corresponding single particle gluon spectrum was computed in~\cite{Krasnitz:2001qu,Krasnitz:2002mn,Lappi:2003bi,Lappi:2009xa,Lappi:2011ju}. \\

While the transition from the initial state of strong color fields to a strongly correlated quasi-particle system already occurs at the leading order classical level, the longitudinal boost invariance of the (LO) classical fields in eq.~(\ref{C3eq:IC1}) is preserved throughout the (LO) classical evolution. This leads to an effectively 2+1 dimensional dynamics~\cite{Krasnitz:1998ns,Lappi:2003bi} making thermalization and isotropization inaccessible at this order. The next-to-leading order CGC result includes quantum fluctuations of the initial state. These fluctuations depend on the longitudinal rapidity and therefore explicitly break the longitudinal boost-invariance of the leading order classical solution. The structure of the solution changes qualitatively requiring one to study the full 3+1 dimensional Yang Mills dynamics.\\

In the NLO resummed prescription, the initial conditions for the classical evolution at $\tau=0^{+}$ are given by~\cite{Dusling:2011rz,Epelbaum:2013waa}
\begin{equation}
{A}^{\rm init.}_\mu (x) = {A}^{cl}_\mu(x)+ \int \dInt\mu_{_K}\;\Big[c_{_K}\,a^{_K}_\mu(x)+c_{_K}^*\,a^{_K*}_{\mu}(x)\Big] \, ,
\label{eq:quantum1}
\end{equation}
and similarly for the canonical conjugate momenta. Here the initial gauge field ${A}^{\rm init.}_\mu$ receives a contribution from the leading order classical solution ${A}_{cl}$ and the next-to-leading order vacuum fluctuations $a_{_K}$, with $K$ collectively denoting the quantum numbers labeling the basis of solutions. The average over the coefficients $c_{_K}$, defines a classical-statistical field theory, namely a classical field theory with a statistical distribution of initial field configurations. In practice this average can be evaluated by a Monte-Carlo procedure, where $c_{_K}$ are Gaussian-distributed complex random numbers~\cite{Dusling:2011rz,Epelbaum:2013waa}.\\

The {\it ab initio} results in eq.~(\ref{eq:quantum1}) are of great significance for understanding non-equilibrium features of the Glasma at very early times $\tau \lesssim \OneOverQ$. In heavy ion collisions, they play an important role in understanding the structure of long range rapidity correlations~\cite{Dumitru:2008wn, Dusling:2009ni} and topological effects such as the Chiral Magnetic Effect~\cite{Fukushima:2008xe,Kharzeev:2009pj,Kharzeev:2010gd,Basar:2010zd}. With regard to thermalization which occurs at relatively late times, the role of these initial conditions is unclear and may depend crucially on the coupling constant.\footnote{This point is discussed further in sec.~\ref{sec:QuoVadis}.}\\ 

Specifically, in the weak coupling limit, quantum fluctuations grow rapidly as $\exp(\sqrt{Q\tau})$ due to plasma instabilities~\cite{Romatschke:2005pm,Romatschke:2006nk,Romatschke:2005ag,Fukushima:2011nq,Berges:2012cj}. They become of the order of the (LO) classical fields on time scales $\tau \sim \OneOverQ \lnSqrOneOverAlpha$~\cite{Berges:2012cj}. In the weak coupling limit, the logarithm becomes large and there is a ``clean'' separation of the time scales $\tau \sim \OneOverQ$ and $\tau \sim \OneOverQ\lnSqrOneOverAlpha$. During this time, the exponential growth of instabilities can readily be observed, while the dynamics of the leading order classical fields remains largely unchanged~\cite{Romatschke:2005pm,Romatschke:2006nk,Romatschke:2005ag,Fukushima:2011nq,Berges:2012cj}.\footnote{Note that refs.~\cite{Romatschke:2005pm,Romatschke:2006nk,Romatschke:2005ag,Fukushima:2011nq,Berges:2012cj} employed a different spectrum of fluctuations at initial time. However, in the limit of very weak coupling, one may argue that the details of the spectrum of fluctuations become less relevant as long as a sufficiently large range of modes is covered. The instability then naturally selects the most unstable modes to exhibit the fastest growth and thus dominate the dynamics at later times.} However, at times $\tau \sim \OneOverQ \lnSqrOneOverAlpha$, the impact of unstable fluctuations on the dynamics of the background field is no longer negligible and the system enters a regime with much slower dynamics~\cite{Romatschke:2005pm,Romatschke:2006nk,Romatschke:2005ag,Fukushima:2011nq,Berges:2012cj}. The longitudinal pressure of the system also becomes manifestly non-zero at times $\tau \sim \OneOverQ \lnSqrOneOverAlpha$. The detailed properties of the system and the amount of pressure generated at these early times may depend on the value of the coupling constant.

\subsection{\texorpdfstring{Initial conditions at late times $(\tau \gtrsim \OneOverQ \lnSqrOneOverAlpha)$: Overoccupied plasma}{Initial conditions at late times: Overoccupied plasma}}
\label{sec:ExpandingIC}
The extremely rapid action of instabilities suggests that the initial conditions at times $\tau \lesssim \OneOverQ$ will lead to a state of the system that looks quite different on a time scale that is parametrically only $\lnSqrOneOverAlpha$ later. In particular, for each element of the ensemble of initial conditions in eq.~(\ref{eq:quantum1}), the quantum fluctuations have grown to be of the order of the classical field by the time $\tau\sim\OneOverQ \lnSqrOneOverAlpha$. Because the initial seed for the quantum fluctuations is statistically distributed, these fluctuations add to the classical field with comparable amplitude and arbitrary phase by the time $\tau\sim \OneOverQ \lnSqrOneOverAlpha$.\footnote{Note that both the classical field and the quantum fluctuations will be redshifted by the expansion over this time scale.} It has been shown for scalar theories, that adding such quantum fluctuations will lead to phase decoherence of the system on relatively short time scales~\cite{Dusling:2010rm}. It is therefore a reasonable expectation that the dynamics will become fluctuation dominated at these later times. Specifically, phase decoherence will mean that the ensemble average of the classical field will no longer be $\langle A \rangle \sim 1/g$. However, one can expect $\langle AA \rangle \sim 1/g^2$. \\

The ensuing overpopulated plasma provides the starting point for different weak coupling thermalization scenarios in kinetic theory.\footnote{Note that the logarithmic factor $\lnSqrOneOverAlpha$ is frequently neglected in the literature on thermalization, where conventionally only powers of the coupling constant are considered in parametric estimates.} Since the occupation numbers are $1 \ll f \ll 1/\alpha_s$, a dual description of the overoccupied non-Abelian plasma in terms of either the classical-statistical framework or the language of kinetic theory applies~\cite{Mueller:2002gd,Jeon:2004dh}. Implicit to this correspondence is the assumption that the plasma consists of strongly correlated quasi-particle excitations (with $\langle A\rangle\sim 0$ and $\langle AA\rangle \sim 1/g^2$) as opposed to a system with the genuinely field like properties of the Glasma at early times. In the CGC picture, the existence of quasi-particle dynamics requires that the initially strong Glasma fields decay into quasi-particle excitations. Based on the previous discussion, we expect that a kinetic description of hard excitations becomes applicable at times $\tau \sim \OneOverQ \lnSqrOneOverAlpha$ when the growth of plasma instabilities saturates.\\

Thus to understand the dynamics of weakly coupled plasmas at later times ($\tau \gtrsim \OneOverQ \lnSqrOneOverAlpha$), it may be sufficient to formulate initial conditions which capture the quasi-particle dynamics that possibly leads to isotropization and thermalization of the system. These initial conditions are sensitive to the value of the coupling constant and uncertainties from higher order contributions in the way they affect the properties of the system at times $\tau_0 \sim \OneOverQ \ln(\alpha_s^{-1})$. Most importantly, the resulting uncertainties will affect the degree of initial anisotropy $\xi_0$ of the single particle distribution in momentum space, and the magnitude of the initial overoccupancy $n_0$ of the plasma. To capture a wide range of different initial conditions, the single particle distribution characterizing the initial conditions for the evolution of the strongly correlated plasma can be modeled as\footnote{The factor two in the denominator is from our choice of convention for normalizing the initial conditions.} 
\begin{equation}
\label{eq:TurbIC}
f(\pt,\pz,\tau_0) = \frac{n_0}{2g^2}\, \Theta\!\left( Q - \sqrt{\pt^2+(\xi_0 \pz)^2}\right)\;.
\end{equation}
This describes the overpopulation of gluon modes, averaged over spin and color degrees of freedom up to the momentum $Q$ -- at the initial time $Q\tau_0\sim\lnSqrOneOverAlpha$ -- after the unstable decay of the initial Glasma fields.\footnote{The momentum scale $Q$ is of comparable magnitude, albeit non-trivially related, to the saturation scale $Q_s$ in the nuclear wavefunctions.}\\

The initial conditions in eq.~(\ref{eq:TurbIC}) provide a rather simple model of the evolving Glasma. However a precise matching to the Glasma is inessential if the non-equilibrium evolution at late times shows a universal attractor, with different initial conditions belonging to the same basin of attraction. The existence of such an attractor solution is implicit to all weak coupling kinetic thermalization scenarios where a certain characteristic behavior is predicted without precise knowledge of the underlying initial conditions. Here we will explicitly verify the existence of a universal attractor by use of classical-statistical lattice simulations for different initial conditions. The wider the set of initial conditions studied, the greater one's confidence that this universal attractor is a unique one relevant for the dynamics of the overpopulated strongly correlated plasma. This will then {\it a posteriori} justify the use of a simplified model for the initial conditions.\\

The initial conditions in eq.~(\ref{eq:TurbIC}) can be implemented in a straightforward way to initialize the gauge field configurations in classical-statistical Yang-Mills simulations. Because the equal-time correlation functions of the (gauge fixed) gauge fields $A_{\mu}^{a}$ and their canonical conjugate momenta $E^{\mu}_a$ can be related to the occupation number defined in a suitable quasi-particle picture (c.f. sec.~\ref{sec:OccupationNumbers}), we simply choose the configurations to reproduce the single particle distribution at initial time. Since the interpretation of quasi-particle excitations requires an additional gauge fixing, we impose the Coulomb type gauge condition $\partial_iA_i+\tau^{-2}\partial_\eta A_\eta=0$ at initial time $\tau_0$. The field configurations are then initialized as a superposition of transversely polarized quasi-particle modes. The gauge fields at initial time $\tau_0$ take the form 
\begin{eqnarray}
\label{eq:quantum2}
A_{\mu}^a(\tau_0,\xvect,\eta)&=&\sum_{\lambda=1,2}\int \frac{\dInt^2 \kvect}{(2\pi)^2}\, \frac{\dInt \nu}{2\pi} \,
\sqrt{f(\kvect,\nu,\tau_0)}\, \\
&&\times\left[c_{\lambda,a}^{\kvect\nu}\, \xi^{(\lambda)\kvect\nu+}_{\mu}(\tau_0)\,
e^{i \kvect\xvect}\,e^{i\nu\eta}
+c.c.\right]\;, \nonumber
\end{eqnarray}
and similarly one finds
\begin{eqnarray}
\label{eq:quantum2E}
E^{\mu}_a(\tau_0,\xvect,\eta)&=&-\tau_0~g^{\mu\nu}\sum_{\lambda=1,2}\int \frac{\dInt^2 \kvect}{(2\pi)^2}\, \frac{\dInt \nu}{2\pi} \,
\sqrt{f(\kvect,\nu,\tau_0)}\,\nonumber \\
&&\times\left[c_{\lambda,a}^{\kvect\nu}\, \dot{\xi}^{(\lambda)\kvect\nu+}_{\nu}(\tau_0)\,
e^{i \kvect\xvect}\,e^{i\nu\eta}
+c.c.\right]\;. \nonumber \\
\end{eqnarray}
for the conjugate momenta. Here $\xi^{(\lambda)\kvect\mu+}_{\mu}(\tau)$ denote the (time dependent) transverse polarization vectors in the non-interacting theory (given in appendix~\ref{app:ModeVectors}) and $c.c.$ denotes complex conjugation. The statistical ensemble is defined by the distribution of the coefficients $c_{\lambda,a}^{\kvect\nu}$, which satisfy the relations
\begin{eqnarray}
\langle c_{\lambda,a}^{\kvect\nu} c_{\lambda',b}^{*\kvect'\nu'}\rangle=\delta_{\lambda\lambda'}\delta_{ab}~(2\pi)^3~\delta^{(2)}(\kvect-\kvect')\delta(\nu-\nu')\;,   \nonumber \\
\end{eqnarray}
whereas $\langle c_{\lambda,a}^{\kvect\nu}c_{\lambda',b}^{\kvect'\nu'}\rangle=\langle c_{\lambda,a}^{*\kvect\nu}c_{\lambda',b}^{*\kvect'\nu'}\rangle=0$. This can be implemented by choosing the coefficients $c_{\lambda,a}^{\kvect\nu}$ as complex Gaussian random numbers in every simulation. \\

Before we proceed to a discussion of the numerical solution of the Yang-Mills equations, we will briefly digress to comment on the range of validity of the classical-statistical method. Specifically, we will address the limitations of 
this method in studying the approach to thermal equilibrium at weak coupling. 

\subsection{Breakdown of classical-statistical dynamics}
There is a restricted class of problems where the dynamics of bosonic quantum fields can be accurately mapped onto
a classical-statistical problem. The most intuitive criteria for this can be formulated in situations where a kinetic description in terms of quasi-particle excitations is applicable~\cite{Mueller:2002gd,Jeon:2004dh}. As noted previously, the system exhibits classical dynamics whenever the typical occupation numbers are much larger than unity. If occupation numbers fall below unity, quantum mechanical processes will
dominate the dynamics. This is clearly seen in a Boltzmann transport framework where classical scattering
processes are sub-leading to quantum mechanical ones for occupation numbers smaller than unity~\cite{Mueller:2002gd,Jeon:2004dh}.\\

It is also possible to formulate more general criteria (which do not rely on a quasi particle picture) in the Schwinger-Keldysh formalism of non-equilibrium quantum field theory,~\cite{Berges:2004yj,Jeon:2013zga,Aarts:2001yn,Berges:2007ym}. This ``classicality condition" is met whenever anti-commutator expectation values for typical bosonic field modes are much larger than the corresponding commutators \cite{Aarts:2001yn,Berges:2007ym}. Stated differently, this concerns the large field or large occupancy limit, which is relevant for important phenomena such as non-equilibrium instabilities or wave turbulence encountered in our study. 
The classicality condition has been discussed in detail in the context of scalar quantum field dynamics \cite{Aarts:2001yn,Berges:2007ym,Arrizabalaga:2004iw,Berges:2008wm}. The coupling to fermions~\cite{Berges:2010zv,Berges:2013oba} and extensions to non-Abelian gauge theories follow along the same lines.\\

We emphasize that the condition for a system to exhibit classical dynamics is in general time dependent. In particular, the approach to complete thermal equilibrium is not accessible within the classical-statistical framework. As a consequence of the Rayleigh-Jeans divergence, the classical thermal state is only well defined in conjunction with an ultraviolet cutoff.\footnote{This cutoff can for instance be implemented by lattice regularization.} In contrast, thermal equilibrium is a genuine quantum state which cannot be reached within classical-statistical field theory. Nevertheless, the classical-statistical regime may extend over large times (even parametrically large times) such that interesting features of the non-equilibrium evolution are observable with classical-statistical dynamics.\\

The time scale $t_{\rm Quantum}$ for entering the quantum regime depends on the properties of the initial state as well as the dynamics in the classical regime. In weak coupling frameworks, this time scale is typically inversely proportional to a power of the coupling constant $t_{\rm Quantum}\sim \OneOverQ \alpha_s^{-q}$, for initially overoccupied systems. For instance, in the bottom up thermalization scenario~\cite{Baier:2000sb} one finds $t_{\rm Quantum}\sim \OneOverQ \alpha_s^{-3/2}$. The range of validity of classical-statistical techniques is thus naturally confined to weak coupling. As the coupling is increased, its range of validity shrinks rapidly. The use of classical-statistical methods at large couplings therefore requires great care since genuine quantum effects may dominate the dynamics already at rather early times.\\

In our view, the appropriate strategy from a weak coupling perspective is to perform classical-statistical simulations at very weak coupling where the framework is robust. Specifically, the strategy will be to extract the parametric dependence of observables on the coupling, and then extrapolate results to the larger couplings of interest. Conversely, performing simulations at large couplings is bedeviled on two fronts. Firstly, the classical-statistical simulation is breaking down early and uncontrollably. Further, the contamination of observables by ultraviolet quantum modes is large. It is an interesting open question whether weak coupling results extrapolated to large couplings can be matched to results from strong coupling frameworks. 

\section{Classical-statistical lattice gauge theory }
\label{sec:lattice}
We shall now describe the framework to perform classical-statistical real-time simulations of $SU(2)$ Yang-Mills equations.\footnote{The extension to SU(3) is straightforward if cumbersome. Previous studies have shown that there is no qualitative difference in the dynamics if the gauge group is changed from SU(2) to SU(3) (see e.g.~\cite{Berges:2008zt}).} For the longitudinally expanding plasma, we employ the Kogut-Susskind lattice Hamiltonian in Fock-Schwinger ($A_{\tau}=0$) gauge to solve Hamilton's equations for the fields and their momentum conjugate variables as a function of proper time on a spatial lattice. The lattice discretization proceeds along similar lines as in standard vacuum or thermal equilibrium lattice QCD. The spatial points ($\xvect,\eta$) are defined on a grid of size $\Nt\times \Nt \times N_{\eta}$, with the lattice spacing $a_\bot$ in the transverse direction and $a_\eta$ in the longitudinal direction. We employ periodic boundary conditions in the three spatial directions and we will collectively label the spatial and temporal coordinates as $x=(\tau,\xvect,\eta)$.

\subsection{Discretization}
\begin{figure}[t!]						
\includegraphics[width=0.5\textwidth]{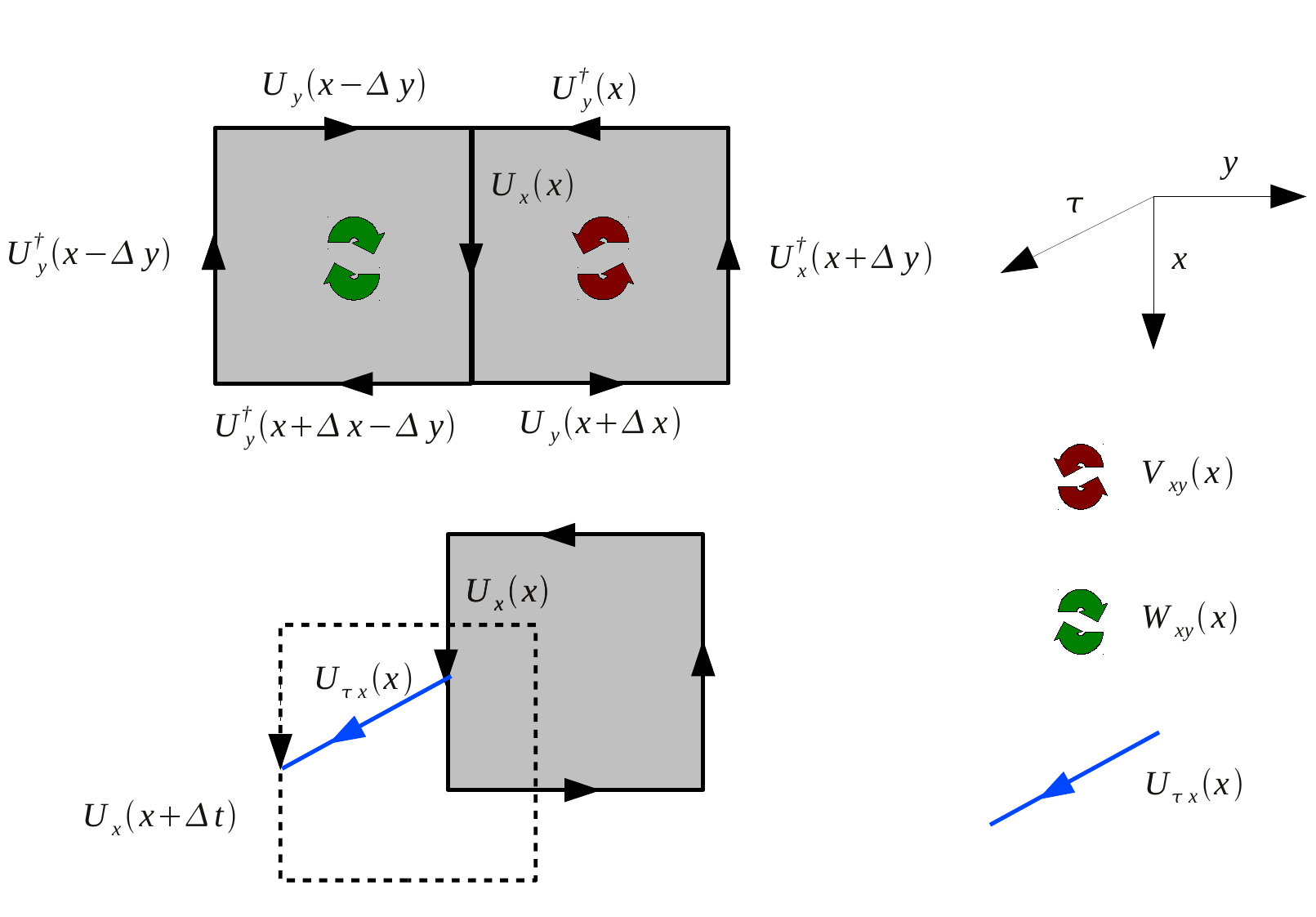}	
\caption{\label{fig:LatticeIllustration} (color online) Illustration of the lattice link and plaquette variables.}	
\end{figure}

In the lattice formulation, the continuum gauge fields $A_{\mu}^a(x)$ are represented in terms of the gauge link variables
\begin{eqnarray}
\label{lat:Link}
U_{i}(x)&=&\exp[iga_\bot A_i^a(x+\ihat/2)\Gamma^a]\;, \nonumber \\
U_{\eta}(x)&=&\exp[iga_\eta A_{\eta}^a(x+\hat{\eta}/2)\Gamma^a]\;,
\end{eqnarray}
where $\Gamma^a=\sigma^{a}/2$ are the generators of the $SU(2)$ Lie algebra in the fundamental representation and the symbol $\hat{\mu}=\hat{x}^1,\hat{x}^2,\hat{\eta}$ denotes the neighboring lattice site in the $\mu$ direction. The gauge link variables can be intuitively understood as approximating the path-ordered Wilson line $\mathcal{P} \exp[ig \int_{x}^{y} A]$, connecting adjacent lattice sites $x$ and $y$ along a straight line path. In particular, the gauge link variables transform as the Wilson lines according to
\begin{eqnarray}
\label{lat:LinkTrafo}
U^{(G)}_{\mu}(x)&=&G(x)U_{\mu}(x)G^{\dagger}(x+\hat{\mu})\;, \nonumber \\ U^{\dagger(G)}_{\mu}(x)&=&G(x+\hat{\mu})U_{\mu}(x)G^{\dagger}(x)\;, 
\end{eqnarray}
under time-independent local gauge transformations $G(x) \in SU(2)$. The classical evolution equations for the gauge link variables can be formulated in terms of the spatial plaquette variables $V_{\mu\nu}(x)$ and $W_{\mu\nu}(x)$, conventionally defined as 
\begin{eqnarray}
\label{lat:Plaquette}
V_{\mu\nu}(x)&=&U_{\mu}(x)U_{\nu}(x+\hat{\mu})U^{\dagger}_{\mu}(x+\hat{\nu})U^{\dagger}_{\nu}(x)\;, \\
W_{\mu\nu}(x)&=&U_{\mu}(x)U^{\dagger}_{\nu}(x+\hat{\mu}-\hat{\nu})U^{\dagger}_{\mu}(x-\hat{\nu})U_{\nu}(x-\hat{\nu})\;, \nonumber
\end{eqnarray}
where $W_{\mu\nu}(x)=U_{\nu}^{\dagger}(x-\hat{\nu})V_{\mu\nu}^{\dagger}(x-\hat{\nu})U_{\nu}(x-\hat{\nu})$. The plaquette variables can be related to the continuum expression of the non-Abelian field strength tensor $\mathcal{F}_{\mu\nu}^{a}(x)$ by virtue of the expansion 
\begin{eqnarray}
\label{lat:PlaquetteContForm1}
V_{\mu\nu}(x)&=&\exp[~~~ig a_{\mu} a_{\nu} \mathcal{F}_{\mu\nu}^a(x+\hat{\mu}/2+\hat{\nu}/2) \Gamma^a+\mathcal{O}(ga^3)] \nonumber \\
\\
W_{\mu\nu}(x)&=&\exp[-ig a_{\mu} a_{\nu} \mathcal{F}_{\mu\nu}^a(x+\hat{\mu}/2-\hat{\nu}/2) \Gamma^a+\mathcal{O}(ga^3)] \nonumber \\ 
\label{lat:PlaquetteContForm2}
\end{eqnarray}
for sufficiently small lattice spacings $a_\bot,a_\eta$. We note that the plaquette variables are defined in the center of the respective Wilson loop as indicated in fig.~\ref{fig:LatticeIllustration}. In this sense, the phase of the plaquette $V_{\mu\nu}(x)$ and $W_{\mu\nu}(x)$ corresponds to the magnetic flux through the area spanned by the Wilson loop to leading order in the lattice spacing. \\

In order to construct the lattice equations of motion, we also discretize the time direction while keeping the temporal lattice spacing $a_\tau\ll (a_\bot,\tau,\tau a_\eta)$ sufficiently close to the continuum limit. We can then introduce the time-like plaquette variables $U_{\tau\mu}(x)$
\begin{eqnarray}
\label{lat:TimeLinkFull}
U_{\tau\mu}(x)=U_{\tau}(x)U_{\mu}(x+\hat{\tau})U_{\tau}^{\dagger}(x+\hat{\mu})U_{\mu}^{\dagger}(x)\;,
\end{eqnarray}
where $\hat{\tau}$ denotes the neighboring lattice site in the temporal direction. In Fock-Schwinger $(A_{\tau}=0)$ gauge, where the temporal links are trivial $(U_{\tau}=\dblone)$, the time-like plaquette variables have the simple structure
\begin{eqnarray}
\label{lat:TimeLink}
U_{\tau\mu}(x)=U_{\mu}(x+\hat{\tau})U_{\mu}^{\dagger}(x)\;.
\end{eqnarray}
Similarly to eqns.~(\ref{lat:PlaquetteContForm1}) and (\ref{lat:PlaquetteContForm2}), the temporal plaquette variables can be related to the (dimensionless) electric field variables $\tilde{E}^{\mu}_a(x)$ on the lattice as 
\begin{eqnarray}
\label{lat:EFieldTimeLink}
\tilde{E}^i_a(x)&=&-2\frac{\tau}{a_\tau}\text{tr}\left[i\Gamma^a U_{\tau i}(y))\right]\;, \nonumber \\ \tilde{E}^{\eta}_a(x)&=&-2\frac{a_\bot^2}{\tau a_\tau a_\eta}\text{tr}\left[i\Gamma^a U_{\tau \eta}(y))\right]\;,  
\end{eqnarray}
for the transverse components $U_{\tau i}(x)$ and longitudinal components $U_{\tau\eta}(x)$ respectively. In the limit of small temporal lattice spacing $(a_{\tau}\to0)$, they are then related to the corresponding continuum fields $E^{i}_a(x)$ by
\begin{eqnarray}
\label{lat:EField}
\tilde{E}^i_a(x)&=&ga_\bot E^{i}_a(x+\ihat/2+\hat{\tau}/2)\;,\nonumber \\  \tilde{E}^{\eta}_a(x)&=&ga_\bot^2E^{\eta}_a(x+\hat{\eta}/2+\hat{\tau}/2)\;.
\end{eqnarray}
Similarly, eq.~(\ref{lat:EFieldTimeLink}) can be inverted to express the time-like plaquette variables in terms of the electric field variables as\footnote{Strictly speaking, this expression is only accurate to next-to-leading order in the temporal lattice spacing $\mathcal{O}(a_\tau^2)$. However the stability of the numerical solution typically requires very small values of $a_\tau\ll (a_\bot,\tau,\tau a_\eta)$ such that in practice the inversion is exact to machine accuracy.}
\begin{eqnarray}
\label{lat:TimeLinkEField}
U_{\tau i}(x)&=&\exp\left[i\frac{a_\tau}{\tau}\tilde{E}^i_a(x)\Gamma^a\right]\;, \nonumber \\ U_{\tau\eta}(x)&=&\exp\left[i\frac{a_\tau\tau a_{\eta}}{a_\bot^2}\tilde{E}^\eta_a(x)\Gamma^a\right]\;.
\end{eqnarray} 
In analogy to the spatial plaquette variables $V_{\mu\nu}(x)$ and $W_{\mu\nu}(x$), the time-like plaquette variables $U_{\tau\mu}(x)$ are defined at half-integer time-steps $x+\hat{\mu}/2+\hat{\tau}/2$, as indicated by blue arrows in fig.~\ref{fig:LatticeIllustration}. We will see shortly that this choice corresponds to the leap-frog discretization scheme employed in the discretized version of the evolution equations. \\

An important feature of the plaquette variables is the fact that these objects transform covariantly under time-independent gauge transformations,
\begin{eqnarray}
\label{lat:PlaquetteTrafo}
V_{\mu\nu}^{(G)}(x)&=&G(x)V_{\mu\nu}(x)G^{\dagger}(x)\;, \nonumber \\
W_{\mu\nu}^{(G)}(x)&=&G(x)W_{\mu\nu}(x)G^{\dagger}(x)\;, \nonumber \\
U_{\tau\mu}^{(G)}(x)&=&G(x)U_{\tau\mu}(x)G^{\dagger}(x)\;.
\end{eqnarray}
This property implies that the trace of a plaquette variable or in general any closed Wilson loop is gauge invariant. It can be used to construct gauge invariant observables as we will discuss below.\\

\subsection{Evolution equations}
The lattice evolution equations can be derived from the lattice Hamiltonian. The evolution equations for the electric field variables can be expressed as 
\begin{eqnarray}
\label{lat:EFieldEOM}
\tilde{E}^{i}_{a}(x)-\tilde{E}^{i}_{a}(x-\hat{\tau}) &=& 2\frac{a_\tau \tau}{a_\bot^2}\sum_{j\neq i}\text{tr}\left[i\Gamma^a(V_{ij}(x)+W_{ij}(x))\right] \nonumber \\
&&+ 2\frac{a_\tau}{\tau a_\eta^2}\text{tr}\left[i\Gamma^a(V_{i\eta}(x)+W_{i\eta}(x))\right] \;, \nonumber \\
\tilde{E}^{\eta}_{a}(x)-\tilde{E}^{\eta}_{a}(x-\hat{\tau})&=&2 \frac{a_\tau}{\tau a_\eta}\sum_{i}\text{tr}\left[i\Gamma^a(V_{\eta i}(x)+W_{\eta i}(x))\right]\;. \nonumber \\
\end{eqnarray}
Similarly eqns.~(\ref{lat:TimeLink}) and (\ref{lat:TimeLinkEField}) can be used to construct the evolution equation for the link variables as
\begin{eqnarray}
\label{lat:LinkEOM}
U_{i}(x+\hat{\tau})&=&\exp\left[i\frac{a_\tau}{\tau}\tilde{E}^i_a(x)\Gamma^a\right]U_{i}(x)\;, \nonumber \\ 
U_{\eta}(x+\hat{\tau})&=&\exp\left[i\frac{a_\tau\tau a_{\eta}}{a_\bot^2}\tilde{E}^\eta_a(x)\Gamma^a\right]U_{\eta}(x)\;.
\end{eqnarray}
We note that the electric field variables are defined at half-integer time steps according to eq.~(\ref{lat:EField}), such that the gauge force on the right hand side of eq.~(\ref{lat:EFieldEOM}) is effectively calculated at the mid-point of the considered time interval. Similarly the time-like plaquettes $U_{\tau\mu}(x)$ on the right hand side of eq.~(\ref{lat:LinkEOM}) are defined at the mid-point $x+\hat{\mu}/2+\hat{\tau}/2$ of the considered time interval. This realization of the evolution equation corresponds to the leap-frog discretization scheme.\\

The discretized version of the Gauss law constraint takes the form
\begin{eqnarray}
\label{lat:GaussLaw}
&&\left(\frac{a_\bot^2}{a_\tau\tau a_\eta}\right)\text{tr}[i\Gamma^a (U_{\tau \eta}(x)-D_{\tau\eta}(x-\hat{\eta}))] \\
&+&\left(\frac{\tau a_{\eta}}{a_\tau}\right)\sum_{i}\text{tr}[i\Gamma^a (U_{\tau i}(x)-D_{\tau i}(x-\ihat))]=0\;, \nonumber
\end{eqnarray}
and needs to be satisfied separately for all color components $a$ at each position $x$ in spacetime. Here we denote
\begin{eqnarray}
D_{\tau\mu}(x-\hat{\mu})=U_{\mu}^{\dagger}(x-\hat{\mu})U_{\tau \mu}(x-\hat{\mu})U_{\mu}(x-\hat{\mu})\;. 
\end{eqnarray}
Since the current on the left hand side of eq.~(\ref{lat:GaussLaw}) is conserved by the equations of motion in the continuum limit, the Gauss law constraint is a non-dynamical constraint that physically meaningful initial conditions have to satisfy. However, during the course of numerical lattice simulations, eq.~(\ref{lat:GaussLaw}) will be violated due to rounding and discretization errors. In practice, the Gauss law constraint is implemented at the initial time and then monitored throughout the subsequent time evolution to ensure that discretization errors are sufficiently small.\\

\subsubsection{Minkowski coordinates}
In addition to simulations for a longitudinally expanding plasma, we will also discuss in sec.~\ref{sec:StaticBox} classical-statistical simulations for a ``static box''~\cite{Berges:2007re,Berges:2008mr,Berges:2008zt}. In this case, it is convenient to employ the Minkowski metric. The lattice setup is similar to that for the expanding plasma. We perform a similar discretization of the spatial coordinates $(i=1,2,3)$ with lattice spacing $a$ and periodic boundary conditions and we employ the temporal axial gauge $(A_t=0)$ instead of Fock-Schwinger gauge $(A_\tau=0)$. The gauge links are related to the continuum fields as 
\begin{eqnarray}
\label{lat:LinkMinkowski}
U_i(x)&=&\exp[iga~A_i^a(x+\ihat/2) \Gamma^a]\;,
\end{eqnarray}
while for electric field variables on the lattice one finds
\begin{eqnarray}
\label{lat:EFieldMinkowski}
\tilde{E}^{i}_a(x)&=&ga^2 E^{i}_a(x+\ihat/2+\hat{t}/2)\;,
\end{eqnarray}
and
\begin{eqnarray}
U_{t i}(x)=\exp\left[i\frac{a_t}{a}\tilde{E}^i_a(x)\Gamma^a\right]\;,
\end{eqnarray}
for all spatial Lorentz indices $(i=1,2,3)$.\\

The lattice evolution equations in Minkowski coordinates can be obtained in analogy to the longitudinally expanding case and take the form
\begin{eqnarray}
\label{lat:EFieldEOMMinkowksi}
\tilde{E}^{i}_{a}(x)-\tilde{E}^{i}_{a}(x-\hat{t}) &=& 2\frac{a_t}{a}\sum_{j\neq i}\text{tr}\left[i\Gamma^a(V_{ij}(x)+W_{ij}(x))\right]\;, \nonumber \\
\end{eqnarray}
and 
\begin{eqnarray}
\label{lat:LinkEOMMinkowksi}
U_{i}(x+\hat{t})&=&\exp\left[i\frac{a_t}{a}\tilde{E}^i_a(x)\Gamma^a\right]U_{i}(x)\;,
\end{eqnarray}
while the Gauss law constraint in Minkowski coordinates reads
\begin{eqnarray}
\label{lat:GaussLawMinkowski}
\left(\frac{a}{a_t}\right)\sum_{i}\text{tr}[i\Gamma^a (U_{t i}(x)-D_{t i}(x-\ihat))]=0\;,
\end{eqnarray}
with
\begin{eqnarray}
D_{ti}(x-\ihat)=U_{i}^{\dagger}(x-\ihat)U_{t i}(x-\ihat)U_{i}(x-\ihat)\;,
\end{eqnarray}
in analogy to the discussion of the longitudinally expanding case.\\

We note that these results can easily be recovered by the replacement $a_\bot\to a,~\tau\to a,~a_\eta\to1,~a_{\tau}\to a_t$ in the evolution equations (\ref{lat:EFieldEOM}) and (\ref{lat:LinkEOM}) in $(\tau,\eta)$ coordinates.\footnote{ A reinterpretation of the longitudinal rapidity coordinate $\eta$ as the Minkowski coordinate $z$ and the proper time coordinate $\tau$ as the Minkowski time $t$ is implied. Note also that the relation between continuum and lattice variables in eqns.~(\ref{lat:Link},\ref{lat:EField}) changes to the expressions in eqns.~(\ref{lat:LinkMinkowski},\ref{lat:EFieldMinkowski}).} This is particularly convenient for tests of the numerical implementation. 

\subsection{Observables}
The non-equilibrium time evolution of the expectation values of suitable observables $\mathcal{O}(x)$ are evaluated as an ensemble average over a statistical distribution of initial values formally expressed as 
\begin{eqnarray}
\label{IntroEq:CSMonteCarloInt}
\langle \mathcal{O}(x) \rangle = \int \mathcal{D}A_0~\mathcal{D}E_0~W_0[A_0,E_0]~\mathcal{O}[A^{0}_{\text{cl}},E^{0}_{\text{cl}};~x]\;. \nonumber \\
\end{eqnarray}
Here $W_0[A_0,E_0]$ denotes the phase-space density of initial conditions. The functional integral extends over all possible field configurations at initial time $\tau_0$. The notation $\mathcal{O}[A^{0}_{\text{cl}},E^{0}_{\text{cl}};~x]$ implies that the observable of interest is to be evaluated as a functional of the classical field solutions $A^{0}_{\text{cl}}$ and $E^{0}_{\text{cl}}$ at the point $x$ in spacetime.\\

While the discretization on a spacetime lattice renders the integration in eq.~(\ref{IntroEq:CSMonteCarloInt}) finite dimensional, the remaining high dimensional integral is evaluated by the following Monte Carlo procedure.
We first generate an ensemble of initial conditions according to the initial phase space density $W_0[A_0,E_0]$. In practice, this corresponds to generating different sets of Gaussian random numbers which enter the initial field configurations in eqns.~(\ref{eq:quantum2},\ref{eq:quantum2E}). For each configuration, the evolution equations (\ref{lat:EFieldEOM}) and (\ref{lat:LinkEOM}) are then solved numerically by a stepwise update. The electric field variables at the next time step are calculated from the previous ones by use of eq.~(\ref{lat:EFieldEOM}). Subsequently, the gauge links at the next point in time are computed by solving eq.~(\ref{lat:LinkEOM}).\footnote{The matrix valued exponential in eq.~(\ref{lat:LinkEOM}) can be calculated explicitly for the $SU(2)$ gauge group according to $\exp[i\alpha^a \Gamma^a]=\cos(a/2)\dblone+2 i~{\sin(a/2) \over a}~\alpha^a\Gamma^a$, where $a=\sqrt{\alpha_a\alpha^a}$.} By iterating this update procedure multiple times, one obtains the solution of the classical field equations at any time of interest. The classical-statistical expectation value of any observable can then be calculated by evaluating the observable separately for each configuration and subsequently averaging over the ensemble of initial conditions.

\subsubsection{Energy-Momentum Tensor}
A central gauge invariant quantity that can be computed using the classical-statistical method outlined above is the energy-momentum tensor~\cite{Romatschke:2006nk}
\begin{eqnarray}
T^{\mu\nu}(x)&=&-g^{\nu\alpha}(x)\mathcal{F}^{\mu\delta}_a(x)\mathcal{F}_{\alpha\delta}^{a}(x) \nonumber \\
&&+\frac{1}{4}g^{\mu\nu}(x)\mathcal{F}^{\gamma\delta}_{a}(x)\mathcal{F}_{\gamma\delta}^{a}(x) \;.
\end{eqnarray}
In particular, we will be interested in the diagonal components of the energy-momentum tensor. These can be associated with the energy density $\epsilon$ and the longitudinal and transverse pressure densities $P_L$ and $P_T$. In comoving coordinates, they are expressed as~\cite{Romatschke:2006nk}
\begin{eqnarray}
\epsilon(x)&=&\langle T^{\tau}_{~\tau}(x)\rangle\;, \qquad P_L(x)=-\langle T^{\eta}_{~\eta}(x)\rangle\;, \nonumber \\
 && P_T(x)=-\frac{1}{2}\langle T^{x}_{~x}(x)+T^{y}_{~y}(x)\rangle\;.
\end{eqnarray}
The above definitions are normalized such that for locally isotropic systems one obtains the relation $P_L=P_T=\epsilon/3$, whereas for anisotropic systems the ratio $P_L/P_T$ can be used to quantify the bulk anisotropy of the plasma. The corresponding lattice expressions are written compactly in terms of the electric and magnetic components as\footnote{Note that all expressions are given in lattice units. The conversion to physical units is achieved by multiplication with appropriate powers of the dimensionless factor $Qa_{\bot}$ and subsequently relating the momentum scale $Q$ to the relevant physical momentum scale.}
\begin{eqnarray}
\label{lat:EnergyPressureEB} 
(g^2 a_\bot^4)~~~\epsilon(x)&=&\frac{1}{2}\left[ \mathcal{B}_{\eta}^2(x) +\mathcal{E}_{\eta}^2(x) + \mathcal{B}_T^2(x) + \mathcal{E}_T^2(x)\right]\;, \nonumber \\
(g^2 a_\bot^4)~P_T(x)&=& \frac{1}{2}\left[\mathcal{B}_{\eta}^2(x) + \mathcal{E}_{\eta}^2(x)\right]  \;, \\
(g^2 a_\bot^4)~P_L(x)&=&  \frac{1}{2}\left[\mathcal{B}_T^2(x) + \mathcal{E}_T^2(x) -\mathcal{B}_{\eta}^2(x) - \mathcal{E}_{\eta}^2(x)\right]\;, \nonumber
\end{eqnarray}
where the individual electric and magnetic components can be obtained as 
\begin{eqnarray}
\mathcal{B}_{\eta}^2(x)&=&4~\langle\text{tr}[\dblone-V_{xy}(x)]\rangle\;, \qquad \mathcal{E}_{\eta}^2(x)=\sum_{a=1}^{N_c^2-1} \langle[\tilde{E}^{\eta}_a(x)]^2\rangle\;, \nonumber \\
\mathcal{B}_{T}^2(x)&=&4\left(\frac{a_\bot^2}{\tau^2a_{\eta}^2}\right) \sum_{i} \langle\text{tr}[\dblone-V_{i\eta}(x)]\rangle\;, \nonumber \\
\mathcal{E}_{T}^2(x)&=&\left(\frac{a_\bot^2}{\tau^2}\right)\sum_{i}\sum_{a=1}^{N_c^2-1} \langle[\tilde{E}^{i}_a(x)]^2\rangle\;. 
\end{eqnarray}
Since the different plaquette variables are formally defined at different half-integer positions on the spacetime lattice, a higher order accurate expression can be obtained by an interpolation of the result with the neighboring plaquette variables. However, since we will primarily be interested in volume averages of the energy density $\epsilon(\tau)$ according to
\begin{eqnarray}
\label{lat:TotalEnergy}
\epsilon(\tau)=\frac{1}{\Nt^2 N_\eta} \sum_{\xvect,~\eta} \epsilon(x)\;, 
\end{eqnarray}
and similarly for the longitudinal and transverse pressures $P_{L/T}$, there is no need to apply this procedure for the purposes of this study. 

\subsubsection{Hard scales}
\label{sec:HardScaleDef}
Besides the energy-momentum tensor, additional gauge invariant observables can be constructed by considering higher dimensional operators such as the covariant derivatives of the field strength tensor~\cite{Kurkela:2012hp}
\begin{eqnarray}
\label{lat:DFDef}
\mathcal{H}^{\mu}_{~\mu}(\tau)=4\int \frac{\dInt^2\xvect}{V_\bot} \frac{\dInt\eta}{L_{\eta}}~ D_{\alpha}^{ab}(x)\mathcal{F}^{\alpha\mu}_{b}(x)~D^{\beta}_{ac}(x)\mathcal{F}_{\beta\mu}^{c}(x)\;, \nonumber \\
\end{eqnarray}
(no summation over $\mu$) where summation over \textit{spatial} Lorentz indices $\alpha,\beta=x,y,\eta$ and color indices $a,b,c=1,...,N_c^2-1$ is implied.\footnote{By virtue of the equations of motion one can equivalently express the right hand side of eq.~(\ref{lat:DFDef}) in terms of time derivatives of the electric fields.} The corresponding expressions on the lattice are
\begin{align}
(g^2 a_\bot^6)&~\mathcal{H}^x_{~x}(\tau)=\frac{16}{\Nt^2 N_\eta}
\sum_{\xvect,\eta}\sum_{a=1}^{N_c^2-1} \Big[\text{tr}[i\Gamma^a(V_{xy}(x) \\
&+W_{xy}(x))]+\frac{a_\bot^2}{\tau^2 a_\eta^2}\text{tr}[i\Gamma^a(V_{x\eta}(x)+W_{x\eta}(x))]\Big]^2\;, \nonumber \\
(g^2 a_\bot^6)&~\mathcal{H}^y_{~y}(\tau)=\frac{16}{\Nt^2 N_\eta} \sum_{\xvect,\eta}\sum_{a=1}^{N_c^2-1}\Big[\text{tr}[i\Gamma^a(V_{yx}(x) \\
&+W_{yx}(x))]+\frac{a_\bot^2}{\tau^2 a_\eta^2}\text{tr}[i\Gamma^a(V_{y\eta}(x)+W_{y\eta}(x))]\Big]^2\;, \nonumber \\ 
(g^2 a_\bot^6)&~\mathcal{H}^\eta_{~\eta}(\tau)=\frac{16}{\Nt^2 N_\eta} \sum_{\xvect,\eta}\frac{a_\bot^2}{\tau^2a_\eta^2}\sum_{a=1}^{N_c^2-1}\Big[\text{tr}[i\Gamma^a(V_{\eta x}(x) \nonumber \\
&+W_{\eta x}(x))]+\text{tr}[i\Gamma^a(V_{\eta y}(x)+W_{\eta y}(x))]\Big]^2\;. 
\end{align}
We will be interested in the quantity
\begin{eqnarray}
\Lambda^2(\tau)&=& \frac{\langle\mathcal{H}^{x}_{~x}(\tau)\rangle+\langle\mathcal{H}^{y}_{~y}(\tau)\rangle+\langle\mathcal{H}^{\eta}_{~\eta}(\tau)\rangle}{\epsilon(\tau)}\;,
\end{eqnarray}
as well as $\Lambda_L^2$ and $\Lambda_T^2$, which are defined as the longitudinal and transverse projections of $\mathcal{H}^{\mu}_{\mu}$ according to
\begin{eqnarray}
\label{lat:LambdaDef}
\Lambda_T^2(\tau)&=& \frac{\langle\mathcal{H}^{\eta}_{~\eta}(\tau)\rangle}{\epsilon(\tau)}\;,\\
\Lambda_L^2(\tau)&=& \frac{\langle\mathcal{H}^{x}_{~x}(\tau)\rangle+\langle\mathcal{H}^{y}_{~y}(\tau)\rangle-\langle\mathcal{H}^{\eta}_{~\eta}(\tau)\rangle}{\epsilon(\tau)}\;, \nonumber
\end{eqnarray}
such that $\Lambda^2=2\Lambda_T^2+\Lambda_L^2$.\\ 

The observables $\Lambda_L$ and $\Lambda_T$ are the characteristic longitudinal and transverse momentum scales of hard excitations. They contain additional information about the evolution of the system beyond what is contained in the energy-momentum tensor. While this interpretation follows immediately from dimensional analysis, it is nevertheless insightful to evaluate the perturbative expressions for $\Lambda_T^2$ and $\Lambda_L^2$. Considering only the Abelian part of the field strength tensor, one obtains 
\begin{eqnarray}
\label{lat:LambdaApprox}
\Lambda_{T}^2(\tau)&\simeq&\frac{\int \dInt^2\pvect~\dInt\pz~2\pt^2~\omega_p~f(\pvect,\pz,\tau)}{\int \dInt^2\pvect~\dInt\pz~\omega_p~f(\pvect,\pz,\tau)}\;, \nonumber \\
\Lambda_{L}^2(\tau)&\simeq&\frac{\int \dInt^2\pvect~\dInt\pz~4\pz^2~\omega_p~f(\pvect,\pz,\tau)}{\int \dInt^2\pvect~d\pz~\omega_p~f(\pvect,\pz,\tau)}\;,
\end{eqnarray}
where $f(\pvect,\pz,\tau)$ denotes the single particle gluon distribution as a function of longitudinal and transverse momenta, and $\omega_p\simeq \pt$ is the relativistic quasi-particle energy in the limit $\pt\gg\pz$. The details of this calculation are presented in appendix~\ref{app:HardScale}.

\subsubsection{Occupation numbers}
\label{sec:OccupationNumbers}
It is also useful to consider a specific set of gauge-dependent quantities in studying the non-equilibrium dynamics of weakly coupled plasmas. However, to extract meaningful quantities, we need to fix the residual gauge freedom (in $A^\tau=0$ gauge) to perform time independent gauge transformations. We do so by implementing the Coulomb type gauge condition
\begin{eqnarray}
\label{lat:CG}
\tau^{-2}\partial_{\eta}A_{\eta}(x)+\sum_{i}\partial_iA_i(x)=0 
\end{eqnarray}
independently at each time $\tau$, where we extract gauge-dependent observables. This can be achieved by use of standard lattice gauge-fixing techniques (see e.g.~\cite{Cucchieri:1995pn}), as discussed in appendix~\ref{app:GaugeFix}.\\

Since the gluon distribution function $f(\pvect,\pz,\tau)$, has a direct analog in kinetic theory, it is particularly useful to establish a direct comparison between the different methods. It can be extracted from equal time (two point) correlation functions in Coulomb gauge and different definitions have been employed in the literature~\cite{Berges:2012ev,Kurkela:2012hp,Schlichting:2012es}. Here we use the Fock state projection, 
\begin{eqnarray}
\label{lat:ParticleNumber}
&&f(\pvect,\pz,\tau)=\frac{\tau^2}{N_g V_{\bot}L_{\eta}}\sum_{a=1}^{N_c^2-1}\sum_{\lambda=1,2} \\
&&\Big<\Big|~g^{\mu\nu} \Big[ \Big(\xi^{(\lambda){\pvect\nu+}}_{\mu}(\tau)\Big)^*
\stackrel{\longleftrightarrow}{\partial_{\tau}} 
A^{a}_{\nu}(\tau,\pvect,\nu)
\Big]\Big|^2\Big>_{\text{Coul.~Gauge}}\;, \nonumber
\end{eqnarray}
where the index $\lambda=1,2$ counts the two transverse polarizations and $N_g=2(N_c^2-1)$ is the number of transversely polarized gluon degrees of freedom. The symbol $\xi^{(\lambda){\pvect\nu+}}_{\mu}(\tau)$ denotes the two time dependent transverse polarization vectors in the free theory. The explicit form for the expanding system in Fock-Schwinger gauge is derived in appendix~\ref{app:ModeVectors}. To evaluate eq.~(\ref{lat:ParticleNumber}) the gauge field $A^{a}_{\nu}(\tau,\pvect,\nu)$ is computed from the plaquette variables by inversion of eq.~(\ref{lat:Link}) and a subsequent fast Fourier transformation to obtain the result in momentum space. Similarly, we obtain the time derivative of the gauge field from the fast Fourier transform of the electric field variables.\\

For the longitudinally expanding system, the longitudinal momentum $\pz$ in eq.~(\ref{lat:ParticleNumber}) is identified as $\pz=\nu/\tau$ from the kinetic term in the field equations. The definition in eq.~(\ref{lat:ParticleNumber}) is such that, in the absence of interactions, $f(\pvect,\pz,\tau)$ is (up to this redshift of longitudinal momenta) exactly conserved by the equations of motion. We also verified this explicitly by performing simulations of the non-interacting theory.\footnote{This can be achieved by initializing a single non-vanishing color component $(a=1)$ in our simulations, whereas all other components vanish identically. In this case one recovers the dynamics of a (compact) $U(1)$ gauge theory.}

\section{Non-expanding non-Abelian plasma}
\label{sec:StaticBox}
We will first address the problem of thermalization from a more general perspective by looking at a class of systems that are simpler than those in heavy ion collisions, but nevertheless share important features. The simplest such system is a non-Abelian plasma in a ``static box'' -- a non-expanding system which is isotropic at all times and initially overoccupied. The discussion in this section will set the stage for further analysis of the similarities and differences to the longitudinally expanding case, where an anisotropy between longitudinal and transverse pressures is naturally generated. This will be discussed in sec.~\ref{sec:ExpandingBox}.\\

The single particle gluon distribution in a homogeneous, isotropic system in weak coupling can be parametrized in a simple way as
\begin{eqnarray}
\label{C1Eq:initial}
f(\p)\sim \alpha_s^{-c}\;\text{for}\;|\p|<Q\;, \quad f(|\p|>Q)\ll1,
\end{eqnarray}
where $Q$ is the scale separating the region of high occupancy from the low occupancy region beyond. Since the coupling constant is small, $\alpha_s(Q) \ll 1$, the exponent $0<c<1$ quantifies the degree to which the system is initially overoccupied. For these values of $c$, at sufficiently small coupling, the occupancy is $f(|\p|\simeq Q)\gg1$ for typical momenta. The plasma exhibits classical-statistical dynamics as long as this condition holds for the occupation numbers of hard excitations. In this classical regime, one can employ the classical-statistical lattice gauge-theory techniques introduced in sec.~\ref{sec:lattice} to study non-equilibrium dynamics. If in addition $f(|\p|\simeq Q)\ll\alpha_s^{-1}$, an equivalent description of the thermalization process may also be achieved within the framework of kinetic theory.\\

The kinetic evolution of systems with the distribution in eq.~(\ref{C1Eq:initial}) has been studied previously in refs.~\cite{Kurkela:2011ti,Blaizot:2011xf} and classical-statistical lattice studies of non-Abelian plasmas in a static box have been performed previously in refs.~\cite{Berges:2008mr,Berges:2012ev,Kurkela:2012hp,Schlichting:2012es}. Here we will discuss the results of numerical studies on larger lattices than previously studied. Before we present these results, we shall first discuss general features of an overoccupied non-Abelian plasma in a static box and subsequently, the microscopic kinetic theory analysis. We will emphasize striking features of the lattice results and show that these indeed have a kinetic description. We will end this section with a discussion of classicality and thermalization for non-expanding non-Abelian plasmas.

\subsection{General considerations}
\begin{figure}[t!]
\centering
\includegraphics[width=0.48\textwidth]{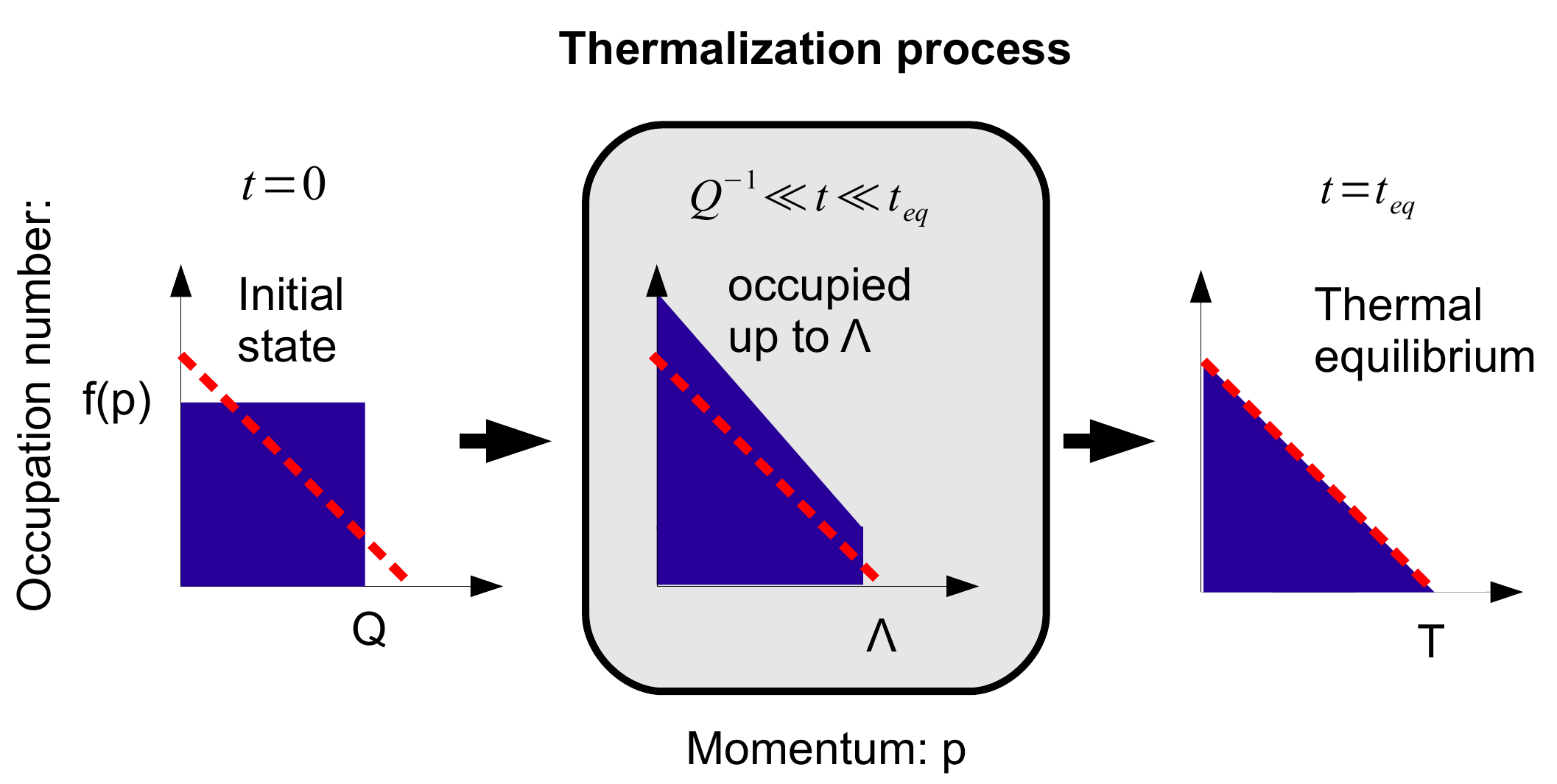}
\caption{\label{fig:KineticCartoon} (color online) Sketch of the thermalization process for a non-expanding isotropic system in kinetic theory. The different plots represent the different stages of the evolution. The red dashed line indicates the final thermal state. Once hard excitations are affected by interactions, a dynamical scale $\Lambda(t)$ develops. The time evolution of the hard scale $\Lambda(t)$ then characterizes the transport of energy towards higher momenta.} 
\end{figure}

Since the energy density $\epsilon \simeq \int \dInt^3\p~\omega_{\p}~f(\p)$ is conserved for a closed system in a static box, one can immediately determine the final state temperature of the system as
\begin{eqnarray} 
T_{\text{Final}}\sim\epsilon^{1/4}\sim\alpha_s^{-c/4}Q\;.
\end{eqnarray}
While in the initially overoccupied state $(0<c<1)$, the energy density is concentrated at the characteristic momentum scale $Q$, the energy density of the final state is dominated by modes with much higher momenta of the order of the temperature $T_{\text{Final}}\gtrsim Q$. This is illustrated in the left panel of fig.~\ref{fig:KineticCartoon}, where we show the initial state (overoccupied up to $Q$) in comparison to the thermal final state indicated by the red dashed line. During the thermalization process the typical momentum of hard excitations $\Lambda(t)$ increases as energy is transported towards the ultraviolet to reach the thermal equilibrium state. This is illustrated in the central panel of fig.~\ref{fig:KineticCartoon}.\\

Energy conservation also requires that the increase of $\Lambda(t)$ in time is accompanied by a decrease in the typical occupancy of hard excitations. For instance, for modes with momenta $\sim Q$ one finds that $f(|\p|\simeq Q,t)$ decreases during the thermalization process from the initial value $f_{\text{Initial}} (|\p|\simeq Q,t)\sim\alpha_s^{-c}$ to the final equilibrium value $f_{\text{Final}} (|\p|\simeq Q,t)\sim T_{\text{Final}}/Q \sim \alpha_s^{-c/4}$. Interestingly, this decrease in occupancy is accompanied by a change of the overall particle number density $N(t)\sim\int \dInt^3\p~f(\p,t)$. Since the particle number density in the final state is parametrically smaller than initially,
\begin{eqnarray}
N_{\text{Final}}\sim T_{\text{Final}}^{3}\sim \alpha_s^{-3c/4} Q^3~\lesssim~N_{\text{Initial}}\sim\alpha_s^{-c}Q^3 ,
\nonumber \\ 
\end{eqnarray}
ultimately number changing inelastic interactions will reduce the total number of excitations. Of course the latter, in contrast to the energy density, is not a conserved quantity.\\

From these general considerations, the central questions with regard to the thermalization process are 
\begin{itemize}
 \item How does energy transport towards the ultraviolet occur? 
 \item How do inelastic interactions reduce the overall particle number during the thermalization process?
\end{itemize}
These questions were first discussed in refs.~\cite{Kurkela:2011ti,Blaizot:2011xf} based on a kinetic theory analysis of elastic and inelastic scattering processes. Here we will provide a slightly different perspective on the kinetic theory discussion. This perspective is influenced by results from the classical-statistical lattice simulations in refs.~\cite{Berges:2008mr,Berges:2012ev,Kurkela:2012hp,Schlichting:2012es} and those we shall discuss shortly.

\subsection{Kinetic theory of turbulent thermalization}
\label{sec:TurbTherm}
To analyze the thermalization process in kinetic theory, one needs to include the effects of elastic and inelastic scattering. We begin with a discussion of elastic scattering. While previous works assumed elastic interactions to be dominated by scattering of hard particles with small momentum transfer \cite{Kurkela:2011ti,Blaizot:2011xf}, we shall employ a different approach. Our strategy will be to directly investigate the possibility of self-similar scaling solutions. Such solutions were shown in~\cite{Schlichting:2012es} to characterize the evolution at late times $t\gg \OneOverQ$. As we will discuss in sec.~\ref{sec:SBLattice}, these solutions emerge as non-thermal fixed points of the evolution such that after a short transient regime very different initial conditions show the same characteristic scaling behavior at late times.\footnote{More precisely, one finds that characteristic quantities such as the hard scale $\Lambda(t)$ exhibit a power-law dependence~\cite{Kurkela:2011ti}. The associated scaling exponents are universal and take identical values for different initial conditions. The amplitudes of the power-law solutions are non-universal and reflect the properties of the initial state, such as the value of $c$ in eq.~(\ref{C1Eq:initial}), as well as the dynamics of the transient regime. The parametric dependence of these non-universal amplitudes has been studied analytically in~\cite{Kurkela:2011ti} and numerically in~\cite{Schlichting:2012es}. However this will be of little relevance to our discussion since we will focus on the universal dynamics of the scaling regime.}\\

The kinetic theory analysis is then analogous to the discussion of turbulent thermalization in scalar field theories, previously investigated in ref.~\cite{Micha:2004bv} in the context of early universe cosmology. We shall follow the same procedure as ref.~\cite{Micha:2004bv} and search for self-similar fixed-point solutions for the gluon distribution function of the form, 
\begin{eqnarray}
\label{C1Eq:DistScaling}
f(\p,t)=(Qt)^{\alpha}~f_S\Big((Qt)^{\beta}\p\Big)\;, 
\end{eqnarray}
where $f_S(x)$ is a \emph{stationary} distribution. The functional form  of $f_S$ characterizes the time-independent shape of the attractor. The prefactor $(Qt)^{\alpha}$ in eq.~(\ref{C1Eq:DistScaling}) characterizes the overall decrease of the amplitude of the distribution in time. The factor $(Qt)^{\beta}$ in the argument describes the evolution of the hard momentum scale $\Lambda(t)\propto Q (Qt)^{-\beta}$. The parametrization on the right hand side of eq.~(\ref{C1Eq:DistScaling}) thus amounts to measuring momenta $\p$ in units of the hard momentum scale $\Lambda(t)$ at a given time $t$ of the evolution.\\

The kinetic evolution is described in terms of a Boltzmann equation of the generic form
\begin{eqnarray}
\label{C1Eq:GenBoltzmann}
\partial_t f(\p,t)=C[f](\p,t)\;,
\end{eqnarray}
where $C[f](\p,t)$ denotes the collision integral including the relevant $n\leftrightarrow m$ scattering processes. As noted, we shall focus on elastic $2\leftrightarrow2$ and inelastic $2\leftrightarrow3$ scattering processes that are expected to drive the evolution in the self-similar regime~\cite{Kurkela:2011ti}. The collision integral for elastic scattering takes the form
\begin{eqnarray}
\label{C1Eq:CollisionIntegral}
&&C[f](\p,t)= \nonumber \\
&&\frac{1}{2}\int_{\textbf{q},\textbf{k},\textbf{l}} 
\frac{|M(\p,\textbf{q},\textbf{k},\textbf{l})|^2}{2\omega_p~2\omega_q~2\omega_k~2\omega_l}~(2\pi)^4~\delta^{(4)}(q+k-l-p) \nonumber \\
&&\times\Big[(1+f_{\p})(1+f_{\textbf{l}})f_{\textbf{q}}f_{\textbf{k}}-f_{\p}f_{\textbf{l}}(1+f_{\textbf{q}})(1+f_{\textbf{k}})\Big]\;,
\end{eqnarray}
where $\delta^{(4)}(q+k-l-p)=\delta(\omega_q+\omega_k-\omega_l-\omega_p)~\delta^{(3)}(\textbf{q}+\textbf{k}-\textbf{l}-\p)$
and the scattering matrix element in the non-relativistic normalization is given by (see e.g. \cite{PeskinSchroeder})
\begin{eqnarray}
\label{eq:LASMatEl}
|M(\p,\textbf{q},\textbf{k},\textbf{l})|^2=128\pi^2\alpha_s^2N_c^2\left(3-\frac{tu}{s^2}-\frac{su}{t^2}-\frac{ts}{u^2}\right)\;, \nonumber \\
\end{eqnarray}
for non-Abelian $SU(N_c)$ gauge theories.\\

Strictly speaking the above expression is only meaningful for scatterings with large momentum transfer. In contrast, for small momentum transfers the vacuum matrix element in eq.~\ref{eq:LASMatEl} diverges and one needs to consider in medium screening effects which regulate the divergence. However, as we will discuss shortly, these medium modifications do not change the scaling properties of the scattering process in time which is the essential ingredient in our scaling analysis. We will therefore first focus on large angle scatterings, where the above matrix element can be used, and subsequently discuss the effect of small angle elastic scatterings.\\

We now follow the turbulence analysis of ref.~\cite{Micha:2004bv} and insert the scaling ansatz in eq.~(\ref{C1Eq:DistScaling}) into the Boltzmann equation (\ref{C1Eq:GenBoltzmann}). Adopting the notation $\Qt=Qt$, the left hand side of the Boltzmann equation is then given by
\begin{eqnarray}
\label{C1Eq:LHSScaling}
\partial_t f(\p,t)&\equiv &Q~\Qt^{\alpha-1}\left[\alpha f_S(\tilde{\p})+\beta~\tilde{\p}~ \bigtriangledown_{\tilde{\p}}f_S(\tilde{\p})\right]_{\tilde{\p}=\Qt^{\beta}\p}\;. \nonumber \\
\end{eqnarray}
To analyze the scaling properties of the right hand side, we first note that the differential cross-section obeys the following transformation properties
\begin{align*}
\label{C1Eq:CrossSectionScaling}
&\int \dInt\Omega^{2\leftrightarrow2}(\p,\textbf{q},\textbf{k},\textbf{l})=\int_{\textbf{q},\textbf{k},\textbf{l}} 
\frac{|M(\p,\textbf{q},\textbf{k},\textbf{l})|^2}{2\omega_p~2\omega_q~2\omega_k~2\omega_l} \nonumber \\
&\qquad \qquad \qquad \qquad \qquad~(2\pi)^4~ \delta^{(4)}(q+k-l-p)\;, \nonumber \\
&(\text{Substitute:}~\tilde{\textbf{q}}=\Qt^{\beta}\textbf{q},~\tilde{\textbf{k}}= \Qt^{\beta}\textbf{k},~\tilde{\textbf{l}}=\Qt^{\beta}\textbf{l}~| \nonumber \\
&~\text{Express}: \p=\Qt^{-\beta} (\Qt^{\beta}\p)~)\;, 
\end{align*}
\begin{align*}
&=\Qt^{-9\beta}\int_{\tilde{\textbf{q}},\tilde{\textbf{k}},\tilde{\textbf{l}}} 
\frac{|M(\Qt^{-\beta} (\Qt^{\beta}\p),\Qt^{-\beta}\tilde{\textbf{q}},\Qt^{-\beta}\tilde{\textbf{k}},\Qt^{-\beta}\tilde{\textbf{l}})|^2}{ 2\omega_{p}~2\omega_{\Qt^{-\beta}\tilde{q}}~2\omega_{\Qt^{-\beta}\tilde{k}}~2\omega_{\Qt^{-\beta}\tilde{l}}} \nonumber \\
&\qquad \qquad ~(2\pi)^4 \delta^{(4)}(\Qt^{-\beta}(\tilde{q}+\tilde{k}-\tilde{l}-\Qt^{\beta}p))\;, \nonumber \\
&(\text{Use:}~\omega_{sp}=|s|\omega_p,~\delta(sx)=|s|^{-1}\delta(x), \nonumber \\
&~|M(sp,sq,sk,sl)|^2=|M(p,q,k,l)|^2~) 
\end{align*}
\begin{align*}
&=\Qt^{-\beta}\int_{\tilde{\textbf{q}},\tilde{\textbf{k}},\tilde{\textbf{l}}} 
\frac{|M(\Qt^{\beta}\p,\tilde{\textbf{q}},\tilde{\textbf{k}},\tilde{\textbf{l}})|^2}{ 2\omega_{\Qt^{\beta}p}~2\omega_{\tilde{q}}~2\omega_{\tilde{k}}~2\omega_{\tilde{l}}} \nonumber \\
&~\qquad\qquad(2\pi)^4~ \delta^{(4)}(\tilde{q}+\tilde{k}-\tilde{l}-\Qt^{\beta}p)\;, \nonumber \\
&(\text{Rename:}~\tilde{\textbf{q}}\rightarrow \Qt^{\beta}\textbf{q},~\tilde{\textbf{k}}\rightarrow \Qt^{\beta}\textbf{k},~\tilde{\textbf{l}}\rightarrow \Qt^{\beta}\textbf{l}~| \nonumber \\
&~\text{Identify with the first line}~) 
\end{align*}
\begin{align}
&=\Qt^{-\beta}\int \dInt\Omega^{2\leftrightarrow2}(\Qt^{\beta}\p,\Qt^{\beta}\textbf{q},\Qt^{\beta}\textbf{k},\Qt^{\beta}\textbf{l})\;.
\end{align}
In the classical regime $(f(\p,t)\gg1)$, the collision integral in eq.~(\ref{C1Eq:CollisionIntegral}) can then be expressed as
\begin{align*}
\label{C1Eq:CollisionIntegralScaling}
&C[f](\p,t)=\frac{1}{2}\int \dInt\Omega^{2\leftrightarrow2}(\p,\textbf{q},\textbf{k},\textbf{l})
~f_{\p}f_{\textbf{l}} f_{\textbf{q}} f_{\textbf{k}} \nonumber \\ 
&\qquad\qquad\quad\qquad\Big[f^{-1}_{\p}+f^{-1}_{\textbf{l}}-f^{-1}_{\textbf{q}}-f^{-1}_{\textbf{k}}\Big]\;. \nonumber \\
&(\text{Use eq.}~(\ref{C1Eq:DistScaling})~\text{to express}~f(\p,t)~\text{in terms of}~f_S(\Qt^{\beta}\p)~) \nonumber
\end{align*}
\begin{align*}
&=\Qt^{3\alpha}~\frac{1}{2}\int \dInt\Omega^{2\leftrightarrow2}(\p,\textbf{q},\textbf{k},\textbf{l})
~f_S(\Qt^{\beta}\p) f_S(\Qt^{\beta}\textbf{l})  \nonumber \\
&\qquad \quad f_S(\Qt^{\beta}\textbf{q}) f_S(\Qt^{\beta}\textbf{k})\Big[f_S^{-1}(\Qt^{\beta}\p)+f_S^{-1}(\Qt^{\beta}\textbf{l}) \nonumber \\
&\qquad \quad -f_S^{-1}(\Qt^{\beta}\textbf{q})-f_S^{-1}(\Qt^{\beta}\textbf{k})\Big]\;, \nonumber \\
&(\text{Use eq.}~(\ref{C1Eq:CrossSectionScaling})~\text{to transform} ~\dInt\Omega^{2\leftrightarrow2}(\p,\textbf{q},\textbf{k},\textbf{l})~) \nonumber
\end{align*}
\begin{align*}
&=\Qt^{3\alpha-\beta}~\frac{1}{2}\int \dInt\Omega^{2\leftrightarrow2}(\Qt^{\beta}\p,\Qt^{\beta}\textbf{q},\Qt^{\beta}\textbf{k},\Qt^{\beta}\textbf{l}) \nonumber \\
&\qquad \qquad f_S(\Qt^{\beta}\p) f_S(\Qt^{\beta}\textbf{l}) f_S(\Qt^{\beta}\textbf{q}) f_S(\Qt^{\beta}\textbf{k}) \nonumber \\
&\qquad \qquad \Big[f_S^{-1}(\Qt^{\beta}\p)+f_S^{-1}(\Qt^{\beta}\textbf{l}) \nonumber \\ 
&\qquad \qquad \quad -f_S^{-1}(\Qt^{\beta}\textbf{q})-f_S^{-1}(\Qt^{\beta}\textbf{k})\Big]\;, \nonumber \\
&(\text{Identify with the first line}~) \nonumber
\end{align*}
\begin{align}
&=\Qt^{3\alpha-\beta}~C[f_S](\Qt^{\beta}\p)\;, \qquad \qquad \qquad \qquad
\end{align}
By use of eqns.~(\ref{C1Eq:LHSScaling}) and (\ref{C1Eq:CollisionIntegralScaling}), the Boltzmann equation (\ref{C1Eq:GenBoltzmann}) can be decomposed into a condition for the fixed point solution $f_S(\p)$
\begin{eqnarray}
\label{C1Eq:TurbFixedPointForm}
\alpha f_S(\p)+\beta~\p\bigtriangledown_{\p}f_S(\p)=Q^{-1}~C[f_S](\p)\;,
\end{eqnarray}
and the scaling relation
\begin{eqnarray}
\label{C1Eq:DynScalingRelation}
\alpha-1=3\alpha-\beta\;.
\end{eqnarray}
The non-trivial solutions of eq.~(\ref{C1Eq:TurbFixedPointForm}) characterize the functional form of the fixed point solutions $f_S(\p)$, whereas the scaling relation (\ref{C1Eq:DynScalingRelation}) constraints the evolution of the system on the fixed point trajectory.\\

While the previous derivation was carried out in the limit of large angle scatterings, where screening effects can be ignored, the scaling relation in eq.~(\ref{C1Eq:DynScalingRelation}) turns out to be much more general. Based on the results in \cite{Kurkela:2011ti,Blaizot:2011xf}, it is straightforward to show that the scaling relation (\ref{C1Eq:DynScalingRelation}) also holds (up to logarithmic corrections) in the limit of (screened) small angle scatterings, where medium modifications of the scattering matrix element need to be considered. While the rate of these processes is parametrically enhanced, the changes in the distribution functions due to individual collisions are parametrically small. Consequently, large cancellations between gain and loss terms occur which finally lead to the same overall scaling behavior in time as for large angle scattering processes~\cite{Kurkela:2011ti,Blaizot:2011xf}.\\

Similarly, it was shown in refs.~\cite{Kurkela:2011ti,Blaizot:2011xf} that, for non-Abelian $SU(N_c)$ gauge theories, inelastic processes also exhibit the same parametric dependence as elastic scattering~\cite{Arnold:2002zm,Kurkela:2011ti,Blaizot:2011xf}. In this case the reduced probability for additional gluon emissions/absorptions is compensated for by the large rate of accompanying small angle scatterings~\cite{Kurkela:2011ti,Blaizot:2011xf}. Thus taking inelastic processes into account again does not affect the overall scaling properties of the collision integral in time that lead to eq.~(\ref{C1Eq:DynScalingRelation}). Consequently, the scaling relation in eq.~(\ref{C1Eq:DynScalingRelation}) applies for particle number conserving large and small angle scatterings as well as particle number changing inelastic interactions~\cite{Kurkela:2011ti,Kurkela:2012hp}.\\

While eq.~(\ref{C1Eq:DynScalingRelation}) does not uniquely determine the scaling exponents, further constraints on the evolution can be derived from conservation laws. Energy conservation in the static box implies that the left hand side of
\begin{eqnarray}
\epsilon&\simeq&\int_{\p} \omega_\p~f(\p,t)=\Qt^{\alpha} \int_{\p} \omega_\p~f_S(\Qt^{\beta}\p)  \nonumber \\
&=&\Qt^{\alpha-3\beta} \int_{\tilde{\p}} \omega_{\Qt^{-\beta}\tilde{\p}}~f_S(\tilde{\p})
\simeq(Qt)^{\alpha-4\beta} \epsilon_0\;.
\end{eqnarray}
should be identical to the right hand side. One therefore obtains the additional scaling relation
\begin{eqnarray}
\label{C1Eq:EnergyScalingRelation}
\alpha-4\beta=0\;.
\end{eqnarray}
While particle number is also conserved for elastic scattering processes, this is clearly no longer the case when inelastic interactions are taken into account. The fact that particle number changing processes exhibit the same scaling behavior in time is therefore essential to adjust the overall occupation numbers during the turbulent thermalization process.\footnote{In situations where inelastic processes are highly suppressed as compared to elastic scattering, a possible excess of particles may be absorbed into the soft momentum sector. This relaxes the constraint of particle conservation in the hard sector, where energy remains the only relevant conserved quantity~\cite{Blaizot:2011xf}. Interestingly, it has been shown in scalar field theories that this interplay of the dynamics of hard and soft modes leads to the formation of a (transient) Bose-Einstein condensate which also modifies the dynamics of the hard sector~\cite{Khlebnikov:1996mc,Berges:2012us}.}\\

Combining the scaling relations obtained from the analysis of the Boltzmann equation in eq.~(\ref{C1Eq:DynScalingRelation}) and the energy conservation constraint in eq.~(\ref{C1Eq:EnergyScalingRelation}) yields the scaling exponents
\begin{eqnarray}
\label{eq:SBScalingExponents}
\alpha=-4/7\;, \qquad \qquad \beta=-1/7\;. 
\end{eqnarray}
This scaling behavior for a non-expanding non-Abelian plasma was previously predicted in~\cite{Kurkela:2011ti} and~\cite{Blaizot:2011xf}. However in these discussions the concepts of self-similarity and turbulent thermalization were not discussed. The fact that~\cite{Kurkela:2011ti} and~\cite{Blaizot:2011xf} obtained the same correct scaling behavior based on a small angle approximation is a consequence of the universality of the scaling exponents and not unique to the particular assumptions employed.

\subsection{Universality \& Turbulence}
It is important to note that details of the underlying field theory such as the number of colors, the coupling constant, or other details of the scattering matrix element, do not enter the above scaling analysis. Instead the dynamical scaling exponents $\alpha$ and $\beta$ are only sensitive to the conserved quantities of the system and the canonical scaling dimensions of the collision integral with respect to powers of the distribution function and momenta. Since these are identical for many different implementations of the interactions, the scaling behavior in eq.~(\ref{eq:SBScalingExponents}) provides a particularly robust prediction of the kinetic theory analysis. Indeed, the above scaling relations were previously derived as part of a more general analysis for scalar field theories in the context of cosmology~\cite{Micha:2004bv}. Since the physical conditions that lead to the scaling relations in eqns.~(\ref{C1Eq:DynScalingRelation}) and (\ref{C1Eq:EnergyScalingRelation}) can be realized in many different ways, one may well expect to observe such a self-similar scaling behavior at yet very different energy scales. The scaling exponents $\alpha,\beta$ are therefore universal in the classical sense that they can be shared by a large variety of strongly correlated many body systems.\\

Our derivation also clearly shows the nature of the self-similar scaling solution as a non-thermal fixed point of the classical evolution. Specifically, there is a striking analogy between the self-similar scaling solution in eq.~(\ref{C1Eq:DistScaling}) and the ubiquitous phenomenon of turbulence. While we considered the thermalization process as the transport of a conserved quantity (energy) towards the ultraviolet, the phenomenon of {\it stationary turbulence} describes the stationary transport of conserved quantities in systems coupled to a source and a sink. The most prominent example is hydrodynamic turbulence. Constant energy injection on large scales results in a stationary turbulent spectrum within an inertial range. In this region, according to Kolmogorov's theory, energy is conserved and transported to smaller and smaller scales until viscous dissipation occurs at microscopic scales~\cite{Frisch}.\\

Similar phenomena occur in a wide class of systems such as waves on a fluid surface, where the turbulent dynamics can often be described in terms of a wave kinetic description~\cite{Falkovich}. This phenomenon is called {\it weak wave turbulence} and is even more reminiscent of the physical situation encountered during the thermalization process studied here. The crucial difference between the thermalization process in a closed system far from equilibrium and the phenomenon of stationary wave turbulence is the fact that in our case neither a source nor a sink are present to inject/deposit energy. In contrast to a stationary turbulent solution, one therefore observes a quasi-stationary self-similar evolution. The latter is additionally characterized by the dynamical exponents $\alpha$ and $\beta$ which describe energy transport towards the ultraviolet, thereby driving the evolution towards thermal equilibrium. In distinction to the {\it driven} case where an active source and sink exist, this situation is also referred to as {\it free} turbulence~\cite{Micha:2004bv}.\\ 

Despite this important difference, it was observed previously~\cite{Micha:2004bv,Berges:2010ez,Nowak:2011sk}, that the functional form of the stationary distribution $f_S$ may still share the universal scaling properties - described by the scaling exponent $\kappa$ -- of the Kolmogorov-Zakharov spectra associated with stationary (weak) wave turbulence in a driven system. The heuristic argument in this context relies on the existence of an inertial range of momenta where the transport of conserved quantities locally remains ``close to turbulent" even for the free turbulence in a closed system. Since the value of the spectral exponent $\kappa$ depends on the nature of the underlying interaction, different proposals have been put forward with regard to non-Abelian gauge theories~\cite{Arnold:2005ef,Arnold:2005qs,Mueller:2006up,Berges:2008mr,Berges:2012ev}.\\

Following the proposal of Bose-Einstein condensation in ref.~\cite{Blaizot:2011xf}, and similar observations in scalar field theories~\cite{Micha:2004bv,Berges:2012us}, it was shown in ref.~\cite{Berges:2012ev} that (in the presence of a large number of very soft excitations) stationary turbulent solutions with the exponent $\kappa=3/2$ could also be obtained in non-Abelian gauge theories. However, when applied to closed systems, classical-statistical lattice simulations revealed that such behavior persists only for a transient time of the evolution. At later times, the excess of soft excitations decays and a different exponent $(\kappa=4/3)$ was found to be realized~\cite{Berges:2012ev}. The question of whether an initial overoccupation may lead to Bose-Einstein condensation was also studied in the classical-statistical lattice simulations of ref.~\cite{Kurkela:2012hp}. The authors concluded that the presence of condensates is unlikely because these configurations decay on short time scales.\\

The spectral exponents $\kappa=4/3$ and $\kappa=5/3$ associated with a stationary particle and energy cascade respectively were originally proposed in ref.~\cite{Berges:2008mr}, based on an analysis of ordinary elastic scattering processes within 2PI effective action techniques. The authors of ref.~\cite{Berges:2008mr} also demonstrated the appearance of a $\kappa=4/3$ spectrum in classical-statistical lattice simulations of closed systems. While this behavior was confirmed independently by simulations in refs.~\cite{Kurkela:2012hp,Schlichting:2012es}, it remains an interesting open question how the observation of $\kappa=4/3$ is compatible with inelastic interactions playing a central role in the thermalization process.

\subsection{Lattice results}
\label{sec:SBLattice}
We will now present results from classical-statistical lattice simulations that confirm the above scenario of turbulent thermalization for a non-expanding non-Abelian plasma. We shall focus on the results relevant to the thermalization process at late times and frequently refer to previous studies~\cite{Kurkela:2012hp,Berges:2012ev,Schlichting:2012es}, where further aspects of the evolution have been discussed in more detail. As noted, the results presented here are for significantly larger lattices than those considered previously.

\subsubsection{Initial conditions}
We employ Gaussian initial conditions based on the single particle distribution
\begin{eqnarray}
f(\p,t_0)=\frac{1}{g^2} \Theta(Q-|\p|)\;, 
\end{eqnarray}
which corresponds to the $c=1$ case in eq.~(\ref{C1Eq:initial}). Similarly to the longitudinally expanding case (discussed in sec.~\ref{sec:ExpandingIC}), the field configurations at initial time are constructed to be 
\begin{eqnarray}
\label{eq:MinkowskiIC}
A_{\mu}^a(t_0,\bf{x})&=&\sum_{\lambda=1,2}\int \frac{\dInt^3 \bf{k}}{(2\pi)^3}\,
\sqrt{f({\bf k},t_0)}\, \\
&&\qquad \times \left[c_{\lambda,a}^{\bf{k}}\, \xi^{(\lambda)\bf{k}+}_{\mu}(t_0)\,
e^{i \bf{k}\bf{x}}
+c.c.\right]\;, \nonumber
\end{eqnarray}
\begin{eqnarray}
E^{\mu}_a(t_0,\bf{x})&=&-\eta^{\mu\nu}\sum_{\lambda=1,2}\int \frac{\dInt^3 \bf{k}}{(2\pi)^3}\,
\sqrt{f({\bf k},t_0)}\,\\
&&\qquad \times\left[c_{\lambda,a}^{\bf{k}}\, \dot{\xi}^{(\lambda)\bf{k}+}_{\nu}(t_0)\,
e^{i \bf{k}\bf{x}}
+c.c.\right]\;. \nonumber
\end{eqnarray}
Here $\eta_{\mu\nu}=\eta^{\mu\nu}=\text{diag}(1,-1,-1,-1)$ denotes the Minkowski metric and $\xi^{(\lambda)\bf{k}+}_{\nu a}(t_0)$ denotes the transverse polarization vectors in temporal axial gauge ($A_t=0$), where the Coulomb gauge condition $({\bf \triangledown} \cdot {\bf A}=0)$ is satisfied at initial time $t_0$. The coefficients $c_{\lambda,a}^{\bf{k}}$ are taken as uncorrelated Gaussian random numbers, similar to those defined in sec.~\ref{sec:ExpandingIC}). 

\begin{figure}[t!]
\centering						
\includegraphics[width=0.5\textwidth]{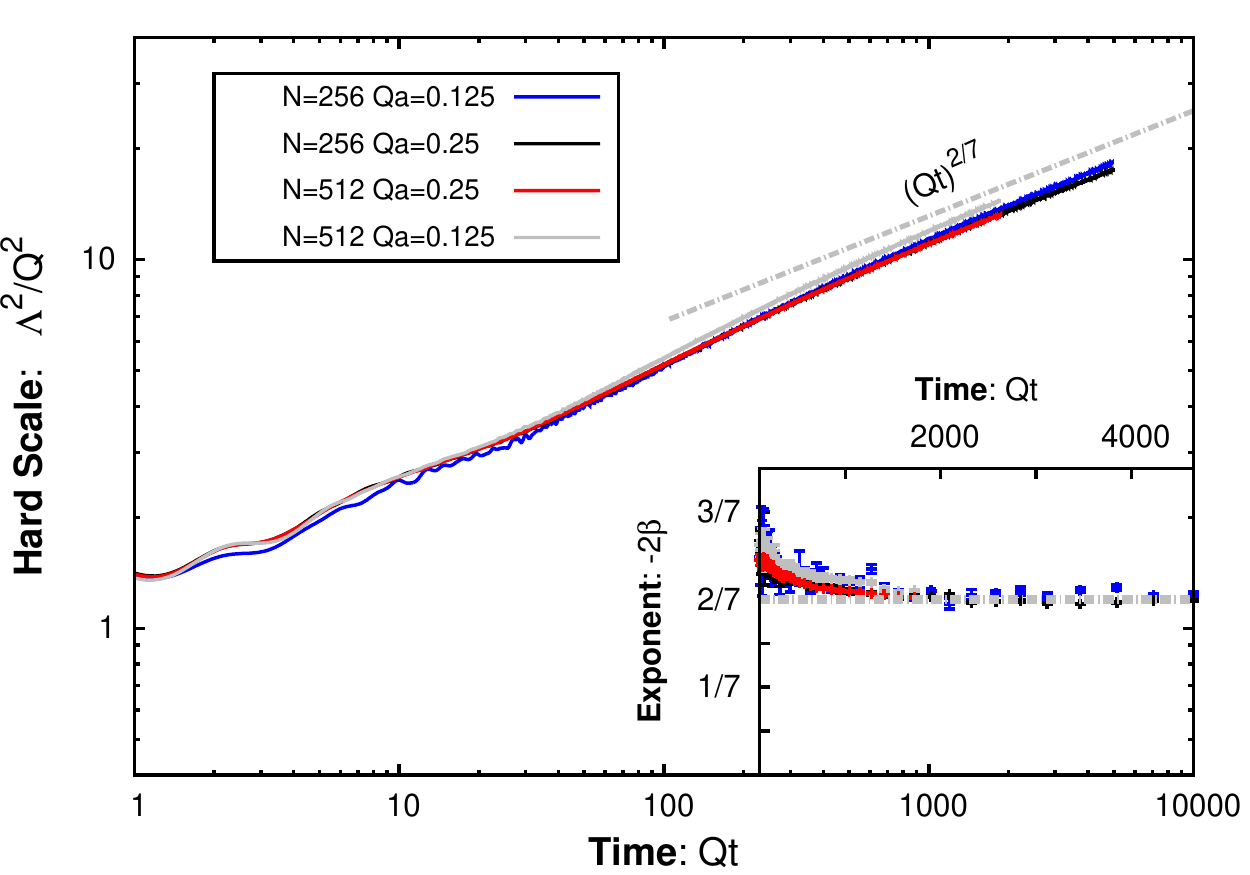}	
\caption{\label{fig:BoxHardScale} (color online) Time evolution of the characteristic momentum of hard excitations. After a transient regime, one observes a clear scaling behavior $\Lambda^2(t) \propto Q^2~(Qt)^{2/7}$. The associated scaling exponent $-2\beta$ can be extracted from the logarithmic derivative shown in the inset. One observes good agreement with $-2\beta=2/7$ as predicted by the kinetic theory analysis. See text for a discussion of the characteristic time scales on the x-axis.}	
\end{figure}							

\subsubsection{Scaling}
We shall first study the evolution of the hard scale $\Lambda(t)$ which characterizes the typical momentum of hard excitations. In the classical-statistical lattice simulations, this quantity is directly accessible in terms of the gauge invariant observable $\Lambda^2$ introduced in sec.~\ref{sec:HardScaleDef}. To facilitate the comparison with the kinetic description, it is useful to consider first the perturbative expression
\begin{eqnarray}
\Lambda^2(t)\simeq\frac{\int_\p ~4\p^2~\omega_\p~f(\p,t)}{\int_\p~\omega_\p~f(\p,t)}\;.
\end{eqnarray}
By use of the self-similarity assumption in eq.~(\ref{C1Eq:DistScaling}), one can then directly obtain the expected scaling behavior as
\begin{eqnarray}
\Lambda^2(t)\propto Q^2 (Qt)^{-2\beta}. 
\end{eqnarray}
To investigate whether this scaling behavior is realized within classical-statistical simulations, we follow ref.~\cite{Kurkela:2012hp} and study the time evolution of the characteristic momentum scale. The results for $\Lambda^2(t)$ are shown in fig.~(\ref{fig:BoxHardScale}), for different lattice discretizations. After a transient regime characterized by a rapid increase of the hard scale, the time evolution of $\Lambda^2(t)$ approaches a clear $\propto (Qt)^{2/7}$ power-law dependence as indicated by the (gray) dashed line. One observes identical behavior for different lattice discretizations indicating the convergence of our results. The inset of fig.~(\ref{fig:BoxHardScale}) shows the (double) logarithmic derivative
\begin{eqnarray}
-2\beta(t)=\frac{d \log(\Lambda^2(t))}{d \log(t)} 
\end{eqnarray}
as a function of time. In the scaling regime one observes good agreement with the value $\beta=-1/7$ obtained from the kinetic theory analysis. \\ 

The time scale $t_{\text{scaling}}$ for the transition to the scaling regime depends on the initial conditions of the evolution and occurs around $Qt_{\text{scaling}}\approx500$ for our initial conditions. These numbers may seem very large because typically in a ``real world'' heavy ion collision one anticipates the lifetime of the plasma to be $Qt\approx50$. In this context, there are two important points to consider. Firstly, as we shall soon discuss, $\alpha_s$ in our simulations has to be extremely small for classical dynamics to be cleanly realized. The values of $Q$ associated with these couplings are orders of magnitude larger than those realized in heavy ion experiments. Hence even time scales corresponding to several hundreds of $\OneOverQ$ are much smaller than the typical lifetimes of heavy ion collisions (controlled by the sizes of the colliding nuclei) at very high energies. The other important point is that for these values of $\alpha_s$, time scales of the order $Qt_{\text{scaling}}\approx500$ are much shorter when compared to the lifetime of the classical regime, which is orders of magnitude larger. Indeed, it is this clean separation of time scales in weak coupling asymptotics that enables one to isolate universal from transient dynamics in these systems. How to extrapolate the weak coupling asymptotics to realistic collider energies will be discussed in sec.~\ref{sec:QuoVadis}.
\\

The dynamics of the transient regime has been investigated in more detail in ref.~\cite{Berges:2012ev} for a similar class of initial conditions. The onset of scaling behavior for a larger class of initial conditions was shown explicitly in ref.~\cite{Kurkela:2012hp}, while ref.~\cite{Schlichting:2012es} investigated the dependence of $t_{\text{scaling}}$ on the initial overoccupancy. These results add to the observation that the scaling behavior at late times is a generic feature of the thermalization process, independent of the underlying initial conditions.

\begin{figure}[t!]
\centering						
\includegraphics[width=0.5\textwidth]{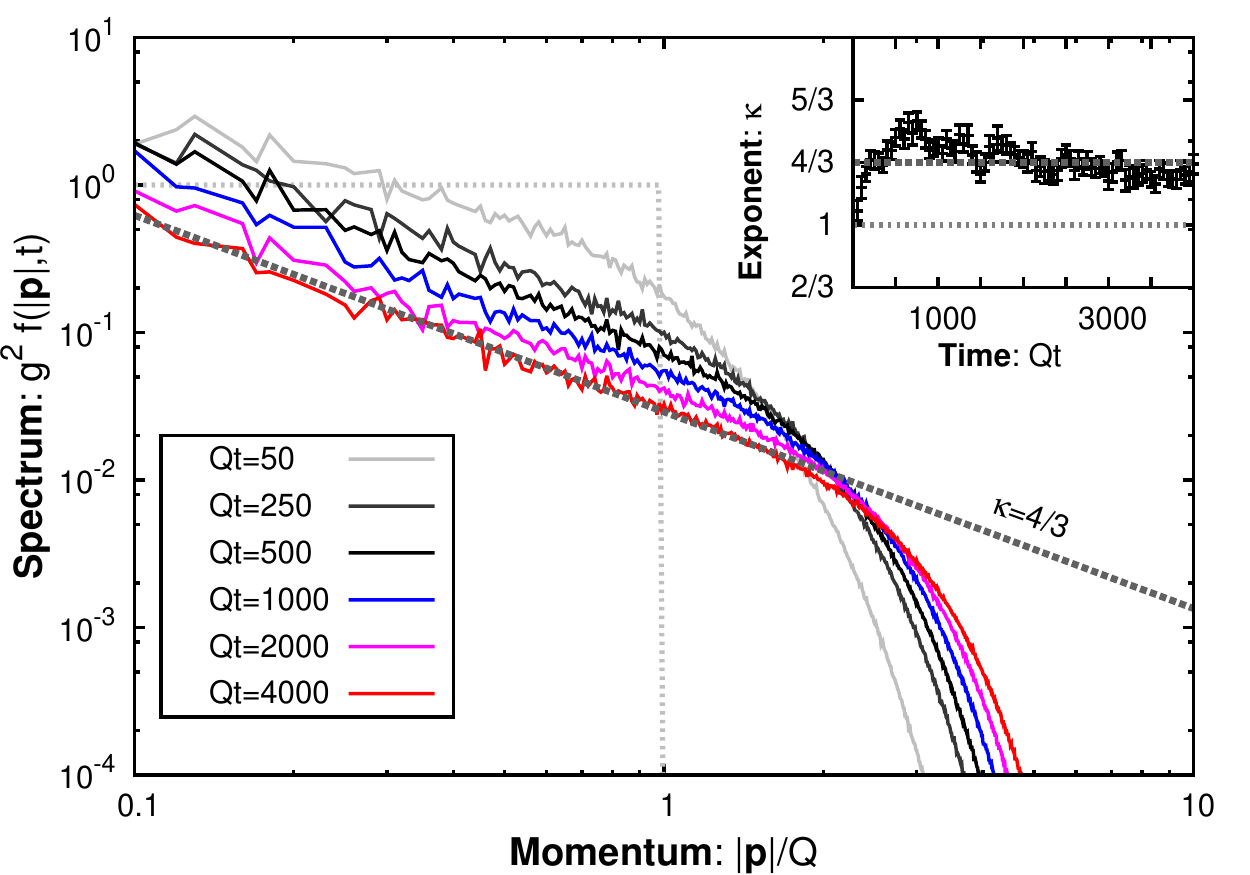}	
\caption{\label{fig:BoxSpectra} (color online) Single particle spectrum at different times $Qt$ of the evolution. The initial distribution is indicated by a gray dashed line. At later times, one clearly observes the emergence of a turbulent power-law spectrum $f(p)\sim p^{-4/3}$ for a large range of momenta. The spectral exponent $\kappa$ can be extracted from a fit to the spectra at different times of the evolution as shown in the inset. The observed exponent is consistent with the value $\kappa=4/3$, characteristic for wave turbulence induced by elastic interactions \cite{Berges:2008mr}.}	 
\end{figure}							

\subsubsection{Turbulent spectra}
\label{sec:AlsoDiscussed}
Before we demonstrate the emergence of self-similarity in our simulations, we will now discuss the properties of the single particle spectra. Our focus will be on the question whether scale invariant power-law distributions characteristic of wave turbulence can be observed. We follow previous works~\cite{Berges:2008mr,Berges:2012ev,Kurkela:2012hp,Schlichting:2012es} and compute the single particle distribution. In analogy to the discussion of the longitudinally expanding case (c.f. sec.~\ref{sec:OccupationNumbers}), we define the single particle distribution as a projection on Fock states
\begin{eqnarray}
\label{lat:ParticleNumberMinkowski}
&&f({\bf p},t)=\frac{1}{N_g (Na)^3}\sum_{a=1}^{N_c^2-1}\sum_{\lambda=1,2} \Big({\bf p},\lambda,a\Big|A\Big)\;,
\end{eqnarray}
where
\begin{eqnarray}
\Big({\bf p},\lambda,a\Big|A\Big)=\Big<\Big| \left(\xi^{(\lambda){{\bf p}+}}_{\mu}(t)\right)^*\stackrel{\longleftrightarrow}{\partial_{t}} 
A^{\mu}_{a}(t,{\bf p})
\Big|^2\Big>_{\text{Coul.~Gauge}}\;, \nonumber \\
\end{eqnarray}
denotes the projection of the gauge field evaluated with the Coulomb gauge condition ${\bf \triangledown}\cdot{\bf A}=0$ satisfied at the time $t$ when the spectrum is calculated\footnote{Our definition only includes transversely polarized excitations and does not account for in-medium modifications of the dispersion of low momentum modes. This choice is primarily motivated by the longitudinally expanding case where the structure of transverse excitations already becomes rather involved. We note that a variety of different definitions has been used in the literature~\cite{Berges:2008mr,Berges:2012ev,Kurkela:2012hp,Schlichting:2012es} and we refer to ref.~\cite{Kurkela:2012hp} for a detailed study of equal-time two-point correlation functions.}.\\

In fig.~(\ref{fig:BoxSpectra}), we present the single particle spectrum $g^2 f(|\p|,t)$ at different times $Qt$ of the evolution. The results shown in fig.~(\ref{fig:BoxSpectra}) were obtained on $256^3$ lattices with $Qa=0.25$. Starting from an overoccupied initial condition -- indicated by the gray dashed line -- one observes how the spectrum quickly extends towards higher momenta. At later times, the spectrum is well described by a power-law for momenta $|\p|\lesssim \Lambda$ and a rapid fall-off for momenta $|\p|\gtrsim \Lambda$. One also observes how the dynamical scale $\Lambda$ evolves towards higher momenta, while the amplitude $f(|\p|\simeq Q,t)$ of the distribution decreases with time.\footnote{As seen in fig.~(\ref{fig:BoxSpectra}), we would, in this particular case, require $g^2\lesssim 10^{-4}$ for the occupancy $f(|\p|\simeq \Lambda(t),t)$ to be much larger than unity over the entire evolution.}\\

To analyze the emergent power-law behavior in more detail, we follow previous works~\cite{Berges:2008mr,Berges:2012ev,Schlichting:2012es} and perform a series of fits to extract the associated scaling exponent $\kappa$ at different times of the evolution. The scaling exponent $\kappa(t)$ as a function of time $Qt$ is shown in the inset of fig.~\ref{fig:BoxSpectra} and compared to different values proposed in the literature. The classical thermal value $\kappa=1$ is clearly ruled out. Instead one observes a good overall agreement with $\kappa=4/3$ in the scaling regime.\footnote{The small systematic deviation from below may be attributed to the modification of the in-medium dispersion of quasi-particle excitations. Since this is not captured by our definition of the single particle spectra, deviations from the above scaling behavior occur for low momentum modes. The structure of these soft excitations and their effect on the single particle spectra is discussed in more detail in ref.~\cite{Kurkela:2012hp}.} From this analysis, and the discussion in sec.~\ref{sec:TurbTherm}, we conclude that this scaling is a clear manifestation of turbulent behavior.

\subsubsection{Self-similarity}
Thus far we confirmed the scaling behavior of the characteristic momentum scale in time and the emergence of a turbulent spectrum. However the most striking property of the kinetic theory solution is the self-similar behavior characterizing turbulent energy transport.\\

\begin{figure}[t!]
\centering						
\includegraphics[width=0.5\textwidth]{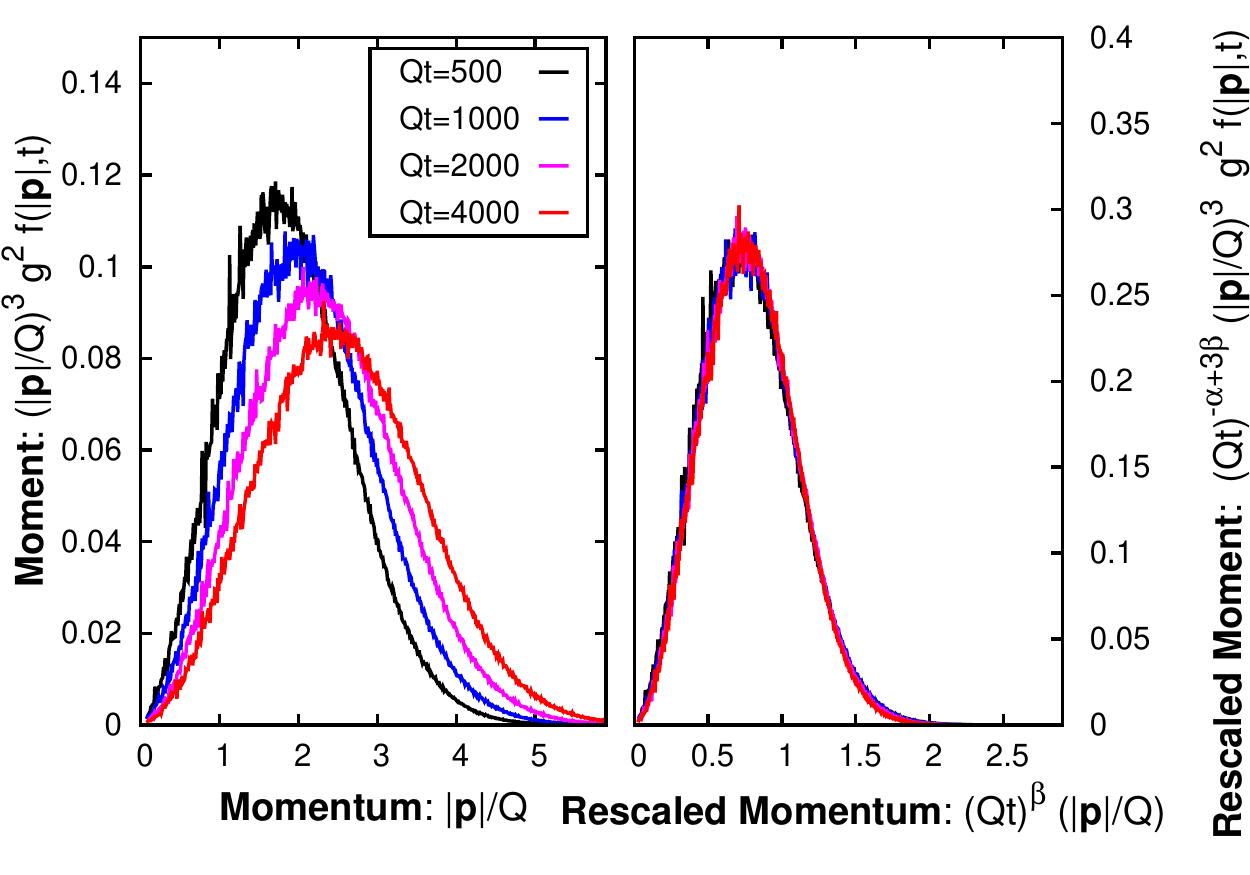}	
\caption{\label{fig:BoxMoments} (color online) (\textbf{left}) Third moment $|\p|^3 f(|\p|,t)$ of the single particle distribution at different times $Qt$ of the evolution. The position of the peak indicates the momentum scale, which dominates the energy density of the system. One clearly observes how energy is transported towards higher momenta. (\textbf{right}) The rescaled moments according to eq.~(\ref{eq:ThirdMomentRescaled}) show a stationary distribution for $\alpha=-4/7$ and $\beta=-1/7$ as in eq.~(\ref{eq:SBScalingExponents}). The fact that all data collapses on a single curve is a striking manifestation of self-similarity.}	
\end{figure}

To demonstrate the emergence of self-similarity in our simulations, we follow ref.~\cite{Schlichting:2012es} and study the time evolution of moments of the single particle distribution. We consider the third moment of the distribution function
\begin{eqnarray}
f^{(3)}(|\p|,t)=|\p|^3 f(|\p|,t)\;,
\end{eqnarray}
which according to the perturbative expression for the energy density 
\begin{eqnarray}
\epsilon\simeq \int \dInt|\p|~\p^2~\omega_\p~f(\p,t)\simeq \int \dInt|\p|~|\p|^3~f(|\p|,t)\;, 
\end{eqnarray}
can be interpreted as the energy density per momentum mode. The distribution $f^{(3)}(|\p|,t)$ is shown in the left panel of fig.~\ref{fig:BoxMoments} as a function of momentum at different times $Qt$ of the evolution. The peak of the distribution corresponds to the momentum scale which dominates the energy density of the system at a given time. One clearly observes how the position of the peak moves towards higher momenta characterizing the transport of energy towards the ultraviolet.\\

The self-similar behavior of energy transport can be observed in the right panel of fig.~\ref{fig:BoxMoments}, where we show a rescaled version of the distribution. According to the self-similarity condition in eq.~(\ref{C1Eq:DistScaling}) one finds that
\begin{eqnarray}
\label{eq:ThirdMomentRescaled}
\left(\frac{\tilde{p}}{Q}\right)^3~f_S(\tilde{p})= (Qt)^{-\alpha+3\beta}~\left(\frac{|\p|}{Q}\right)^{3}~f(|\p|,t)\;,
\end{eqnarray}
yields a stationary distribution when plotted as a function of the rescaled momentum $\tilde{p}=(Qt)^{\beta} |\p|$. This is shown in the right panel of fig.~\ref{fig:BoxMoments}, where we show the right hand side of eq.~(\ref{eq:ThirdMomentRescaled}) as a function of the rescaled momentum. Indeed one observes that with $\alpha=-4/7$ and $\beta=-1/7$ as in eq.~(\ref{eq:SBScalingExponents}) the data obtained at different times of the evolution collapses onto a single curve. This is a striking manifestation of the self-similarity of the evolution. 

\subsection{Classicality and complete thermalization}
Following our observation of self-similarity, one can extrapolate the result to understand the dynamics of thermalization at later times. Indeed, since the theory is scale invariant on the classical level, one expects no deviations to occur within the realm of classical dynamics.\footnote{In practice, the energy transport towards the ultraviolet is limited by the lattice UV cutoff. Indeed previous studies have shown that (classical) thermalization occurs as a cutoff effect~\cite{Moore:2001zf}. However in the classical continuum theory -- which is a well defined theory in the absence of vacuum fluctuations -- scale invariance is preserved at all scales. Thus one expects the turbulent cascade to continue for all times.} This is of course different in the quantum theory where scale invariance is explicitly broken. Since the system becomes more and more dilute as the evolution proceeds, the quantum evolution effects can be efficiently discussed at the level of the Boltzmann equation.\\

The first point to note is that to obtain the self-similar scaling solution in eq.~(\ref{C1Eq:DistScaling}) we considered the classical limit of the Boltzmann equation. This is of course well justified as long as the competing quantum processes are highly suppressed. However, as the turbulent cascade proceeds, the occupation number of hard excitations
\begin{eqnarray}
\label{eq:Sec4nHard}
\nHard=f(|\p|\simeq \Lambda(t),t)
\end{eqnarray}
becomes smaller and smaller and quantum effects become increasingly important. When at the time $t_{\text{Quantum}}$ the occupation number $\nHard$ becomes of order unity, quantum effects become of the same order of magnitude as the classical dynamics and can no longer be neglected.\\

From an extrapolation of the scaling behavior in eq.~(\ref{C1Eq:DistScaling}), it is straightforward to estimate
\begin{eqnarray}
\label{C1Eq:tQuantum}
t_{\text{Quantum}}\sim Q^{-1} \alpha_s^{-7/4}\;. 
\end{eqnarray}
Beyond this point, classical-statistical simulations no longer provide a reliable approximation -- even on a qualitative level. Instead one expects quantum effects to drive the system to the unique thermal fixed point.\\

Since the energy transport towards the ultraviolet is accomplished on the same time scale, one may expect a short time scale for the subsequent approach to complete thermal equilibrium. The estimate in eq.~(\ref{C1Eq:tQuantum}) then also provides a lower bound on the estimate of the thermalization time.

\section{Expanding non-Abelian plasma}
\label{sec:ExpandingBox}
We will now discuss the dynamics of the longitudinally expanding non-Abelian plasma relevant to heavy ion collisions at ultrarelativistic energies. As discussed in sec.~\ref{sec:CGC}, we expect such a weakly coupled plasma to be described as a strongly correlated system of quasi-particles at times $\tau_0\sim \OneOverQ \lnSqrOneOverAlpha$ after the collision. The properties of this initial state can then be modeled in terms of a single particle distribution as in eq.~(\ref{eq:TurbIC}), which we employ as initial conditions for our simulations.\\

To investigate a preferably large range of different initial conditions, we will vary the occupancy parameter $n_0$ and the anisotropy parameter $\xi_0$ and study their effect on the evolution. In our previous publication \cite{Berges:2013eia}, the initial time was chosen as\footnote{In view of the parametric estimate $Q\tau_0\sim \lnSqrOneOverAlpha)$, this choice corresponds to gauge couplings on the order of $\alpha_s\sim10^{-5}$.} $Q\tau_0=100$ to minimize discretization errors (c.f. sec.~\ref{sec:DiscErrors}), while accessing sufficiently late times $\tau\gg \tau_0$ in order to observe universal aspects of the evolution. Here we will also verify that varying the initial time in a range $Q\tau_0 =100$ -- $1000$ only affects the transient evolution but does not change the universal properties at later times.\\

\subsection{General considerations}
Before we discuss the non-equilibrium dynamics based on classical-statistical lattice simulations of this system, we shall briefly note some aspects specific to the dynamics of the longitudinally expanding plasma.\\

A major complication in this context is the fact that the energy density of the system is no longer a conserved quantity due to the longitudinal expansion. Instead the time evolution of the energy density is described by Bjorken's law,
\begin{eqnarray}
\label{eq:BjorkenLaw}
\partial_{\tau} \epsilon=-\frac{\epsilon+P_L}{\tau} \;,
\end{eqnarray}
and depends on the expansion rate $1/\tau$ as well as the dynamical equation of state $P_L(\tau)/\epsilon(\tau)$ of the system. Because the time evolution of $P_L(\tau)/\epsilon(\tau)$ is in general non-trivial, it is thus (in contrast to the non-expanding case) not possible to determine the final temperature of the system without further assumptions about the evolution.\\ 

Moreover, the one-dimensional expansion of the plasma also leads to a redshift of longitudinal momentum modes such that the plasma will be anisotropic on large time scales. Even though interactions naturally compete with this process, it is a non-trivial question whether they are sufficiently strong to maintain a close to isotropic system at all times of the evolution. Indeed several studies within kinetic theory indicate that the momentum broadening due to classical scattering dynamics may not be sufficiently strong to completely counter the redshift~\cite{Baier:2000sb,Kurkela:2011ti,Kurkela:2011ub}. As a consequence, most kinetic thermalization scenarios predict an increase of the anisotropy in the classical regime, while isotropy is restored only during the final stages of the quantum equilibration process.\footnote{A noteworthy exception is the scenario discussed in~\cite{Blaizot:2011xf} which we shall discuss further later.} Nevertheless, the effect of classical interactions competes with the expansion of the system at all times; the induced momentum broadening then leads to a characteristic time evolution very different from simple free streaming behavior. The time evolution of the momentum space anisotropy can thus be used as an observable to distinguish between different thermalization scenarios.\\

In view of the discussion of turbulent thermalization in sec.~\ref{sec:StaticBox}, it is also an interesting question whether the dynamics of the expanding non-Abelian plasma exhibits the universal features associated with a non-thermal fixed point. However, since the longitudinal expansion leads to a dilution of the system and renders the plasma anisotropic on large time scales, it is by no means obvious how the concepts developed in sec.~\ref{sec:StaticBox} apply in this situation.\\

The central questions concerning the evolution of the expanding plasma are 
\begin{itemize}
\item How efficient is momentum broadening compared to the expansion? How does the anisotropy of the system evolve?
\item Is there a universal attractor? What are its universal properties? How is this related to different thermalization scenarios?
\end{itemize}
We will now address these questions using the machinery of real-time classical-statistical lattice simulations. The various discretization parameters used in our studies are listed in table~\ref{tab:Discretization}. We note that the sizes of the lattices are the largest thus far employed in the study of expanding non-Abelian plasmas. As will become clear, lattices of comparable size are essential to resolve key aspects of the dynamics over the lifetime of the evolution.\\ 

The discussion in this section is organized as follows: In sec.~\ref{sec:VUnivScal}, we study the evolution of the anisotropy and demonstrate the emergence of a universal scaling behavior. We analyze the spectral properties of the attractor in sec.~\ref{sec:ExpSelfSimilar} and show the emergence of a self-similar behavior characteristic of ``free'' wave turbulence. The discretization dependence of our results is discussed in sec.~\ref{sec:DiscErrors} and  we determine the dynamical scaling exponents in sec.~\ref{sec:VScalAna}. We also consider variations of the initial time $Q\tau_0$ in sec.~\ref{sec:VLate} and illustrate different routes to reach the universal attractor. A kinetic theory analysis of the turbulent thermalization process is presented in sec.~\ref{sec:TTExp}. We close this section with a comparison of our results to different weak coupling thermalization scenarios in sec.~\ref{sec:TheAttractor} and a discussion of possible implications for heavy ion experiments in sec.~\ref{sec:QuoVadis}.

\subsection{Universal scaling}
\label{sec:VUnivScal}
\begin{figure}[t!]
\centering						
\includegraphics[width=0.5\textwidth]{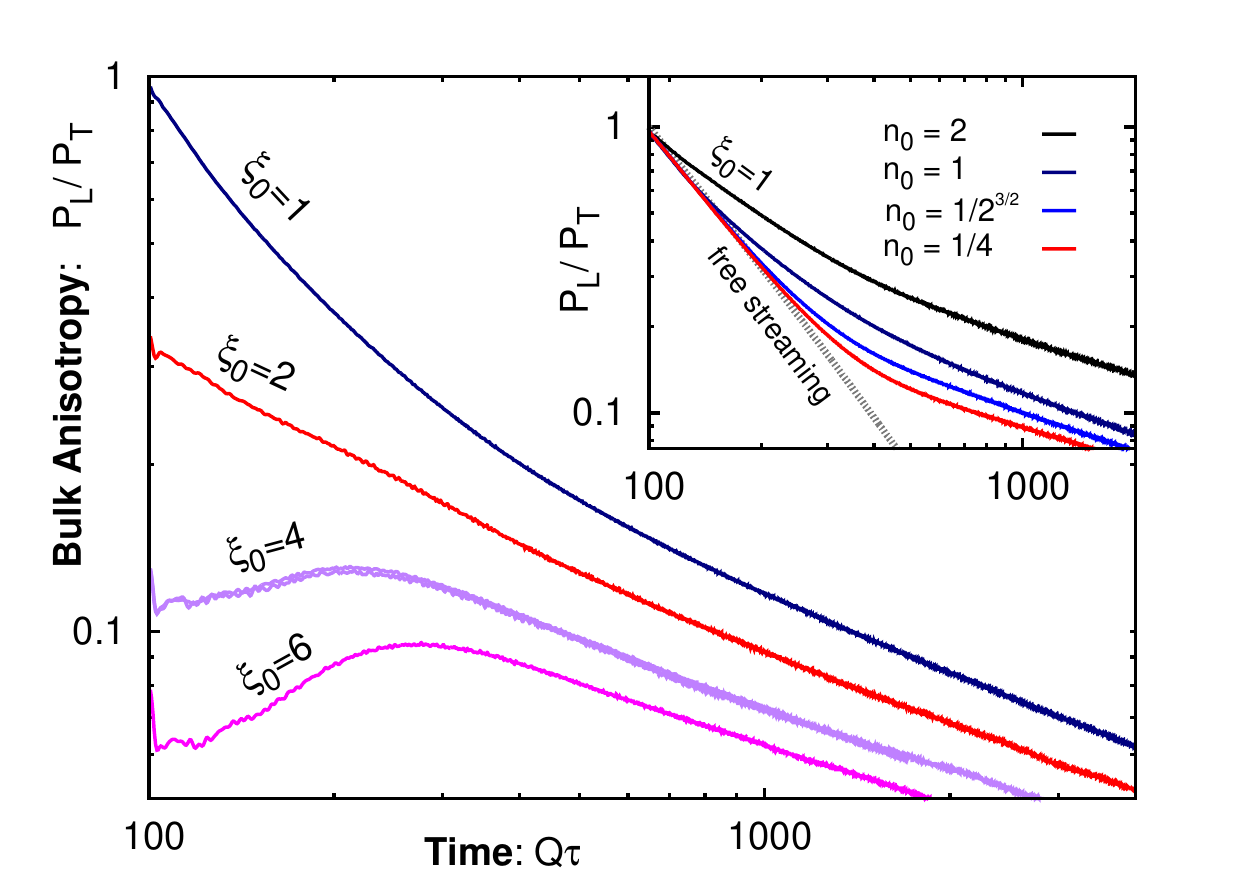}	
\caption{\label{fig:Pressure} (color online) Ratio of longitudinal to transverse pressure as a function of time for different initial anisotropies $\xi_0$ and fixed initial occupancy $n_0 = 1$. The inset shows the same quantity for initially isotropic systems ($\xi_0=1$) and different initial occupancies $n_0$ along with the free streaming (dashed) curve. See text for a discussion of the time scales on the x-axis in this and subsequent figures.}	
\end{figure}							

\begin{table}[b!]
\begin{tabular}{||c|c|c||c|c|c|c||}
\hline \hline
\multicolumn{3}{||c||}{Configuration} & \multicolumn{4}{c||}{Lattice parameters} \\
 $\xi_0$ & $n_0$ & $Q\tau_0$ & $\Nt$ & $N_\eta$ & $Qa_{\bot}$ & $ a_{\eta}$ \\ \hline \hline
$1,~2$ & $1$ & $100$ & $256$ & $2048$ & $1.0$ & $1.25\times 10^{-3}$ \\ \hline
$4,~6$ & 1 & $100$ & $512$ & $1024$ & $1.0$ & $2.5\times 10^{-3}$ \\ \hline
$1$ & $2,~1,~\nicefrac{1}{\sqrt{2}},~\nicefrac{1}{2}$ & $100$ & $256$ & $2048$ & $1.0$ & $1.25\times 10^{-3}$ \\ \hline
$1$ & $\nicefrac{1}{\sqrt{8}},~\nicefrac{1}{4}$ & $100$ & $512$ & $1024$ & $1.0$ & $2.5 \times 10^{-3}$ \\ \hline \hline
$1$ & $1$ & $1000$ & $256$ & $2048$ & $0.5$ & $6.25 \times 10^{-5}$ \\ \hline \hline
\end{tabular}
\caption{\label{tab:Discretization} Discretization parameters for different initial conditions. Unless stated otherwise, these parameters are employed for all classical-statistical lattice simulations.}
\end{table}

We first study the time evolution of the bulk anisotropy of the system. In fig.~\ref{fig:Pressure} we show the ratio of longitudinal to transverse pressure of the system $P_L/P_T$ as a function of time. The curves in the main plot are for different initial anisotropies $\xi_0$ and fixed initial occupancy $n_0 = 1$. The curve for $\xi_0=1$ corresponds to an initially isotropic system, whereas for $\xi_0=2,4,6$ the plasma is already anisotropic at initial time $Q\tau_0=100$. The inset shows the results for different initial occupancies $n_0=2,1,1/\sqrt{8},1/4$ and an initially isotropic system $(\xi_0=1)$.\\

When starting from an isotropic initial distribution ($\xi_0 = 1$), the system is seen in fig.~\ref{fig:Pressure} to become more and more anisotropic with time as a consequence of the longitudinal expansion. Indeed the early time behavior is governed by free streaming whereas at later times the anisotropy of the system increases more slowly as a consequence of interactions. This transition is further elaborated on in the inset where the free streaming (dashed) curve is shown for comparison. One observes that systems that are initially more dilute ($n_0<1$) exhibit a longer period of free streaming behavior before the transition to the scaling behavior occurs. The results for strong initial anisotropies, such as for $\xi_0=4$ and $6$, show a different behavior at early times. More specifically, one observes a short transient regime where the ratio $P_L/P_T$ increases. On a qualitative level, this behavior resembles previous studies of the Glasma evolution in refs.~\cite{Romatschke:2005pm,Romatschke:2006nk,Romatschke:2005ag,Fukushima:2011nq,Berges:2012cj} and may be attributed to plasma instabilities. However the choice of initial conditions and different dynamics at early times does not affect the evolution at later times. Most remarkably, all curves show the same characteristic scaling behavior after the transient regime. \\

\begin{figure}[t!]
\centering						
\includegraphics[width=0.5\textwidth]{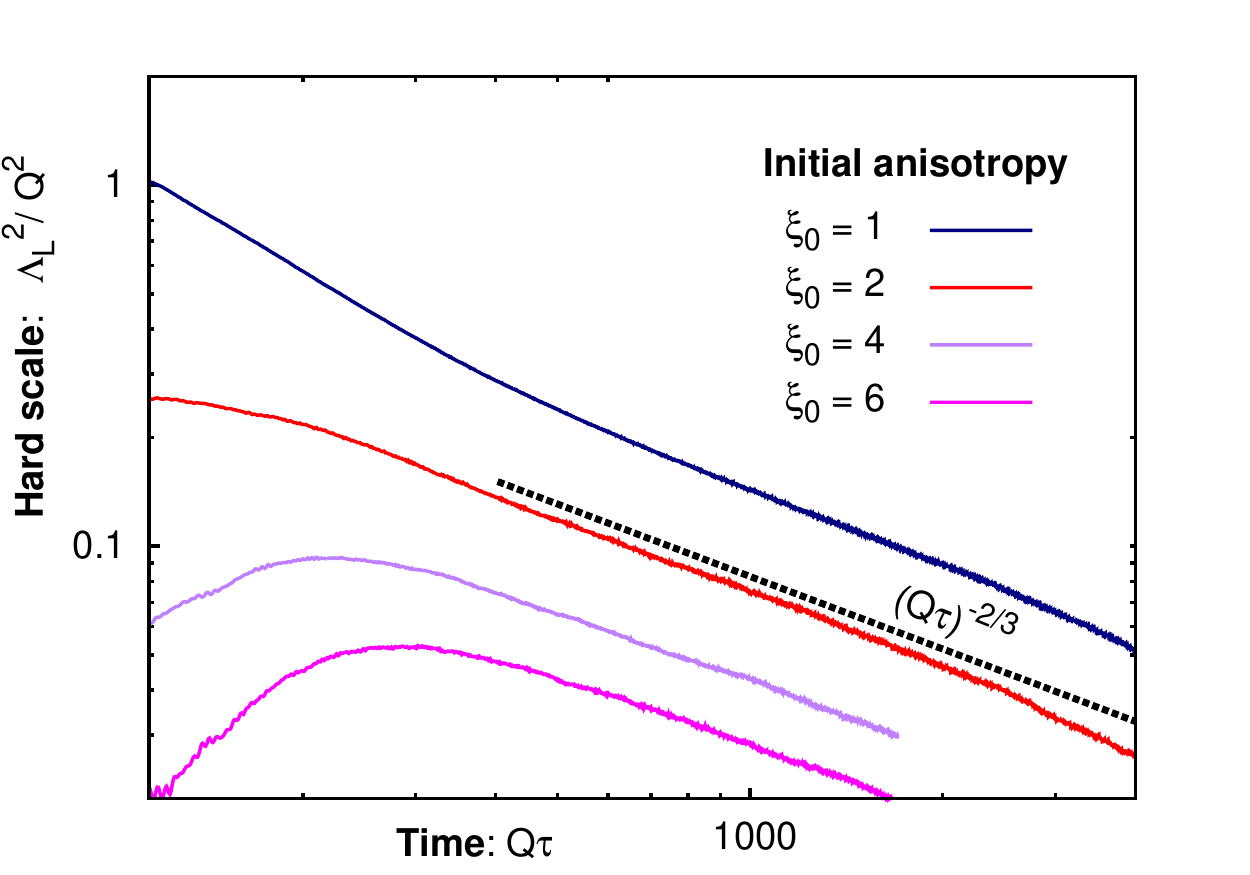}	
\caption{\label{fig:HardScale} (color online) Time evolution of the characteristic longitudinal momentum scale for different initial anisotropies $\xi_0$. At late times all curves exhibit the same scaling behavior.}	
\end{figure}							

\begin{figure}[t!]
\centering						
\includegraphics[width=0.5\textwidth]{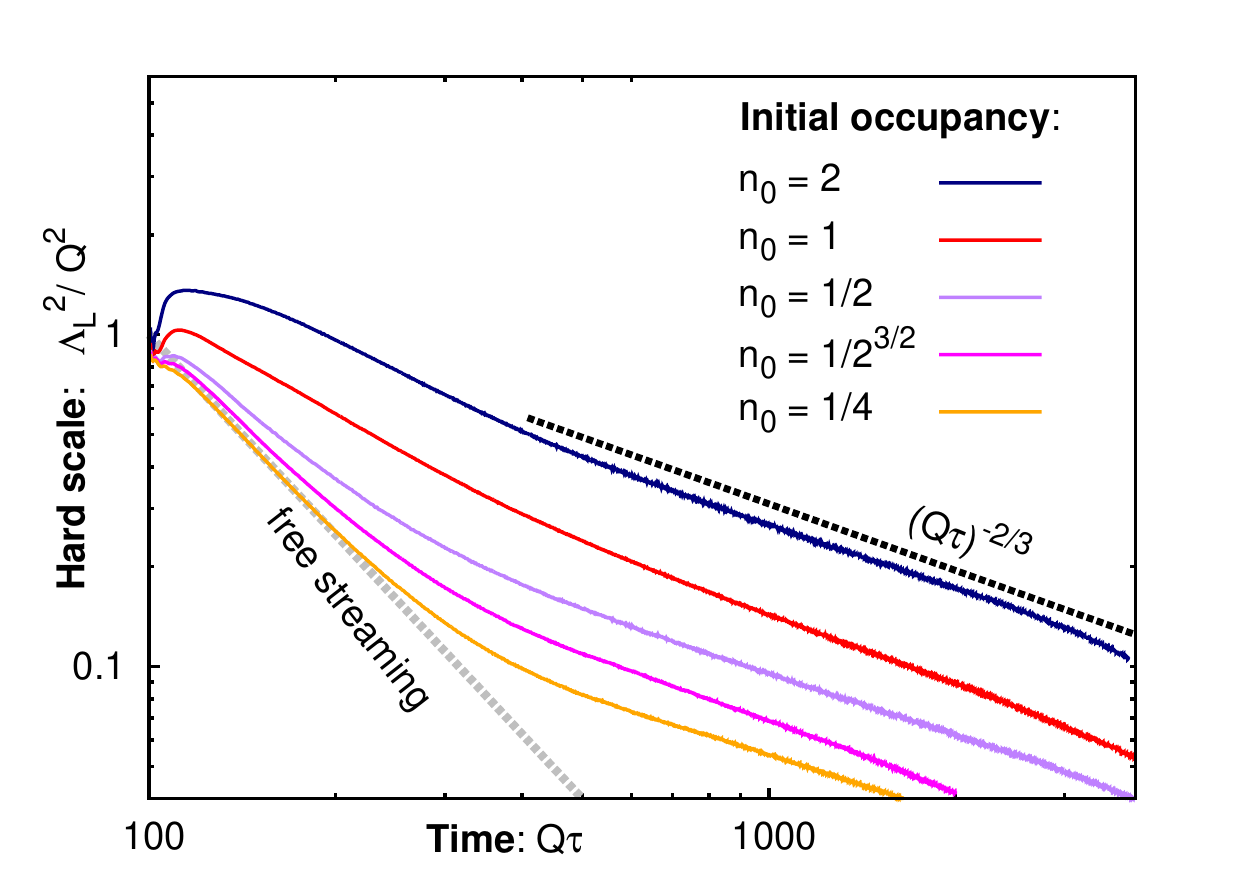}	
\caption{\label{fig:HardScaleN} (color online) Time evolution of the characteristic longitudinal momentum scale for different initial occupancies $n_0$. The curves for small initial occupancies show a transition from initial free streaming behavior to universal scaling at late times.}	
\end{figure}							
To analyze this behavior in greater detail, we study the time evolution of the transverse and longitudinal hard momentum scales $\Lambda_T$ and $\Lambda_L$. As stated previously in sec.~\ref{sec:HardScaleDef}, these gauge-invariant observables characterize respectively the typical transverse and longitudinal momenta of hard excitations. The advantage of the hard scale observables, relative to components of the stress energy tensor, is that they probe harder excitations of the system. This is particularly the case for the longitudinal components: while the longitudinal pressure $P_L$ is dominated by excitations with relatively small transverse momenta, the longitudinal and transverse hard scales $\Lambda_L$ and $\Lambda_T$ probe the system at the same characteristic momentum scale. This allows for a more straightforward interpretation in terms of a kinetic description.\\ 

The time evolution of the longitudinal hard scale $\Lambda_L^2$ is shown in fig.~\ref{fig:HardScale} for different initial anisotropies $\xi_0$ and initial occupancy $n_0 = 1$. In fig.~\ref{fig:HardScaleN}, we show the result for different initial occupancies $n_0$ for an initially isotropic system $(\xi_0=1)$. Again the dynamics at early times is very sensitive to the initial conditions and ranges from an approximate free streaming behavior observed for very dilute systems $(n_0=1/4)$ to a rapid increase of the hard scale observed for very anisotropic systems $(\xi_0=6)$. However, after the transient regime, one clearly observes the emergence of a universal power-law dependence.\\

This universal behavior of the hard scales can be characterized in terms of the scaling exponents $\gamma$ and $\beta$ as
\begin{equation}
\Lambda_L^2(\tau) \propto Q^2~(Q \tau)^{-2 \gamma}\, , \quad \Lambda_T^2(\tau) \propto Q^2~(Q \tau)^{-2 \beta} .
\label{C3eq:scales}
\end{equation}
The comparison to the dashed curves $\propto (Q \tau)^{-2/3}$ in figs.~\ref{fig:HardScale} and \ref{fig:HardScaleN} indicates an approximate value of $\gamma \simeq 1/3$. The quantitative extraction of the scaling exponents is discussed in more detail in sec.~\ref{sec:VScalAna}.\\

The observed value of $\gamma \simeq 1/3$ should be contrasted to a free streaming system where one obtains $\gamma=1$. As also shown in fig.~\ref{fig:HardScaleN}, the redshift due to free streaming would lead to a much faster decrease of longitudinal momenta\footnote{In the free streaming case, the scaling behavior arises not due to universality but simply due to the absence of interactions.}. Therefore the comparatively slow increase of anisotropy associated to the value of $\gamma\simeq1/3$ can only be explained due to the persistence of strong interactions in the system which continuously increase the longitudinal momenta of excitations relative to the free streaming behavior. The observed behavior thus provides a direct verification of a strongly interacting system throughout the entire evolution.\\ 

The transition to the scaling regime occurs around $Q\tau\gtrsim650$ for an initially isotropic system with $\xi_0=n_0=1$. While at first sight this time scale may appear to be very large, the dynamics takes place on a much longer time scale in the limit of weak coupling. Consequently, the above time scale is actually very small compared to the overall extent of the classical regime and even smaller compared to the time scale when thermalization occurs at weak coupling. We also note that the time scale for the transition to the scaling regime depends on the initial conditions. For instance the results in fig.~\ref{fig:HardScaleN} suggest that the higher the occupancy factor $n_0$, the shorter the free streaming regime, and the more rapid the approach to universal behavior. Conversely, the dependence on the initial anisotropy $\xi_0$ observed in fig.~\ref{fig:HardScale} appears to be more complicated. The relevance of these time scales for heavy ion collisions was discussed in sec.~\ref{sec:AlsoDiscussed} and is discussed further in sec.~\ref{sec:QuoVadis}.\\

\begin{figure}[t!]
\centering						
\includegraphics[width=0.5\textwidth]{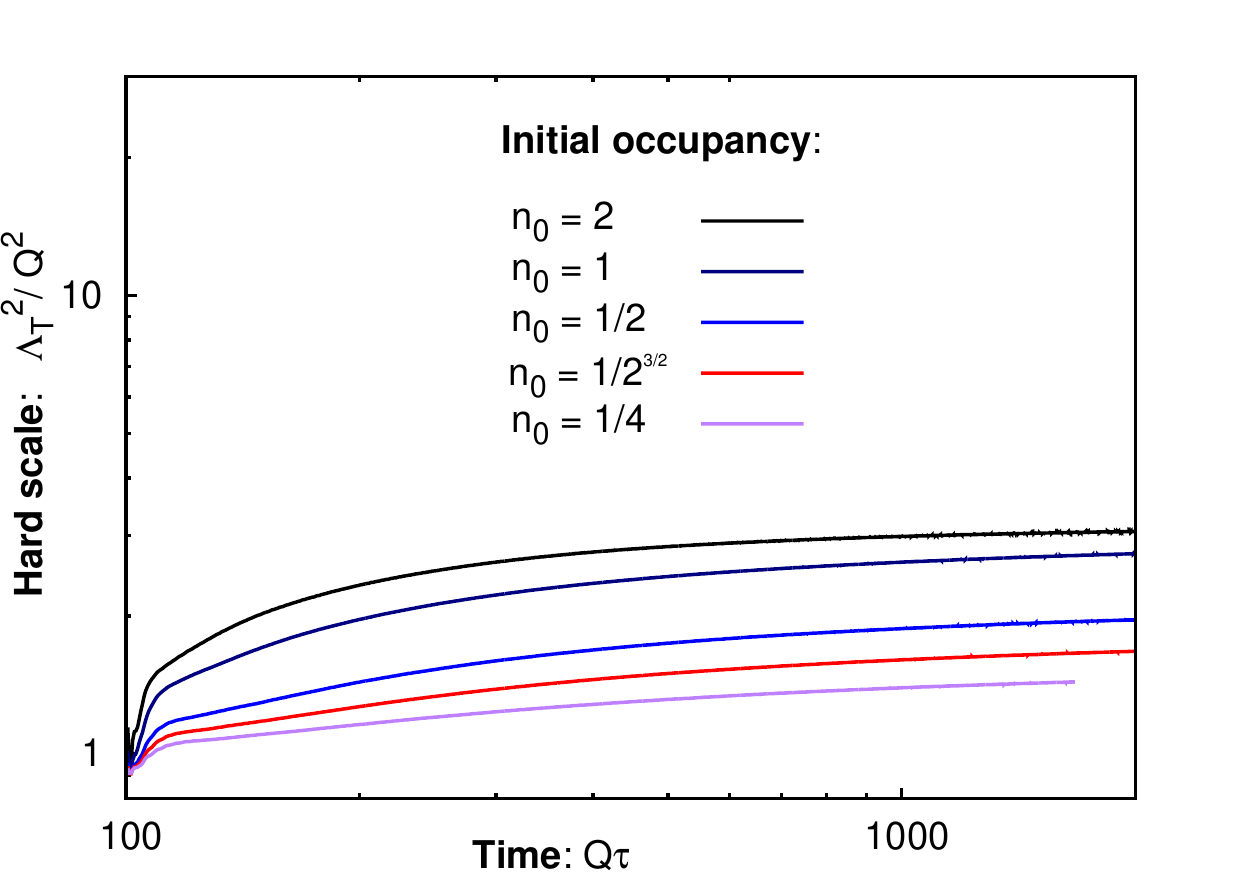}	
\caption{\label{fig:HardScaleTrans} (color online) Time evolution of the characteristic transverse momentum scale for different initial occupancies $n_0$. The curves show an approximately constant behavior in the scaling regime.}	
\end{figure}							

The time evolution of the characteristic transverse momentum scale $\Lambda_T$ is shown in fig.~\ref{fig:HardScaleTrans} for different initial occupancies $n_0$ and fixed initial anisotropy $\xi_0=1$. At very early times one observes a rapid hardening which is more pronounced for higher occupancies. This is a consequence of the large overpopulation and can be understood as a redistribution of the energy into higher momentum modes. Once this is accomplished, one observes that the characteristic transverse momentum scale $\Lambda_T$ stays approximately constant in time. In particular for the scaling regime $(Q\tau\gtrsim 650)$ one finds that the typical transverse momenta of hard excitations do not change appreciably. This corresponds to $\beta\simeq0$ in eq.~(\ref{C3eq:scales}).

\subsection{Spectral properties and self-similarity}
\label{sec:ExpSelfSimilar}
\begin{figure}[t!]
\centering						
 \includegraphics[width=0.5\textwidth]{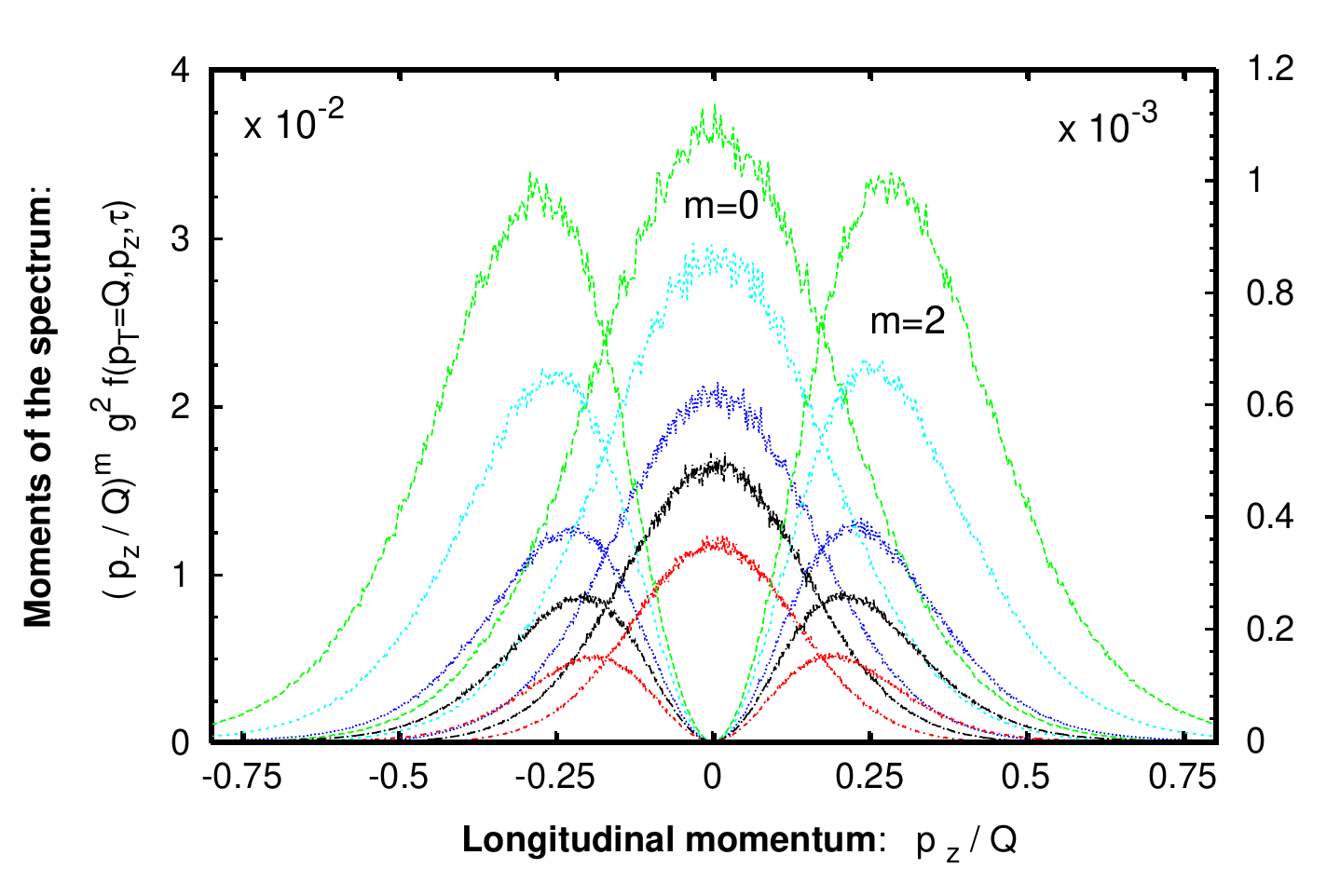}
 \includegraphics[width=0.5\textwidth]{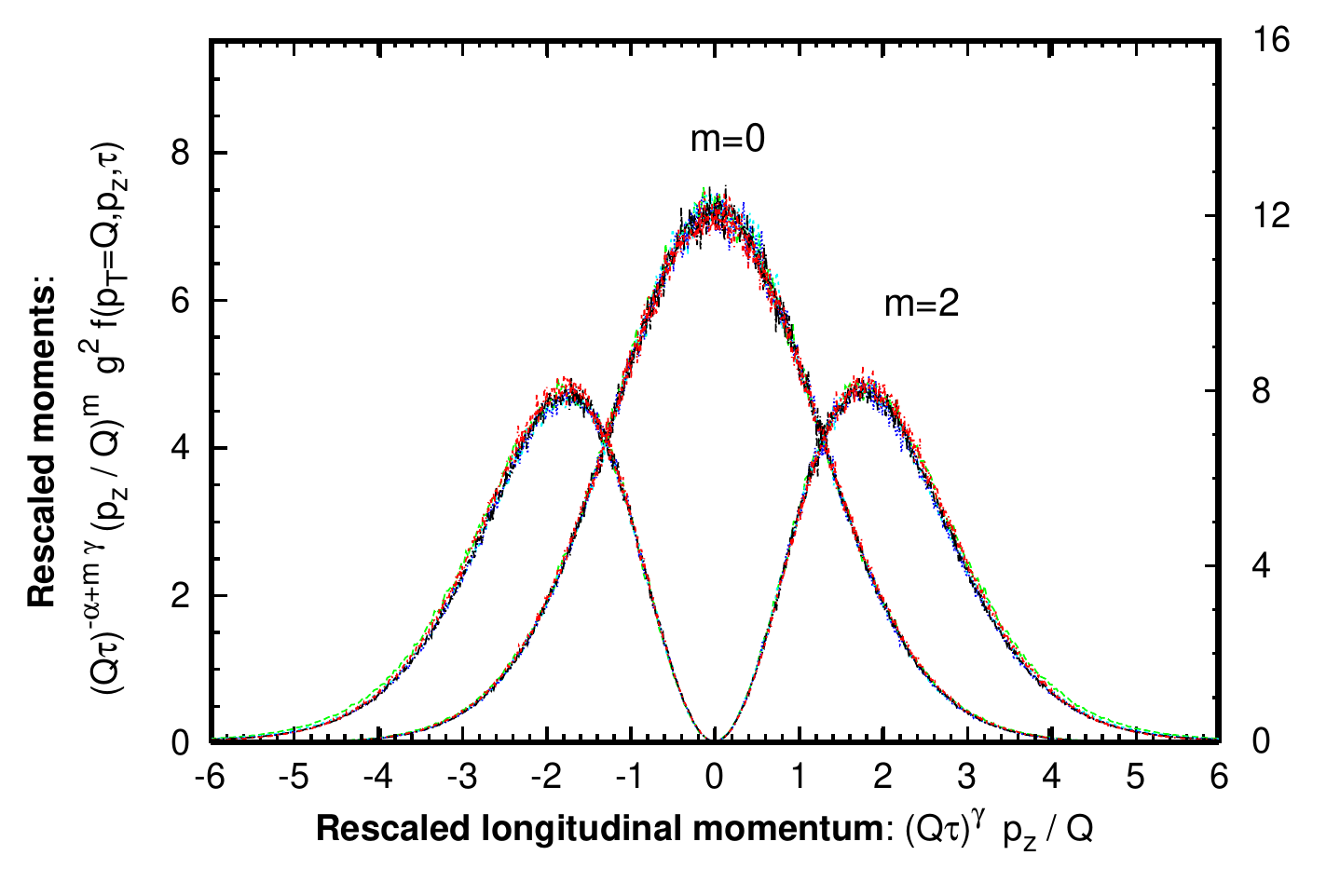}
  \caption{\label{fig:self-similar} (color online) (\textbf{top}) Moments of the single particle distribution function as a function of longitudinal momenta. Note that the longitudinal spectra are evaluated at transverse momentum $\pt\simeq Q$. Here we used $\Nt=256,~N_{\eta}=4096$ lattices with spacings $Qa_{\bot}=1.0$ and $a_{\eta}=6.25\times 10^{-4}$. The different curves correspond to different times $Q\tau=750,~1000,~1500,~2000,~3000$ (top to bottom) of the evolution. (\textbf{bottom}) The rescaled moments of the distribution function are found to collapse onto a single curve when plotted as a function of the rescaled longitudinal momentum variable.}
\end{figure}
While the gauge invariant observables clearly point to a universal scaling behavior of the different momentum scales in the problem a microscopic understanding of these phenomena can be obtained from the single particle spectra. More specifically, we will now study the time evolution of the gluon distribution function $f(\pt,\pz,\tau)$, which we extract from gauge-fixed equal-time correlation functions as discussed in sec.~\ref{sec:OccupationNumbers}.\\

In the top panel of fig.~\ref{fig:self-similar}, we show the zeroth and second moment of the single particle gluon distribution as a function of longitudinal momentum. The spectra are evaluated at hard transverse momenta $\pt\simeq Q$ and the different curves (top to bottom) correspond to different times $Q\tau=750$ to $3000$ (early to late) in the scaling regime. As a consequence of the longitudinal expansion, the system becomes more and more dilute and the amplitude of the distribution decreases in time. Similarly the typical longitudinal momenta of hard excitations become smaller and smaller, and the width of the maximum decreases in time. Interestingly, the spectral shape of the distribution can be characterized by a (nearly) Gaussian peak structure.\\

As we discussed in sec.~\ref{sec:StaticBox}, a striking property of a turbulent thermalization process is the self-similar temporal evolution of the system. In terms of the gluon distribution function, a self-similar evolution for an expanding system has to fulfill the condition
\begin{eqnarray}
\label{C3eq:scalingf}
f(\pt,\pz,\tau)=(Q\tau)^{\alpha}f_S\Big((Q\tau)^\beta \pt,(Q\tau)^\gamma \pz\Big) ,
\end{eqnarray}
where $f_S$ denotes a {\em stationary} distribution independent of time. In analogy to our prior discussion, the dynamical scaling exponents $\alpha,\beta,\gamma$ are universal and describe the evolution of the system. Since -- unlike the static box case -- longitudinal and transverse momenta can evolve independently in the expanding system, they are characterized by two separate scaling exponents $\beta$ and $\gamma$ in eq.~(\ref{C3eq:scalingf}). The scaling exponents $\beta$ and $\gamma$ agree with the previous definition in eq.~(\ref{C3eq:scales}), as can be verified by evaluating the perturbative expression for the hard scales in eq.~(\ref{lat:LambdaApprox}). The parametrization in eq.~(\ref{C3eq:scalingf}) thus effectively amounts to measuring transverse and longitudinal momenta at a given time in units of the characteristic momentum scales $\Lambda_T(\tau)$ and $\Lambda_L(\tau)$ respectively. As previously, the scaling exponent $\alpha$ describes the overall decrease of the amplitude of the distribution in time.\\ 

To investigate the emergence of self-similarity in our simulations, we follow the same strategy as in sec.~\ref{sec:StaticBox} and study rescaled moments of the single particle distribution at different times. By use of the self-similarity relation (\ref{C3eq:scalingf}) with $\beta\simeq0$ one finds that rescaled moments of the distribution function
\begin{eqnarray}
\left(\frac{\tilde{\pz}}{Q}\right)^{m} f_S(\pt \simeq Q,\tilde{\pz})&=&(Q\tau)^{-\alpha+m\gamma} \left(\frac{\pz}{Q}\right)^{m}\nonumber \\
&&\times f(\pt\simeq Q,\pz,\tau)\;, 
\end{eqnarray}
yield a stationary distribution when plotted as a function of the rescaled longitudinal momentum $\tilde{\pz}=(Q\tau)^\gamma \pz$. This is shown in the lower panel of fig.~\ref{fig:self-similar} where we plot the rescaled moments of the distribution as a function of the rescaled longitudinal momenta. Here we employ the scaling exponents $\alpha=-0.8$ and $\gamma=0.28$ to achieve optimal matching.\footnote{The alert reader may wonder about the comparatively large value of $\alpha=-0.8$, which we employed in fig.~\ref{fig:self-similar}. As we will discuss in sec.~\ref{sec:VScalAna}, the values of $\alpha$ and $\beta$ are strongly correlated. Consequently, small deviations from $\beta=0$, which is implicitly assumed in the analysis of fig.~\ref{fig:self-similar}, already have a significant impact on the observed values of $\alpha$. The combined analysis of all scaling exponents is discussed in more detail in sec.~\ref{sec:VScalAna}.} Indeed, the rescaled data for different times is seen to collapse onto a single curve. As for the static box case, this result is a striking manifestation of self-similarity.\\

\begin{figure}[t!]
\centering						
 \includegraphics[width=0.5\textwidth]{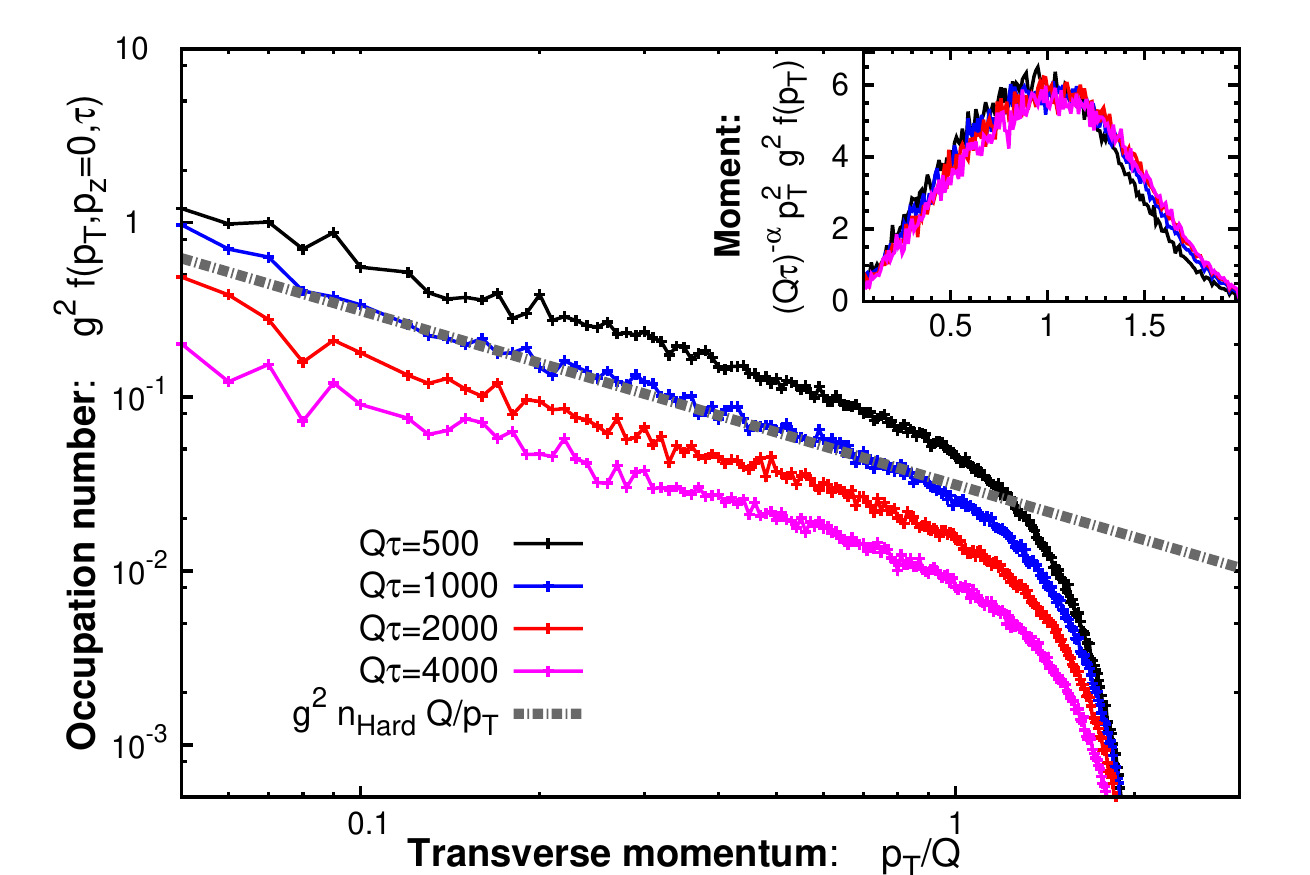}
  \caption{\label{fig:ptSpectra} (color online) Dependence of the gluon distribution function on transverse momentum $\pt$ for $\pz=0$ at different times. The inset shows the self-similar evolution for the rescaled second moment of the distribution as a function of transverse momentum $\pt$ for vanishing longitudinal momentum $\pz=0$.}
\end{figure}

Fig.~\ref{fig:ptSpectra} displays the distribution, now as a function of transverse momentum $\pt$ for vanishing longitudinal momentum ($\pz=0$) at different times. Similarly to the non-expanding case, the spectrum can be characterized as a power-law over a large range of transverse momenta $\pt\lesssim\Lambda_T$ and a rapid fall-off for higher momenta $\pt\gtrsim\Lambda_T$. In the expanding case, the spectral index of the power-law is close to the classical thermal value $\kappa=1$ as indicated by the gray dashed line in fig.~\ref{fig:ptSpectra}. We also verified explicitly that simulations for different initial conditions lead to the same spectral shape of the attractor.\\ 

The spectrum in fig.~\ref{fig:ptSpectra} clearly shows a self-similar evolution with a decreasing amplitude 
\begin{equation}
\label{C3eq:nharddef}
\nHard(\tau)=f(\pt\simeq Q,\pz=0,\tau)\;,
\end{equation}
while the position of the hard ``cut-off scale'' $\Lambda_T$ remains approximately constant in time.\footnote{Note that since $\Lambda_T\simeq Q$ is approximately constant, the definition of the hard scale occupancy is analogous to the one in eq.~(\ref{eq:Sec4nHard}) for the non-expanding system.} This result is further elaborated on in the inset of fig.~\ref{fig:ptSpectra}, where we show the rescaled second moment of the transverse distribution, $(Q\tau)^{-\alpha} \pt^2~f(\pt,\pz=0,\tau)$ on a linear scale. Indeed, the position of the peak is seen to remain at the same position without further rescaling of the momentum axis applied. Taking into account the rescaling of the overall amplitude with $\alpha=-0.8$, the results at different times again collapse onto a single curve to good accuracy.

\subsection{Discretization errors}
\label{sec:DiscErrors}
We will now discuss the discretization dependence of our results and demonstrate the level of convergence of our simulations. The major challenge arises from the fact that the characteristic momentum scales are time dependent and need to be properly resolved on the spatial lattice at all times of the simulation. If this was not the case, simulations will show lattice artifacts, which can accumulate in time. The classical dynamics on the lattice can then become very different from that of the underlying continuum field theory and a reliable extraction of continuum physics becomes impossible.\\

In general, discretization errors appear due to the infrared cut-off set by the (inverse) size of the lattice
\begin{eqnarray}
\Omega_{\bot}^{IR}=\frac{\pi}{\Nt a_{\bot}}\;, \quad \Omega_z^{IR}=\frac{\pi}{N_{\eta} \tau a_{\eta}}\;,
\end{eqnarray} 
and/or the ultraviolet cutoff set by the (inverse) lattice spacing
\begin{eqnarray}
\Omega_{\bot}^{UV}=\frac{\pi}{a_{\bot}}\;, \quad \Omega_{z}^{UV}=\frac{\pi}{\tau a_{\eta}}\;.
\end{eqnarray}
They are more severe for longitudinally expanding non-Abelian plasmas since 
\begin{itemize}
 \item[I)] The longitudinal ultraviolet cutoff scale $\Omega_z^{UV}\sim \tau^{-1}$ is explicitly time dependent.
 \item[II)] The time evolution of the physical scales is much faster in the longitudinally expanding case.
\end{itemize}
With regard to the latter point, the physical scales of the system are given by the transverse and longitudinal hard momentum scales $\Lambda_{T/L}$, and a soft scale of the order of the Debye screening scale $m_D$. In kinetic theory, the soft scale can be estimated to be 
\begin{eqnarray}
\label{eq:PertMDebye}
m^2_D(\tau)=4 g^2 N_c \int \frac{\dInt^2\pvect}{(2\pi)^2}\frac{\dInt\pz}{(2\pi)} \frac{f(\pt,\pz,\tau)}{\sqrt{\pt^2+\pz^2}}\;.
\end{eqnarray}
Based on the discussion in secs.~\ref{sec:ExpSelfSimilar} and \ref{sec:VScalAna}, we can estimate the time evolution in the scaling regime, where one finds that
\begin{eqnarray}
\Lambda_{T}\sim \text{const.} \;, \quad \Lambda_{L}\sim (Q\tau)^{-1/3} \;, \quad \nHard\sim(Q\tau)^{-2/3}\;, \nonumber \\
\end{eqnarray}
and the hard contribution to the screening scale behaves as
\begin{eqnarray}
 m_D\sim \sqrt{\Lambda_L \Lambda_T \nHard} \sim (Q\tau)^{-1/2}\;.
\end{eqnarray}
This temporal behavior of the physical scales is problematic to resolve on the lattice. Firstly, the longitudinal UV cut-off $\Omega_z^{UV}$ decreases faster than the physical momentum scale $\Lambda_L$. This will ensure that, at sufficiently late times, the physical scale $\Lambda_L$ will no longer be resolved properly on the finite size lattice. One therefore has to have a sufficiently large lattice in the longitudinal direction to guarantee that this happens beyond all times of interest. Secondly, the soft scale $m_D$ decreases quickly in time while the transverse IR cut-off $\Omega_\bot^{IR}$ remains constant. Thus in this case as well soft modes $\sim m_D$ will no longer be resolved on a given transverse lattice at late times. If the soft scale were not resolved, processes that may play a key role in thermalization (such as soft splittings and plasma instabilities) will be affected.\\

\begin{figure}[t!]
\centering						
\includegraphics[width=0.5\textwidth]{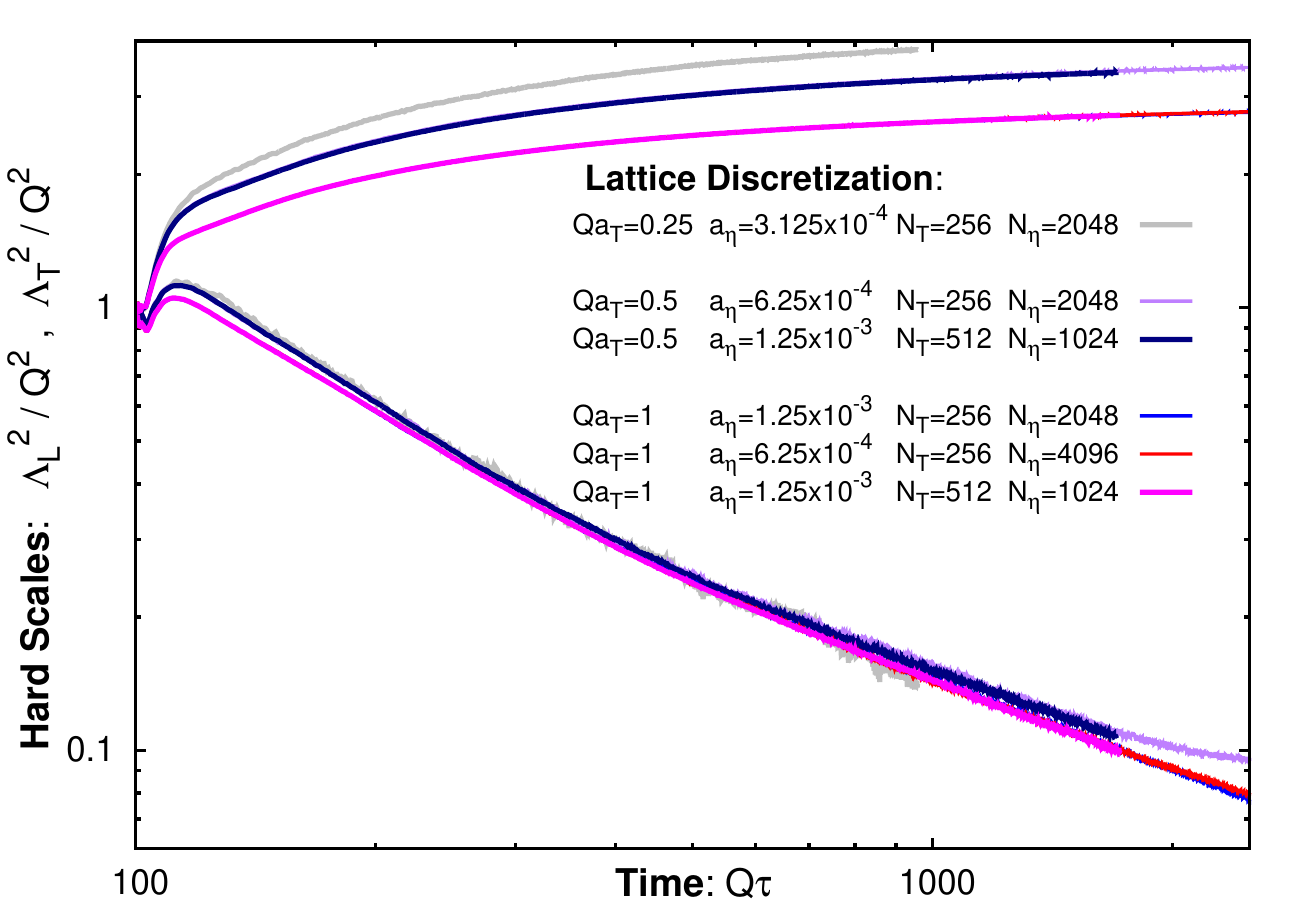}	
\caption{\label{fig:HardScaleDisc} (color online) Time evolution of the characteristic longitudinal and transverse momentum scales for different lattice discretizations.}	
\end{figure}							

Fortunately, it is possible to overcome these challenges on realistic (computer simulation) time scales by performing numerical simulations with sufficiently large transverse volumes and small longitudinal lattice spacings which ensure that physical observables become independent of the size of the system and the lattice spacing at all times of the simulation. To achieve this goal, we monitor the convergence of our results under variation of the lattice spacing and the number of lattice sites.\\

In fig.~\ref{fig:HardScaleDisc}, we show the time evolution of the characteristic longitudinal and transverse momentum scales $\Lambda_{T/L}$ for fixed initial conditions $(\xi_0=n_0=1)$ and different lattice discretizations. The upper curves for $\Lambda_T^{2}$, clearly show a residual dependence of the amplitude on the transverse lattice spacing $a_\bot$. However, the late time behavior of the curves for different $a_\bot$ is identical with $\Lambda_T\sim \text{const.}$ and the results are independent of the transverse size of the system $\Nt a_{\bot}$ as well as the longitudinal discretization. By investigating the (transverse) single particle spectra (as in fig.~\ref{fig:ptSpectra}) for different $a_{\bot}$, we found that the difference in amplitude can be attributed to modifications of the high momentum fall-off. Most importantly, we found that this residual dependence does not change the scaling behavior as has also been observed in previous simulations of non-expanding non-Abelian plasmas~\cite{Kurkela:2012hp}.\\

The lower curves in fig.~\ref{fig:HardScaleDisc} show the time evolution of the longitudinal hard scale $\Lambda_L^{2}$. The results for times $Q\tau\lesssim 1000$ (where data is available from all simulations) appear to have converged well with respect to all discretization parameters. At late times $Q\tau\gtrsim 2000$, one notices a deviation of the purple curve, which corresponds to the results for $Qa_{\bot}=0.5,~a_{\eta}=6.25\times 10^{-4}, \Nt=256$ and $N_{\eta}=2048$ with the smallest transverse lattice volume. This is also characteristic for the behavior we observed for even smaller transverse volumes not shown in fig.~\ref{fig:HardScaleDisc}. We found that they follow the universal scaling behavior for a limited amount of time -- controlled by the transverse lattice size -- before they start to show significant deviations from the large volume behavior. However the fact that $\Lambda_L^{2}$ shows a consistent scaling behavior over a large time scale, and for all curves in fig.~\ref{fig:HardScaleDisc}, leads us to conclude that there is no significant discretization dependence of the results for the largest lattices examined.\\

\begin{figure}[t!]
\centering						
\includegraphics[width=0.5\textwidth]{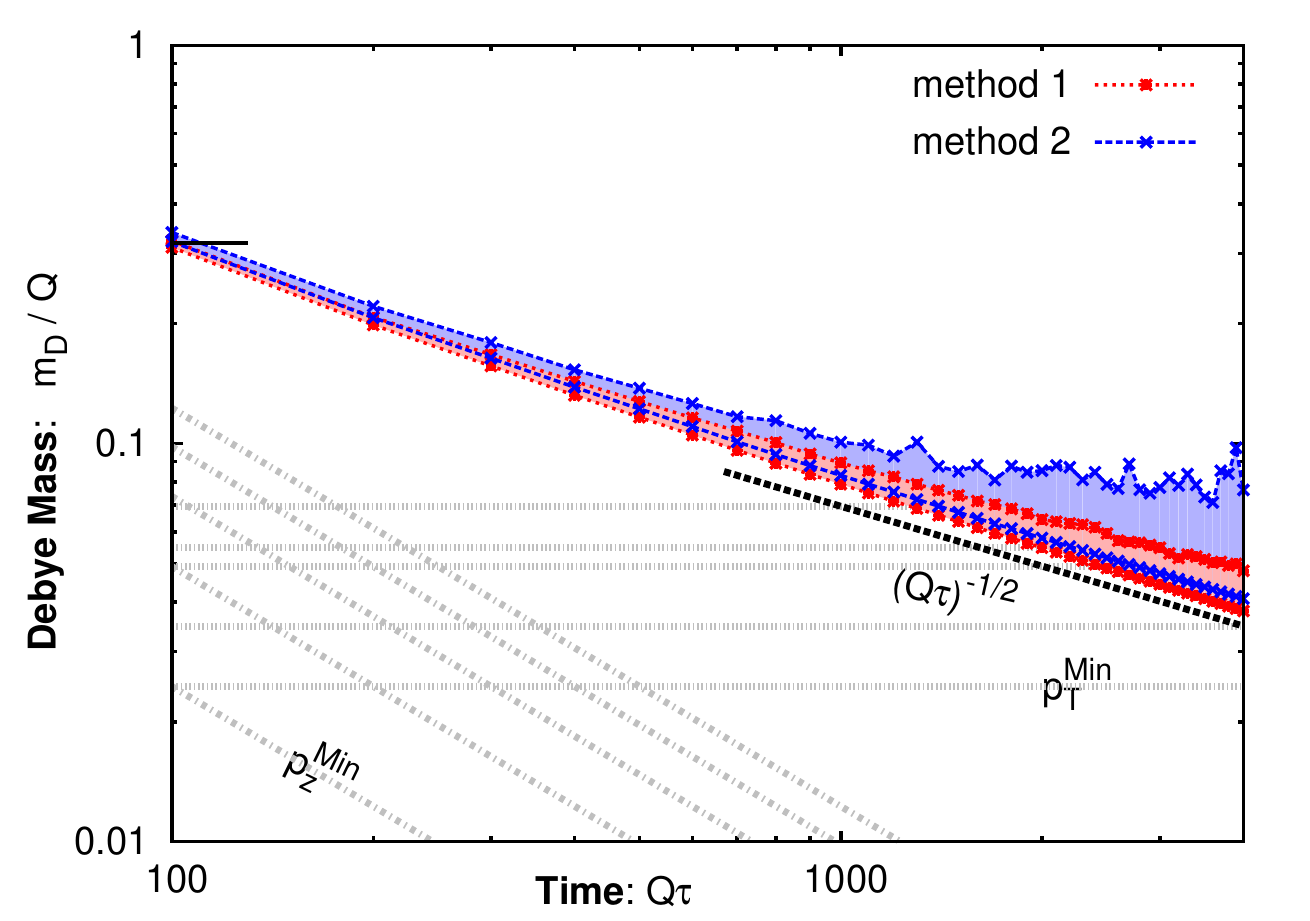}	
\caption{\label{fig:Debye} (color online) Time evolution of the Debye scale, extracted from the perturbative expression in eq.~(\ref{eq:PertMDebye}). The smallest lattice momenta in the longitudinal and transverse directions are also indicated by gray dashed lines.}	
\end{figure}							

We also verified explicitly that the soft Debye scale is resolved on the lattice at all times of the simulation. For this purpose, we evaluate the perturbative expression in eq.~(\ref{eq:PertMDebye}) in two different ways. The first method is to extract the single particle distribution and subsequently evaluate the momentum integral in eq.~(\ref{eq:PertMDebye}) by numerical integration using the continuum formula. The second method is to replace the momentum integral on the right hand side of eq.~(\ref{eq:PertMDebye}) by a sum over all momentum modes on the lattice according to
\begin{eqnarray}
\int \frac{\dInt^3\p}{(2\pi)^3} \to \frac{\tau^{-1}} {V_{\bot} L_{\eta}} \sum_{\pvect,~\pz}\;.
\end{eqnarray}
The results for the first method for $\xi_0=n_0=1$ correspond to the red lines in fig.~\ref{fig:Debye}; the results of the second method are shown as blue lines. Since the definition of quasi-particles is not unambiguous for soft modes, we also evaluated eq.~(\ref{eq:PertMDebye}) excluding all modes with momenta $|\p|<0.1 Q$. This corresponds to the lower curves of each color while the upper curves show the contribution from all modes. The five smallest lattice momenta in the longitudinal and transverse directions for a $\Nt=256,~N_{\eta}=2048$ lattice with spacings $Qa_{\bot}=1,~a_\eta=1.25\times 10^{-3}$ are also indicated in fig.~\ref{fig:Debye} by gray dashed lines. One clearly observes that the Debye scale is resolved at all times of interest, in particular in the regime where scaling is manifest. 
\subsection{Extraction of scaling exponents}
\label{sec:VScalAna}
We will now discuss the quantitative procedure to extract the scaling exponents $\alpha,\beta,\gamma$ from our simulations. This analysis is complicated by the fact that
\begin{itemize}
 \item[I)] scaling only sets in at later times after the transient regime,
 \item[II)] at very late times one is always facing discretization errors. 
\end{itemize}
One can therefore extract the scaling exponents only for a particular range of times which need to be both large enough that one is past the transient regime and small enough that the lattice discretization is sufficiently good. Our strategy to identify this regime is to investigate
\begin{itemize}
\item[a)] the scaling behavior of different initial conditions to see at what time a common scaling behavior is realized,
\item[b)] different lattice discretizations to ensure that the results are independent of the discretization for all times of interest.
\end{itemize}
Based on the results presented in the previous sections, we found these conditions to be realized for times $800\lesssim Q\tau \lesssim2000$. While problems I) and II) can be properly addressed with the above procedure, an additional complication arises when comparing our results to different weak coupling thermalization scenarios in the literature. While the weak coupling thermalization scenarios in~\cite{Baier:2000sb,Bodeker:2005nv,MSW,Kurkela:2011ti,Kurkela:2011ub} are strictly realized only in the limit of very large anisotropy at very late times, our simulations at finite times and finite anisotropy will always show systematic corrections to the predicted scaling behavior. Since different quantities are sensitive to finite time corrections in different ways, a comparison between different extraction methods can be used to estimate these systematic uncertainties. 

\begin{figure}[t!]
\centering						
 \includegraphics[width=0.5\textwidth]{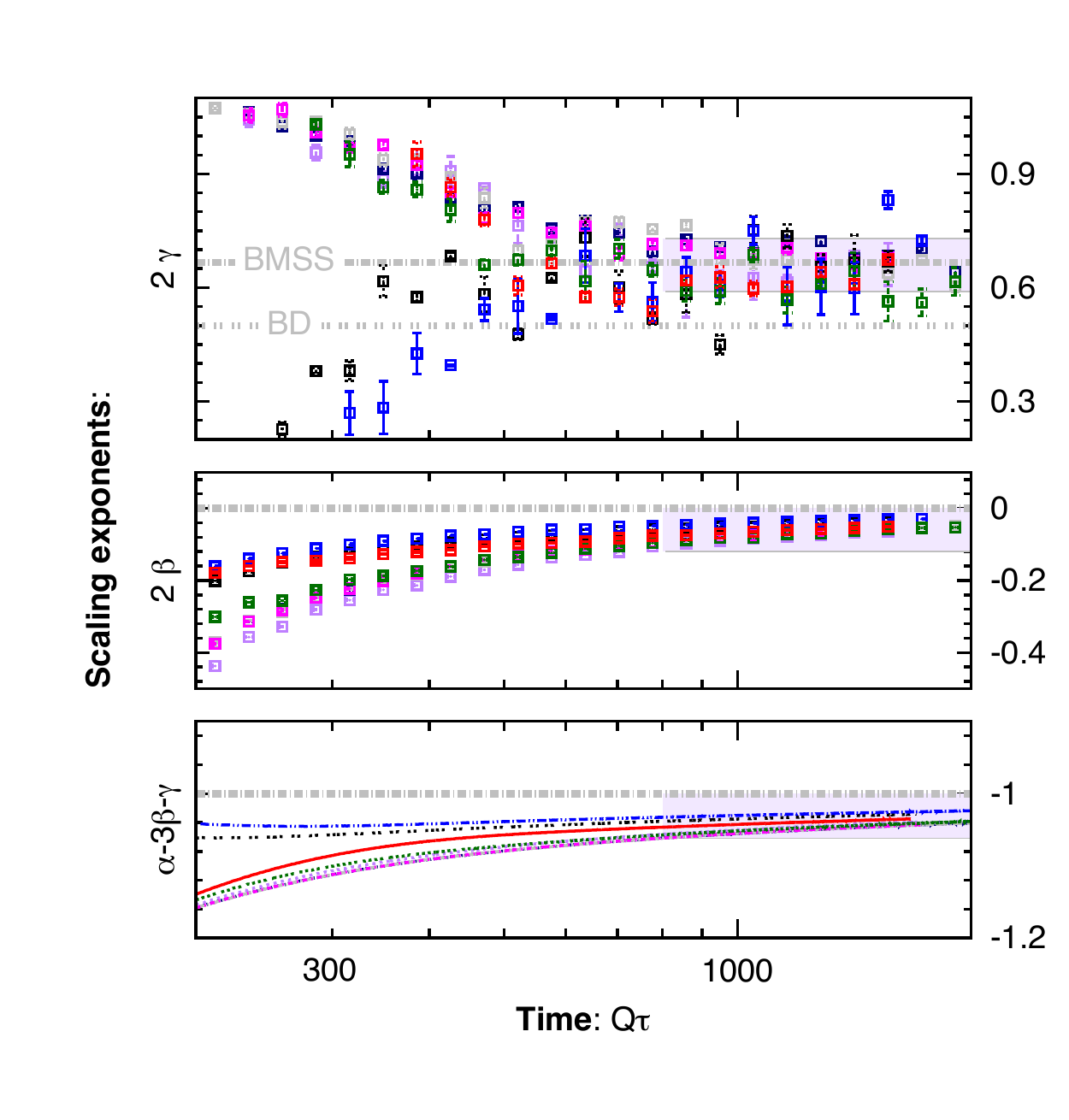}
  \caption{\label{fig:GaugeInvExp} (color online) Local scaling exponents $\beta,\gamma$ and $\alpha-3\beta-\gamma$ as functions of time. The results  are extracted from the logarithmic derivatives of gauge invariant observables for different initial conditions and lattice discretization. In the scaling regime the results for the scaling exponent $\gamma$ for different initial conditions converge to a common scaling exponent. The exponents $\beta$ and $\alpha-3\beta-\gamma$ are subject to systematic scaling corrections which decrease monotonically in time. The gray bands indicate the extracted values in the scaling regime along with their errors. The scaling exponents in the original bottom up scenario~\cite{Baier:2000sb} (BMSS: $\gamma=1/3,\beta=0,\alpha-3\beta-\gamma=-1$) and an instability modified version~\cite{Bodeker:2005nv} (BD: $\gamma=1/4,\beta=0,\alpha-3\beta-\gamma=-1$) are shown for comparison as gray dashed lines.}
\end{figure}

\subsubsection{Gauge invariant analysis}
We will first discuss the extraction of the scaling exponents from the hard scales and the energy momentum tensor. To determine the scaling exponent $\gamma$, we investigate the scaling behavior of the longitudinal hard scale $\Lambda_L^2$. We first divide the data in fig.~\ref{fig:HardScale} in logarithmically equidistant time bins and then locally extract the scaling exponent from the logarithmic derivative
\begin{eqnarray}
2\gamma(\tau)=-\frac{d\log(\Lambda_L^{2}(\tau))}{d \log(\tau)}\;. 
\end{eqnarray}
The result is shown in the top panel of fig.~\ref{fig:GaugeInvExp}, where we present the extracted scaling exponent $2\gamma(\tau)$ as a function of time for a set of four different initial conditions in the range $\xi_0 = 1$ -- $6$ and $n_0 = 0.25$ -- $1$. After the transient regime, where the local exponents are quite different for different initial conditions and subject to large error bars, one observes a clear convergence towards a single value at later times. We also display results from the evolution for $\xi_0=n_0=1$ using four different lattices in the range $\Nt = 256$ -- $512$, $N_\eta = 1024$ -- $4096$ with $Q a_\bot = 0.5$ -- $1$ and $a_\eta = (0.625$ -- $2.5)\cdot 10^{-3}$ to take into account possible discretization dependencies. By averaging over all data points for $800\lesssim Q\tau \lesssim2000$ we obtain the estimate 
\begin{eqnarray}
2\gamma = 0.67\pm0.07\;(\text{stat.})\;,
\end{eqnarray}
as indicated by the gray band in fig.~\ref{fig:GaugeInvExp}. The values of $\gamma$ in the original bottom up thermalization scenario (BMSS)~\cite{Baier:2000sb} ($\gamma=1/3$) and the instability modified bottom up scenario by Boedeker (BD)~\cite{Bodeker:2005nv} ($\gamma=1/4$) are also indicated in fig.~\ref{fig:GaugeInvExp} as horizontal gray dashed lines.\\ 

Similarly, we extract the scaling exponent $\beta$ from the scaling behavior of the transverse hard scale
\begin{eqnarray}
2\beta(\tau)=-\frac{d\log(\Lambda_T^{2}(\tau))}{d \log(\tau)}\;.  
\end{eqnarray}
The result is shown in the middle panel of fig.~\ref{fig:GaugeInvExp}, where we plot $2\beta(\tau)$ as a function of time. One observes that this local scaling exponent, extracted from the lattice data, approaches zero monotonically at late times. The residual deviation from zero is a clear manifestation of the finite time scaling corrections mentioned above. We find that this deviation is 
\begin{eqnarray}
|\beta(\tau)| < 0.06\;,
\end{eqnarray}
for $800\lesssim Q\tau \lesssim2000$ as indicated by the gray band in fig.~\ref{fig:GaugeInvExp}.\\

Finally, we extract the linear combination of scaling exponents $\alpha-3\beta-\gamma$ from the scaling behavior of the energy density
\begin{eqnarray}
\alpha(\tau)-3\beta(\tau)-\gamma(\tau)=\frac{d\log(\epsilon(\tau))}{d\log(\tau)}\;.
\end{eqnarray}
According to Bjorken's law in eq.~(\ref{eq:BjorkenLaw}) one can directly extract this quantity as
\begin{eqnarray}
\label{eq:EDLogDev}
\frac{d\log(\epsilon(\tau))}{d\log(\tau)}=-\left(1+\frac{P_L(\tau)}{\epsilon(\tau)}\right)\;.
\end{eqnarray}
In fig.~\ref{fig:GaugeInvExp}, the right hand side of eq.~(\ref{eq:EDLogDev}) is shown as function of time. One observes that the scaling exponent approaches the anisotropic scaling limit $\alpha-3\beta-\gamma=-1$ monotonically from below. The residual deviation is
\begin{eqnarray}
|\alpha(\tau)-3\beta(\tau)-\gamma(\tau)+1| < 0.05\;, 
\end{eqnarray}
for $800\lesssim Q\tau \lesssim2000$ as indicated by the gray band.

\subsubsection{Self-similarity analysis}
We also performed an alternative analysis to extract the scaling exponents $(\alpha,\beta,\gamma)$ from the self-similar evolution of the single-particle distribution. In this analysis, we compare rescaled moments of the distribution at different times $(Q\tau_{\text{Test}}=1250,1500,1750,2000)$ with those at a reference time $(Q\tau_{\text{Ref}}=1000)$. While for a perfect scaling the different curves should all give the same results, one can attempt to minimize the deviation between the rescaled curves to determine the most appropriate scaling exponents.\\

In practice, deviations from perfect scaling will occur even for the correct set of scaling exponents due to statistical uncertainties of the data as well as systematic deviations from the scaling behavior in eq.~(\ref{C3eq:scalingf}). We quantify these in terms of a likelihood distribution $W(\alpha,\beta,\gamma)$ which allows us to distinguish different sets of scaling exponents. Our method is described in more detail in app.~\ref{app:LikelihoodDist}. \\

\begin{figure}[t!]
\centering						
 \includegraphics[width=0.5\textwidth]{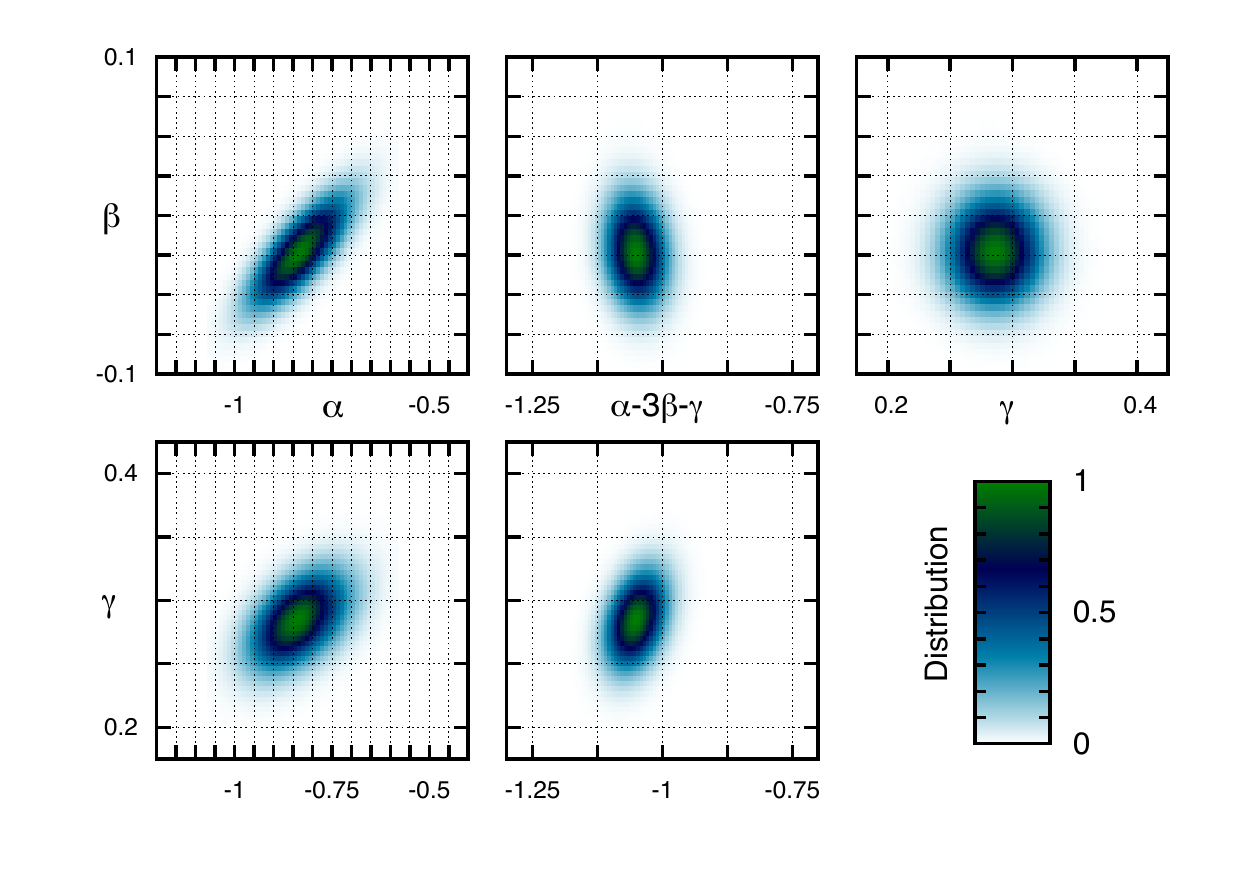}
  \caption{\label{fig:ExpCorr} (color online) Contour plots of the likelihood distribution $W$ as a function of different combinations of the scaling exponents $(\alpha,\beta,\gamma)$. One observes that $\alpha$ is strongly correlated with both $\beta$ and $\gamma$. In contrast the degree of correlation between $\alpha-3\beta-\gamma$ with $\beta$ and $\gamma$ is noticeably smaller. The top right panel also shows that the scaling exponents $\beta$ and $\gamma$ are uncorrelated.}
\end{figure}

\begin{figure}[t!]
\centering						
 \includegraphics[width=0.5\textwidth]{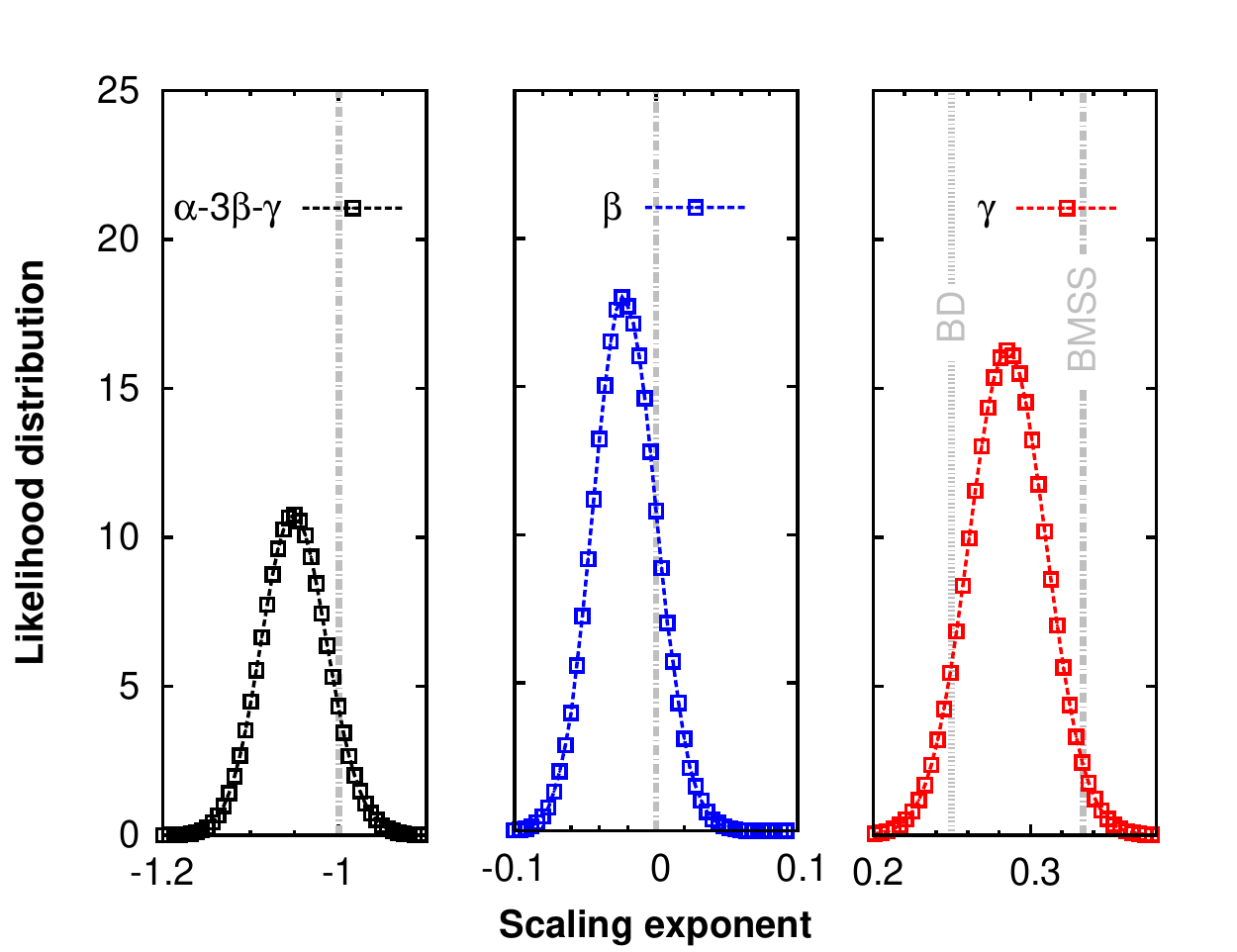}
  \caption{\label{fig:ExpFit} (color online) Likelihood distribution $W$ of the different scaling exponents. The gray vertical lines correspond to the respective values in the bottom up thermalization scenario (BMSS)~\cite{Baier:2000sb} and one of its modified versions (BD)~\cite{Bodeker:2005nv}.}
\end{figure}

The results of our analysis are summarized in figs.~\ref{fig:ExpCorr} and \ref{fig:ExpFit}. In fig.~\ref{fig:ExpCorr} we show contour plots of the likelihood distribution $W$ as a function of different combinations of the scaling exponents $(\alpha,\beta,\gamma)$. These two-dimensional distributions such as $W(\alpha,\beta)$ are obtained by integration over the third exponent not shown in each panel, e.g. $W(\alpha,\beta)=\int \dInt\gamma~W(\alpha,\beta,\gamma)$. They are normalized in such a way that the maximum value is unity in each panel. One observes from fig.~\ref{fig:ExpCorr} that smaller values of $\alpha$ are more likely to appear in combination with smaller values of $\beta$ or $\gamma$ respectively. This correlation between the scaling exponent $\alpha$ and the scaling exponents $\beta$ and $\gamma$ gives rise to the tilted ellipsoids observed in the left panels. However, if one considers their combination $\alpha-3\beta-\gamma$ as shown in the central panels of fig.~\ref{fig:ExpCorr}, the correlation with both $\beta$ and $\gamma$ is significantly reduced. Since $\beta$ and $\gamma$ are also uncorrelated -- as can be inferred from the top right panel -- we conclude that the exponents $\alpha-3\beta-\gamma$, $\beta$ and $\gamma$ can all be determined nearly independently.\\ 

In fig.~\ref{fig:ExpFit}, we show the likelihood distributions for the scaling exponents $\beta$ and $\gamma$ as well as for the linear combination $\alpha-3\beta-\gamma$. The one dimensional distributions are obtained by integration over all other exponents and normalized to yield unity upon integration. The scaling exponents we obtain from fig.~\ref{fig:ExpFit} take the values 
\begin{eqnarray}
\alpha-3\beta-\gamma&=&-1.05\pm0.04\;(\text{stat.})\;, \nonumber \\
\beta&=&-0.02\pm0.02\;(\text{stat.})\;, \nonumber \\
\gamma&=&0.285\pm0.025\;(\text{stat.})\;.
\end{eqnarray}
The corresponding values in the original bottom up thermalization scenario (BMSS)~\cite{Baier:2000sb} ($\gamma=1/3$) and the instability modified bottom up scenario by Boedeker (BD)~\cite{Bodeker:2005nv} ($\gamma=1/4$) are also indicated in fig.~\ref{fig:ExpFit} as vertical gray dashed lines. Since the scaling exponents $\beta=0$ and $\alpha-3\beta-\gamma=-1$ are the same in both scenarios we only show a single line in the left and central panels.

\subsubsection{Discussion}
By extracting the scaling exponents $(\alpha,\beta,\gamma)$ with two different methods we can estimate the systematic uncertainties of our analysis. Concerning the scaling exponents $\beta$ and $\alpha-3\beta-\gamma$ the results of the different methods show very good agreement. The scaling corrections due to  finite time and finite anisotropy effects can be estimated from the gauge invariant analysis and decrease monotonically in time. For the considered range of times, these systematic errors are on the order of $0.06$ for $\beta$ and $0.05$ for the scaling exponent $\alpha-3\beta-\gamma$ of the energy density. Based on the strictly monotonic behavior we can safely extrapolate the values of $\beta$ and $\alpha-3\beta-\gamma$ to the highly anisotropic scaling limit, where
\begin{eqnarray}
\beta \nearrow 0\;,\qquad \alpha-3\beta-\gamma \nearrow -1\;.
\end{eqnarray}
This result is in agreement with the classical evolution in most weak coupling thermalization scenarios.\\

The scaling exponent $\gamma$ can be used to clearly distinguish between different weak coupling thermalization scenarios. As we will discuss in more detail in sec.~\ref{sec:TheAttractor}, the predicted values range from $\gamma=1/8$ in the instability driven scenario by Kurkela and Moore~\cite{Kurkela:2011ub} over a whole family of possible solutions by Mueller, Shoshi and Wong~\cite{MSW} to the value $\gamma=1/3$ in the original bottom up scenario~\cite{Baier:2000sb}. While the result of our gauge invariant analysis  $\gamma \simeq 0.33$, shows a clear preference for the bottom up scenario~\cite{Baier:2000sb} the self-similar scaling analysis points to slightly smaller values of $\gamma \simeq 0.28$, between the values $\gamma=1/3$ in the bottom up scenario~\cite{Baier:2000sb} and $\gamma=1/4$ in the instability modified scenario by Boedeker~\cite{Bodeker:2005nv}. We believe that the deviation between the two methods can be attributed to scaling corrections at finite anisotropy and finite times. While the gauge invariant measurement is certainly more rigorous, these corrections may affect both measurements in a systematic way. Within these uncertainties of our simulations, we are therefore not able to exclude small modifications of the bottom up scenario a la Boedeker.

\begin{figure}[t!]
\centering
\includegraphics[width=0.5\textwidth]{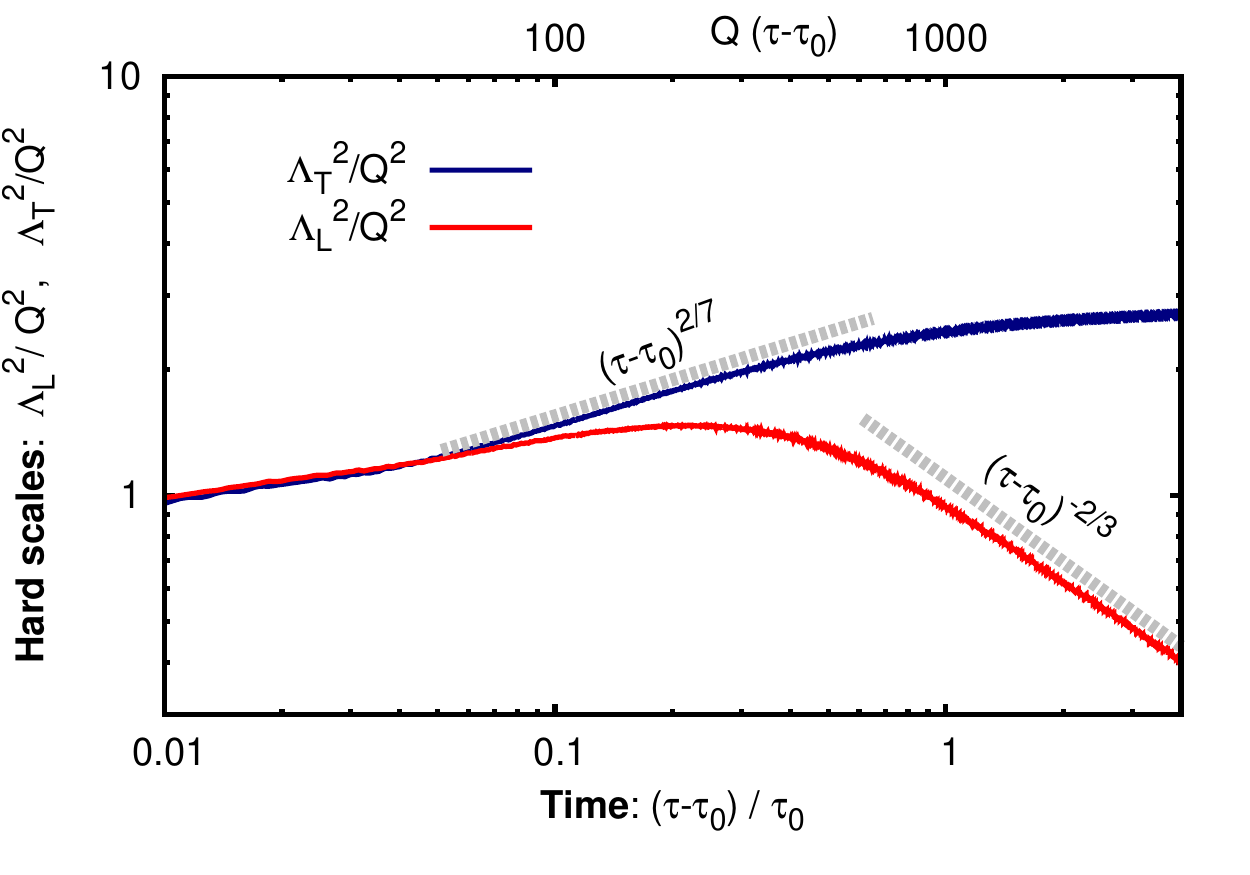}
\caption{\label{fig:Qt1000Lambda} (color online) Time evolution of the transverse and longitudinal hard scales for slowly expanding systems ($Q\tau_0=1000$). One observes a transient regime where the transverse hard scale $\Lambda_T^2$ shows an approximate $(\tau-\tau_0)^{2/7}$ scaling indicated by the gray (dashed) line. At later times, $\Lambda_T^2$ becomes approximately constant whereas $\Lambda_L^2$ exhibits the characteristic $(\tau-\tau_0)^{-2/3}$ scaling.}	
\end{figure}							

\begin{figure}[t!]
\centering
\includegraphics[width=0.5\textwidth]{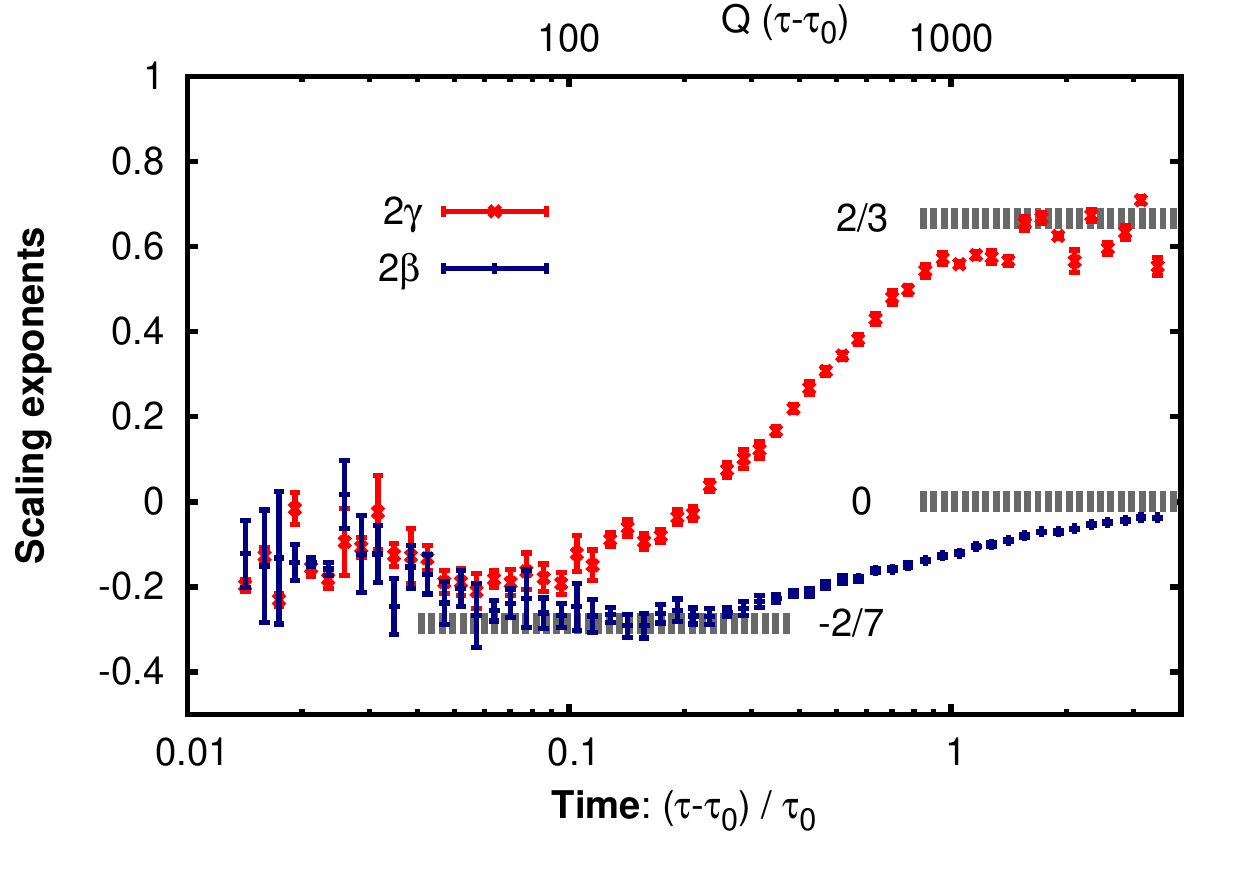}	
\caption{\label{fig:Qt1000Exp} (color online) The local scaling exponents $2\beta$ and $2\gamma$ for a slowly expanding plasma ($Q\tau_0=1000$). One clearly observes the transition between the different scaling regimes for a non-expanding system ($2\beta=2\gamma=-2/7$ -- c.f. sec.~\ref{sec:StaticBox}) and the longitudinally expanding system ($\beta\simeq0$,~$\gamma\simeq2/3$).}
\end{figure}							


\begin{figure}[t!]
\centering
\includegraphics[width=0.5\textwidth]{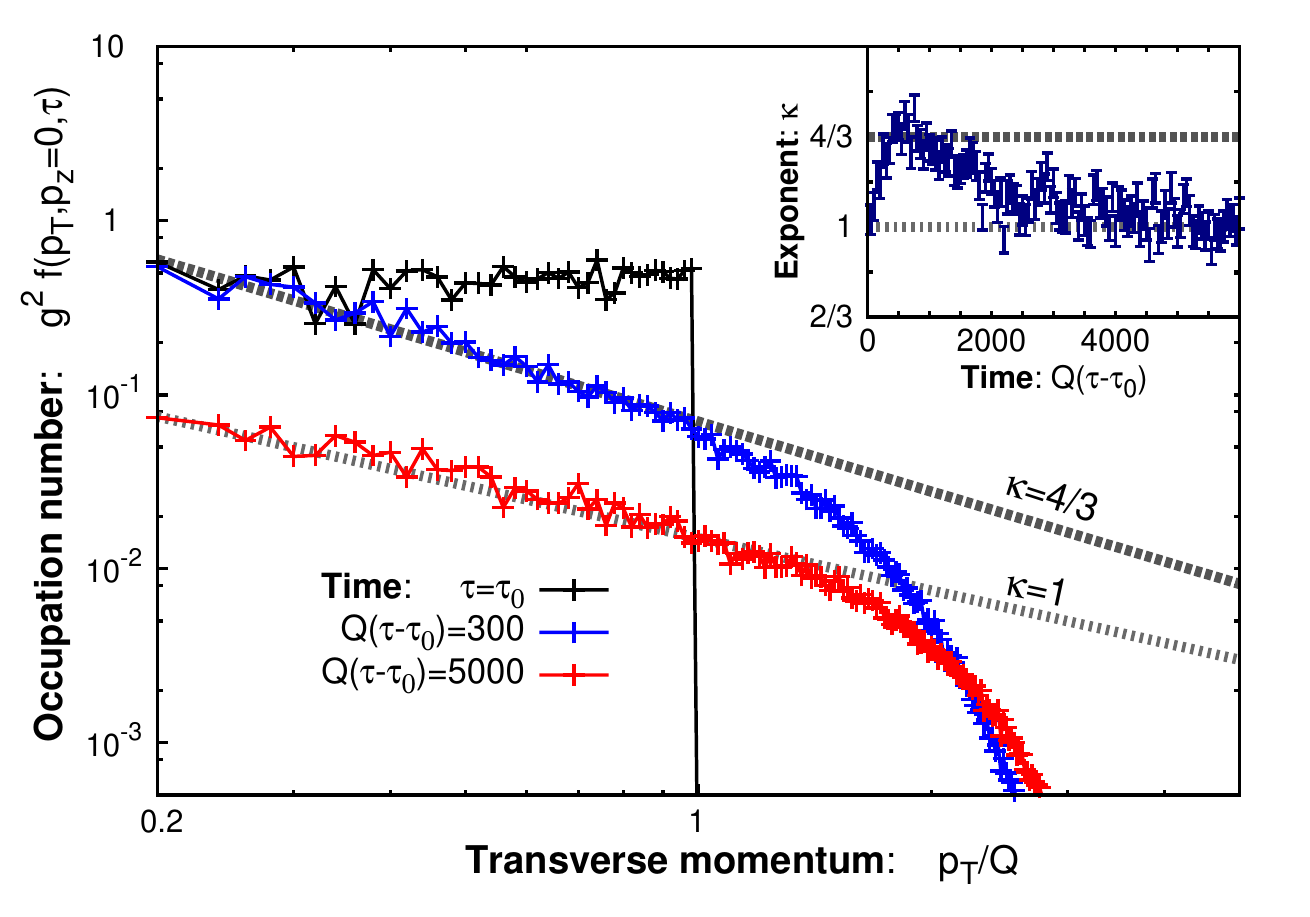}	
\caption{ (color online) \label{fig:Qt1000Spectra} Single particle spectrum as a function of transverse momentum for a slowly expanding plasma ($Q\tau_0=1000$). The inset shows the spectral index of the low momentum power-law as a function of time. At early times, the spectrum resembles the attractor solution for a non-expanding system discussed in sec.~\ref{sec:StaticBox}. At later times, one clearly observes the transition to the attractor for a longitudinally expanding system.}
\end{figure}

\subsection{Interpolation of results for non-expanding and expanding plasmas}
\label{sec:VLate}
We will now also consider variations of the initial time and study the dynamics for larger values of $Q \tau_0\gg1$. In this case the initial scattering rate $\sim Q$ is much larger than the initial expansion rate $\sim1/\tau_0$ of the system. Thus one expects to recover the results for non-expanding systems at early times $(\tau-\tau_0)\sim Q^{-1}$. As the system evolves along the static box attractor, the scattering rate begins to fall more rapidly than the expansion rate and the system evolves away from this attractor solution. However, at even later times $(\tau-\tau_0)\sim\tau_0$, the expansion rate and the scattering rate become of comparable size. One then expects a transition to the universal scaling solution for expanding systems where the two rates evolve at the same speed.\\
 
To investigate this behavior in more detail, we chose a large value of $Q\tau_0=1000$ to achieve a clear separation between different time scales. The results are presented in figs.~\ref{fig:Qt1000Lambda},~\ref{fig:Qt1000Exp} and \ref{fig:Qt1000Spectra}. In fig.~\ref{fig:Qt1000Lambda}, we show the evolution of the longitudinal and transverse hard scales as a function of time. For $(\tau-\tau_0)\lesssim\tau_0$ one observes an approximate scaling of $\Lambda_T^2\propto(\tau-\tau_0)^{2/7}$ as previously observed for non-expanding systems in the results shown in this paper (c.f. sec.~\ref{sec:StaticBox}) and previously~\cite{Schlichting:2012es,Kurkela:2012hp}. This is indicated by the gray dashed line. At later times $(\tau-\tau_0)\gtrsim\tau_0$ the transverse hard scale $\Lambda_T^2$ becomes approximately constant in time, whereas the longitudinal hard scale $\Lambda_L^2$ exhibits an approximate $(\tau-\tau_0)^{-2/3}$ scaling, characteristic of the attractor solution for longitudinally expanding systems\footnote{Note that at late times $\tau\gg \tau_0$ this corresponds to scaling in $(Q\tau)^{-2/3}$ as observed in fig.~\ref{fig:HardScale}.}.\\

The transition between the different attractors is further demonstrated in fig.~\ref{fig:Qt1000Exp} showing the local scaling exponents $2\beta(\tau)$ and $2\gamma(\tau)$ as a function of time. One observes a transient regime where the universal exponent $2\beta\simeq-2/7$ for a non-expanding system is realized. The behavior at late times is similar to the one observed in figs.~\ref{fig:HardScale} and \ref{fig:HardScaleTrans}. The scaling exponent for the longitudinal hard scale takes the value $2\gamma\simeq2/3$ and the exponent $2\beta$ for the transverse hard scale approaches zero monotonically from below.\\

Interestingly, a similar transition can be observed in the single particle spectra. This is shown in fig.~\ref{fig:Qt1000Spectra}, where we show the single gluon distribution as a function of transverse momentum. Starting from an initially overpopulated system indicated by the black lines, one initially observes a transition towards a $f(\pt,\pz=0)\sim\pt^{-4/3}$ power-law spectrum as discussed in sec.~\ref{sec:StaticBox}. This takes place on a time scale on the order of hundreds of $Q^{-1}$ but less than $\tau_0$. At later times, $(\tau-\tau_0)\sim\tau_0$ one observes the transition to the attractor solution for longitudinally expanding systems. As previously observed in fig.~\ref{fig:ptSpectra}, the spectrum at late times is characterized by a low momentum power-law with spectral index $\kappa\simeq1$ and a rapid fall-off at higher momenta. The transition between different attractors can also be deduced from the inset, where we show the (local) spectral index $\kappa$ as a function of time. Our results for large $Q\tau_0$ demonstrate yet another non-trivial route to reach the universal attractor solution for a longitudinally expanding plasma. It adds significantly to our claim of the existence and universality of this attractor solution.

\subsection{Turbulent thermalization of the expanding plasma}
\label{sec:TTExp}
We shall now discuss the turbulent thermalization process in the framework of kinetic theory. For the longitudinally expanding system, the time evolution of the gluon distribution is described by a Boltzmann equation of the form~\cite{Baier:2000sb,Kurkela:2011ub,Blaizot:2011xf}
\begin{eqnarray}
\label{C3eq:Boltzmann}
\left[\partial_\tau-\frac{\pz}{\tau}\partial_{\pz}\right]f(\pt,\pz,\tau)=C[f](\pt,\pz,\tau) \;,
\end{eqnarray}
with a generic collision term $C[f](\pt,\pz,\tau)$ for $n\leftrightarrow m$ scattering processes. The additional term on the left hand side of eq.~(\ref{C3eq:Boltzmann}) represents the redshift of longitudinal momenta and thus captures the effects of the longitudinal expansion. We then follow the same steps as in sec.~\ref{sec:StaticBox} and perform the standard turbulence analysis of~\cite{Micha:2004bv}, to investigate possible self-similar evolution of the type given by eq.~(\ref{C3eq:scalingf}). The scaling behavior of the collision integral is thereby described in terms of the exponent $\mu=\mu(\alpha,\beta,\gamma)$ such that
\begin{equation}
\label{C3eq:scalingC}
C[f](\pt,\pz,\tau)= (Q\tau)^{\mu}\, C[f_S]\Big((Q\tau)^{\beta}\pt,(Q\tau)^{\gamma}\pz\Big)\;.
\end{equation}
The precise form depends on the nature of the underlying interaction. Substituting this form into the Boltzmann equation (\ref{C3eq:Boltzmann}), one finds that the fixed point solution $f_ S(\pt,\pz)$ satisfies the relation 
\begin{eqnarray}
\label{C3eq:attractor}
&&\alpha f_S(\pt,\pz)
+\beta \pt \partial_{\pt} f_S(\pt,\pz) \nonumber \\
&&+\left(\gamma-1\right) \pz \partial_{\pz} f_S(\pt,\pz)=Q^{-1} C[\pt,\pz;f_S]\;, 
\end{eqnarray}
with the scaling relation
\begin{eqnarray}
\label{C3eq:scalingmu}
\alpha-1=\mu(\alpha,\beta,\gamma)\;. 
\end{eqnarray}

As previously argued by Baier,~Mueller,~Schiff and Son (BMSS) in the bottom up thermalization scenario~\cite{Baier:2000sb}, the classical interaction of hard excitations is dominated by elastic scattering with small momentum transfer. This is valid as long as their occupancies are large $(\nHard\gg1)$ and the dominant effect of these collisions on the particle distribution $f(\pt,\pz,\tau)$ is to broaden the longitudinal momentum distribution by multiple incoherent small-angle scatterings. This broadening of the gluon distribution in longitudinal momentum may be characterized by a collision integral of the Fokker-Planck type,
\begin{eqnarray}
\label{C3eq:Celast}
C^{({\rm elast})}[f](\pt,\pz,\tau)=\hat{q}~\partial_{\pz}^2 f(\pt,\pz,\tau)\;,
\end{eqnarray}
where the momentum diffusion parameter $\hat{q}$ in this expression is parametrically given by~\cite{Kurkela:2011ti,Blaizot:2011xf}
\begin{eqnarray}
\label{C3eq:diffusion}
\hat{q}\sim \alpha_s^2 N_c^2\int\frac{\dInt^2\pt}{(2\pi)^2}\int\frac{\dInt\pz}{2\pi}~f^2(\pt,\pz,\tau)
\end{eqnarray}
for $SU(N_c)$ gauge theories in the limit of high occupancies. This approximation is supposed to describe the dominant physics relevant for the overall scaling with time which enters the scaling relation (\ref{C3eq:scalingmu}) considered in the following. However, it does not have to be an accurate approximation for the solution of the fixed-point equation (\ref{C3eq:attractor}). This in general requires more detailed information about the momentum dependence of the collision integral\footnote{This can already be observed from the discussion in sec.~\ref{sec:StaticBox} of the non-expanding plasma. There the small angle approximation reproduces the correct scaling in time~\cite{Schlichting:2012es,Kurkela:2012hp} but fails to describe the spectral properties of the fixed point correctly~\cite{Berges:2008mr}.}.\\

The scaling properties of the collision integral in eq.~(\ref{C3eq:Celast}) lead to $\mu(\alpha,\beta,\gamma) = 3\alpha-2\beta+\gamma$ for the self-similar distribution as in eq.~(\ref{C3eq:scalingf}). The scaling relation in eq.~(\ref{C3eq:scalingmu}) obtained from the Boltzmann equation then takes the form
\begin{eqnarray}
2\alpha-2\beta+\gamma+1 &=& 0\;. \;(\text{Small angle scattering})
\end{eqnarray} 
Since the elastic scattering kernel in eq.~(\ref{C3eq:Celast}) is particle number conserving, a further scaling relation is obtained from integrating the distribution function over transverse momenta $\pt$ and rapidity wave numbers $\nu = \tau \pz$. By use of the scaling form in eq.~(\ref{C3eq:scalingf}), this constraint leads to the scaling relation
\begin{eqnarray}
\label{C3eq:ParticlConsExp}
\alpha-2\beta-\gamma+1&=&0\;. \quad(\text{Number conservation})
\end{eqnarray}
A further scaling relation can be extracted from energy conservation. Taking into account that the mode energy behaves as $\omega(\pt,\pz) \simeq \pt$ in the anisotropic scaling limit, this yields the condition
\begin{eqnarray}
\label{C3eq:EnergyConsExp}
\alpha-3\beta-\gamma+1&=&0\;. \quad (\text{Energy conservation})
\end{eqnarray}
The scaling exponents $\alpha,\beta,\gamma$ are then completely determined by the three scaling relations for particle number conservation, energy conservation and small-angle elastic scattering as incorporated in the BMSS kinetic approach. This yields the set of scaling exponents
\begin{eqnarray}
\alpha=-2/3\;, \qquad \beta=0\;, \qquad  \gamma=1/3 \;,
\end{eqnarray}
as the final result. This result is in excellent agreement with those extracted from our lattice simulations, within the stated statistical and systematical uncertainties.\\

The above scaling solution has the remarkable property that both energy and particle number are conserved in a single turbulent cascade. This is particularly so when contrasted to the non-expanding system discussed in sec.~\ref{sec:StaticBox}. In the latter, enforcing both particle number and energy conservation for a single cascade leads to the scaling relations \footnote{The relations in eq.~(\ref{C3eq:StaticBoxConsLaws}) can be obtained directly from eqns.~(\ref{C3eq:ParticlConsExp}) and (\ref{C3eq:EnergyConsExp}) by setting $\beta=\gamma$ for isotropic systems and dropping the additional summand $1$, which appears solely due to the longitudinal expansion.}
\begin{eqnarray}
\label{C3eq:StaticBoxConsLaws}
\alpha-3\beta&=&0\;, \qquad (\text{Particle number conservation}) \nonumber \\ 
\alpha-4\beta&=&0\;. \qquad (\text{Energy conservation})
\end{eqnarray}
Since the eqns.~(\ref{C3eq:StaticBoxConsLaws}) do not allow for non-trivial solutions, there is no single turbulent cascade conserving both energy and particle number in an isotropic system. Instead, a dual cascade emerges in situations where both conservation laws apply~\cite{Berges:2010ez}. In contrast, in the longitudinally expanding case, the anisotropy of the system allows for different scaling exponents $\beta\neq\gamma$. These then lead to non-trivial solutions of the corresponding scaling relations in eqns.~(\ref{C3eq:ParticlConsExp}) and (\ref{C3eq:EnergyConsExp}).\\

It is also interesting to observe that the exponent $\beta=0$ is entirely fixed by enforcing both conservation laws without further knowledge about the underlying dynamics. We therefore expect the exponent $\beta=0$ to appear also for a larger class of systems, which are dominated by elastic interactions and undergo a longitudinal expansion. This includes in particular $O(N)$ symmetric scalar quantum field theories, where inelastic processes are highly suppressed compared to elastic scattering. The study of these systems is in progress and will be reported elsewhere.

\subsection{The attractor}
\label{sec:TheAttractor}
Besides the BMSS scenario, alternative thermalization scenarios at weak coupling with different attractor solutions have been proposed in the literature. One such scenario, which takes into account the role of plasma instabilities in the classical regime, is the Boedeker (BD) scenario~\cite{Bodeker:2005nv}. In this scenario, it is argued that plasma instabilities lead to an overpopulation $f\sim 1/\alpha_s$ of modes with $|\p|\lesssim m_D$. The coherent interaction of hard excitations with the soft sector then causes an additional momentum broadening such that the longitudinal momenta of hard excitations fall at a slower rate. This evolution leads to the scaling exponents
\begin{eqnarray}
&&\alpha=-3/4\;, \qquad \beta=0\;, \qquad \gamma=1/4\;, \nonumber \\
&&\qquad \qquad \quad (\text{BD scenario})
\end{eqnarray}
which are numerically close to the BMSS solution.
\\

In the Kurkela and Moore (KM) scenario~\cite{Kurkela:2011ub}, it is argued that plasma instabilities play a key role for the entire evolution. The evolution in the classical regime proceeds in a similar way as in the BD scenario. However a larger range of overpopulated modes with $|\pvect|\lesssim m_D$ and $\pz \lesssim (\Lambda_T/\Lambda_L) m_D$ is considered in the KM scenario. This leads to a highly efficient momentum broadening, causing the longitudinal momenta of hard excitations to decrease much slower than in the BMSS evolution. The evolution in the classical regime of the KM scenario can be characterized by the scaling exponents
\begin{eqnarray}
&&\alpha=-7/8\;,\qquad \beta=0\;, \qquad \gamma=1/8\;,\nonumber \\
&&\qquad \qquad \quad (\text{KM scenario}) 
\end{eqnarray}
and the impact of plasma instabilities for the subsequent quantum evolution has also been discussed in~\cite{Kurkela:2011ub}.\\

In the Blaizot,~Gelis,~Liao,~McLerran and Venugopalan (BGLMV) scenario~\cite{Blaizot:2011xf}, elastic scattering is argued to be highly efficient in reducing the anisotropy of the system. This generates an attractor with a fixed anisotropy $\delta_s$ which is treated as a free parameter. The evolution in the BGLMV scenario proceeds with the scaling exponents
\begin{eqnarray}
&&\alpha=-(3-\delta_s)/7\;, \quad \beta=(1+2\delta_s)/7\;, \quad  \gamma=(1+2\delta_s)/7\;, \nonumber \\
&&\qquad \qquad \qquad \qquad (\text{BGLMV scenario})
\end{eqnarray}
such that the momentum space anisotropy of the system $\Lambda_L/\Lambda_T$ remains constant in time. Since in this scenario the exponents $\beta$ and $\gamma$ coincide, it is not possible to simultaneously satisfy the constraints for energy and particle conservation for elastic scattering. However, the authors of~\cite{Blaizot:2011xf} argued that the excess of particles resulting from the initial overoccupation may be absorbed in the soft sector. This would eventually lead to the formation of a transient Bose-Einstein condensate, which ultimately decays due to inelastic processes occurring over longer time scales.\\

\begin{figure}[t!]						
\centering
\includegraphics[width=0.5\textwidth]{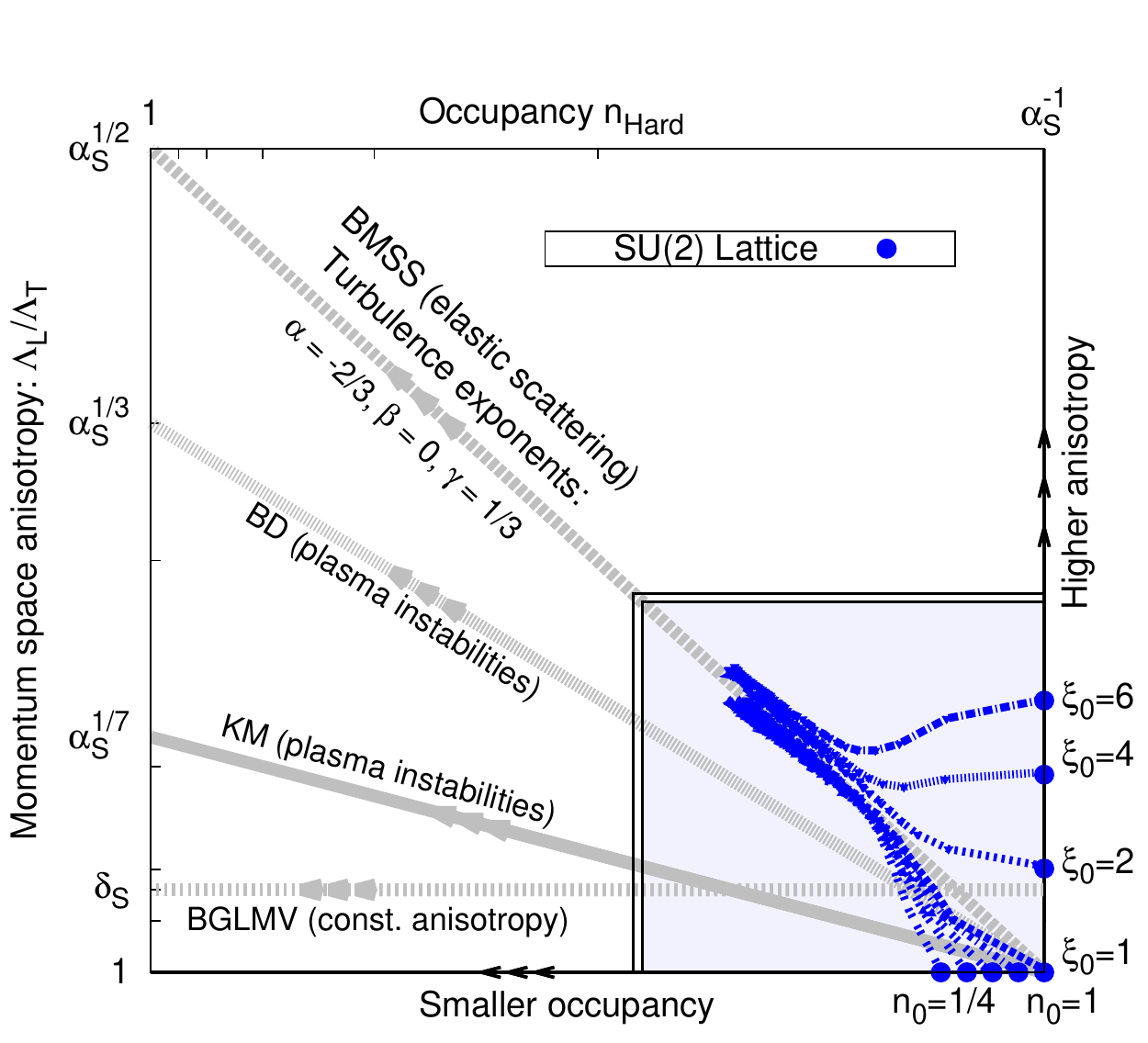}	
\caption{\label{fig:Cartoon} (color online) Evolution in the occupancy--anisotropy plane. Indicated are the thermalization scenarios proposed in (BMSS)~\cite{Baier:2000sb}, (BD)~\cite{Bodeker:2005nv}, (KM)~\cite{Kurkela:2011ub} and (BGLMV)~\cite{Blaizot:2011xf}. The blue lines show the results of classical-statistical lattice simulations for different initial conditions. One clearly observes the attractor property of the BMSS solution.}							
\end{figure}							

In order to clearly distinguish between the different thermalization scenarios, we investigate the evolution of the plasma in the occupancy--anisotropy plane, originally introduced in refs.~\cite{Kurkela:2011ti,Kurkela:2011ub}. Our findings are compactly summarized in fig.~\ref{fig:Cartoon}, where we compare the observed time evolution to the different thermalization scenarios. The horizontal axis shows the characteristic hard scale occupancy
$\nHard(\tau)=f(\pt\simeq Q,\pz=0,\tau)$ as defined in eq.~(\ref{C3eq:nharddef}), in the classical regime of occupancies $\nHard \gg 1$. The vertical axis shows the momentum-space anisotropy, which can be characterized in terms of the ratio of typical longitudinal momenta $\Lambda_L$ to the typical transverse momenta $\Lambda_{T}$. The gray lines in fig.~\ref{fig:Cartoon} indicate the different thermalization scenarios, while the blue lines show a projection of our simulation results to the anisotropy-occupancy plane. The different initial conditions are indicated by blue dots. The advantage is that some of the non-universal amplitude normalizations do not appear in this projection. Instead, one finds that ultimately all curves exhibit a similar evolution along the diagonal, clearly illustrating the attractor property. Again, this result is in excellent agreement with the analytic discussion of the BMSS kinetic equation in the high-occupancy regime.\\

By extrapolating our results to later times, we can also estimate the time scale to enter the quantum regime, where the characteristic occupancies become of order unity. Since initially the occupancy is parametrically given by $\nHard(\tau_0)\sim\alpha_s^{-1}$, and subsequently decreases as $\nHard(\tau)\propto (Q\tau)^{-2/3}$, this leads to the estimate
\begin{eqnarray}
\tau_{\text{Quantum}}\sim Q^{-1}\alpha_s^{-3/2}
\end{eqnarray}
in accordance with the original bottom up thermalization scenario~\cite{Baier:2000sb}. In the quantum regime $Q\tau\gtrsim\alpha_s^{-3/2}$, the classical-statistical framework can no longer be applied and modifications of the above kinetic equations need to be considered. While different scenarios of how thermalization is completed in the quantum regime have been developed based on kinetic descriptions~\cite{Baier:2000sb,Kurkela:2011ub}, it is an outstanding open question how to address the quantum dynamics in non-Abelian gauge theories from first principle simulations.

\subsection{Quo vadis, thermalization?}
\label{sec:QuoVadis}
In the previous subsections, we established the existence of a non-thermal attractor at weak coupling. We observed that, for a wide range of initial conditions, the spacetime evolution of the plasma displays the same scaling behavior. While the emergence of a universal attractor far from equilibrium is remarkable, the question
\begin{itemize}
 \item[a)] How relevant are details of the initial conditions for the dynamics of the thermalization process?
\end{itemize}
-- in particular at larger values of the coupling -- clearly requires a more careful assessment and will be addressed below. The universal scaling solution is consistent with the bottom up thermalization scenario, where neither isotropization nor thermalization are realized in the classical regime. On the contrary, the system becomes increasingly anisotropic and the obvious conceptual question related to this behavior is
\begin{itemize}
 \item[b)] Does the system isotropize and thermalize at all in a dynamical regime where the QCD coupling is still weak?
\end{itemize}
In view of the standard paradigm that the quark-gluon plasma can be described in terms of hydrodynamics it is also interesting to ask
\begin{itemize}
 \item[c)] Can the system still be described by hydrodynamics even if neither isotropization nor thermalization is achieved?
\end{itemize}
From the practical point of view, if the answers to b) and c) are negative, one might be tempted to conclude that weak coupling is not relevant for an understanding of heavy ion collisions even at the highest LHC energies.

\subsubsection{Conceptual issues}
Let us first address the conceptual issues and turn subsequently to their phenomenological consequences. In the 
BMSS scenario, inelastic $2\leftrightarrow 3$ processes begin to play a big role beyond the estimated time scale
$\tau_{\text{Quantum}}\sim Q^{-1}\alpha_s^{-3/2}$ where classical dynamics fails to describe the evolution. These isotropize the bulk of the system on a parametric time scale $\sim Q^{-1}\alpha_s^{-5/2}$. Complete equilibration is achieved quite rapidly thereafter -- the system thermalizes on the time scale $\sim Q^{-1}\alpha_s^{-13/5}$~\cite{Baier:2000sb}. We take this as a ``proof of principle" demonstration that thermalization can indeed occur in the quantum regime. Similar conclusions have been reached in different thermalization scenarios, where isotropization and subsequent thermalization also occur only in the quantum regime~\cite{Kurkela:2011ub}. \\

Concerning the dependence on the initial conditions, a closer look at the classical-statistical dynamics reveals that the non-universal amplitudes are indeed sensitive to transient features of the evolution, such as the initial occupancy $n_0$ and the initial anisotropy $\xi_0$. For instance one finds that in the BMSS scenario -- considering elastic scattering and free streaming only -- the longitudinal hard scale behaves as $\Lambda_L^2 \propto n_0^{4/3}  Q^2/(Q\tau)^{2/3}$. Similarly, other dynamical quantities such as the lifetime of the classical regime will also depend on $n_0$. However, in weak coupling asymptotics ($\alpha_s\rightarrow 0$), details of the initial conditions have a negligible effect. This is because $n_0$ must, by construction, be a number of order unity and is therefore much smaller than inverse powers of the coupling constant. Similarly, the separation of different dynamical regions is ``clean" in the weak coupling limit, where the characteristic time scales $\OneOverQ$, $\OneOverQ \lnSqrOneOverAlpha$, $\OneOverQ \alpha_s^{-3/2}$ etc. can clearly be distinguished. In particular, the time scale to reach the universal attractor $\sim\OneOverQ \lnSqrOneOverAlpha$ is much smaller than the lifetime of the classical regime $\OneOverQ \alpha_s^{-3/2}$.\footnote{In our simulations we employ $\alpha_s \sim 10^{-5}-10^{-6}$ such that $\OneOverQ \lnSqrOneOverAlpha\sim\mathcal{O}(100)$ whereas $\OneOverQ \alpha_s^{-3/2}\sim\mathcal{O}(10^{8})$.} The effect of the transient evolution then becomes negligible and the weak coupling late time asymptotics is universal to high accuracy. However this parametric separation no longer exists for large values of the coupling constant. Consequently, one expects transient features of the evolution to be more pronounced and a detailed knowledge of the initial conditions may be essential. \\

A similar dependence on details of initial conditions is also anticipated in the quantum regime. The BMSS computations (and similarly all weak coupling scenarios) are valid in weak coupling asymptotics. The prefactors in the estimates of the thermalization time have not been computed for any dynamical scenario. This is because one is dealing with a complicated multi scale problem which is very hard to solve even in such asymptotics. \\

One such first principles approach to the problem, starting from the CGC dynamics of the incoming nuclear wavefunctions, was advocated in~\cite{Gelis:2007kn,Dusling:2010rm,Dusling:2011rz,Epelbaum:2013waa}. How the classical fields and small quantum fluctuations around these evolve through the nuclear collision until times $\tau\lesssim \OneOverQ$ was identified. Because the quantum fluctuations grow exponentially after the collision, a method to resum the leading instabilities to all orders was devised. This stabilizes the perturbative series on a time scale $\sim Q^{-1} \ln^2(1/\alpha_s)$. However, there are further sub-leading $\mathcal{O}(\alpha_s)$ corrections,\footnote{These are also very relevant for observables that are sensitive to the early time dynamics such as long range rapidity correlations \cite{Dumitru:2008wn,Dusling:2009ni}.} that also grow and become of the order of the classical fields on a time scale $\sim Q^{-1} \ln^2(1/\alpha_s^2)$, which is parametrically of the same order. Because this is true to all orders in perturbation theory, straightforward weak coupling approaches become increasingly problematic on times that are only $\sim Q^{-1}\ln^2(1/\alpha_s)$. This is particularly so for the problem of thermalization, since the above time scale is parametrically much shorter than $1/\alpha_s^{3/2}$, the lifetime of the classical regime. While this problem is generic to all approaches that have ``secular-like'' divergences, as is well known in the condensed matter theory literature~\cite{Goldenfeld}, it is also conceivable that in practice the applicability of the formalism may be greater than can be deduced from the simple parametric estimates.\\

A way forward is to investigate whether the initial conditions in~\cite{Epelbaum:2013waa,Gelis:2013rba} lead to the same attractor solution at late times. In our view, the attractor solution we have found is robust. The fact that even solutions constructed to lie initially in the well of the static box attractor evolve to those of the expanding non-Abelian plasma is striking. Nevertheless, the broadest array of initial conditions should (and will) be explored. It is interesting in any case to see whether, and on what time scales, the results of~\cite{Epelbaum:2013waa,Gelis:2013rba} evolve to the universal attractor with decreasing $\alpha_s$. How this transition occurs, if it does, will provide further insight into the transient dynamics.\\

The other conceptual issue is whether hydrodynamics is applicable even if isotropization and thermalization is not 
realized in the classical regime. In conventional kinetic theory derivations of hydrodynamic equations, a gradient expansion is performed around the thermal stationary point. Similar kinetic theory inspired hydrodynamic equations can 
be derived when the system is far off-equilibrium; this however requires the existence of a non-thermal attractor as an expansion point. The effective dissipative coefficients computed in such a hydrodynamic scenario, and their interpretation, could in principle be very different from those in near-equilibrium hydrodynamics.

\subsubsection{Practical considerations}
The heavy ion experiments at RHIC and at the LHC find large values of Fourier moments of azimuthally anisotropic particle distributions. Further systematic studies indicate that these can be reproduced by dissipative relativistic hydrodynamic equations with very small values of $\eta/s$ -- the viscosity to entropy density ratio. However, considerable uncertainty surrounds when hydrodynamics needs to be applied after the collision. A successful model of initial conditions that incorporates non-equilibrium Yang-Mills dynamics is the IP-Glasma+MUSIC model~\cite{Schenke:2012wb,Schenke:2012hg,Gale:2012rq}. In this model, Yang-Mills dynamics is matched to viscous hydrodynamics at very early times $\tau_h\sim 0.2-0.6$~fm/c. One should note though that only 2+1-D boost-invariant Yang-Mills evolution is included in this model; thus the longitudinal dynamics is free streaming dynamics.\\

Both the transient and universal dynamics in our 3+1-dimensional simulations accommodate considerable re-scattering. One anticipates thus that more flow is generated in this framework allowing one to extend the matching to hydrodynamics to larger values of $\tau_h$. Applications of the BMSS estimates at face value give isotropization/thermalization times $\sim 2-4$~fm/c~\cite{Baier:2002bt,Baier:2011ap}. Since there are a number of ill-determined prefactors, this time scale can easily be pushed by a factor of two to earlier times.\\

To the best of our knowledge, few models of initial conditions account for significant entropy production due to early time non-equilibrium dynamics. In the BMSS scenario, the entropy grows by a factor of $\alpha_s^{-2/5}$ between the initial state and thermalization~\cite{Baier:2002bt,Baier:2011ap}. Recent computations in the Glasma framework, where particle production is microscopically constrained by HERA data, require additional entropy production in the final state~\cite{SVT-inpreparation}. There is now a tremendous wealth of bulk heavy ion data at a number of different energies and centralities. It is therefore not too far fetched to imagine that an extension of the IP-Glasma framework (to include the 3+1-D Yang-Mills dynamics considered here) may provide important constraints on thermalization -- in particular, the $O(1)$ prefactors in our computations.\\

Finally, we note that anisotropic hydrodynamic models fare quite well in comparison to heavy ion collision data~\cite{Florkowski:2013sk,Florkowski:2013lza}. These models, as implemented, are quite different from the anisotropic hydrodynamics that could be constructed from our attractor simulation. Nevertheless, their success suggests that a consistent matching of our 3+1-D Yang-Mills dynamics to anisotropic hydrodynamics may provide a future direction for the quantitative phenomenology of heavy ion collisions.

\section{Summary and Outlook}
\label{sec:Conclusion}
We studied in this paper the thermalization process in highly occupied non-Abelian plasmas at weak coupling. We argued that the non-equilibrium dynamics of such systems is classical in nature and can be studied from first principles within the framework of real-time lattice simulations.\\ 

We initially used classical-statistical simulations to explore non-equilibrium dynamics in a non-expanding non-Abelian plasma. We presented general arguments that the thermalization process of an overoccupied plasma can be viewed as the energy transport towards the ultraviolet. We observed that -- after a short transient regime -- this energy transport is achieved by a turbulent ultraviolet cascade with a quasi-stationary evolution in time.\\

The self-similar dynamics of this process is characterized by a set of scaling exponents $\alpha$ and $\beta$ and a stationary distribution $f_S$. The dynamical scaling exponents $\alpha$ and $\beta$ are universal and can be determined from a kinetic theory analysis of the relevant scattering processes. The spectral properties are characterized in terms of the stationary distribution $f_S$, which shows a scale-invariant power-law distribution at low momenta. The spectral exponent $\kappa$ of this power-law is non-thermal. Instead the observed value is consistent with the Kolmogorov-Zakharov spectra of (stationary) weak wave turbulence ($\kappa=4/3$).\\ 

This notion of universality far from equilibrium manifests itself in the fact that similar phenomena can be observed across very different energy scales, ranging from early universe cosmology~\cite{Micha:2004bv} to the dynamics of ultra-cold quantum gases~\cite{Semikoz:1994zp,Semikoz:1995rd,Nowak:2011sk}. From this perspective, the results may after all not appear very surprising, since they are shared by a large variety of strongly correlated many-body systems out of equilibrium.\\

We also discussed how the turbulent cascade continues until times $t_{\text{Quantum}}\sim Q^{-1} \alpha_s^{-7/4}$ where the occupation numbers of hard modes become of order unity. We argued that quantum effects are then no longer negligible and the classical-statistical framework breaks down. However, the system already appears to be close to thermal equilibrium at this stage of the evolution. Thus one may use $t_{\text{Therm}}\sim t_{\text{Quantum}}$ as an estimate for the thermalization time.\\

The high precision lattice results also provide important insight into the dynamics of soft excitations. We emphasize that results from classical-statistical lattice simulations can and should be compared to solutions of kinetic equations~\cite{Blaizot:2013lga,Huang:2013lia}. Since both methods have an overlap in the range of validity for occupation numbers $1 \ll f \ll 1/\alpha_s$ a quantitative comparison may provide important constraints on the infrared regularization of scattering matrix elements. This regularization is an essential ingredient of the kinetic approach. However, this requires the inclusion of the classical field/Bose enhanced terms in the Boltzmann equation, which has proven challenging in present transport models.\\

We then studied the non-equilibrium dynamics of longitudinally expanding non-Abelian plasmas, relevant to relativistic heavy ion collisions in the limit of weak coupling at very high collider energies. We discussed the dynamics of such a collision in the CGC framework and argued that the system becomes fluctuation dominated on a time scale $\tau\sim \OneOverQ \lnSqrOneOverAlpha$. The system in this regime can then be described as an overoccupied ensemble of quasi-particle excitations. We characterized this state in terms of the initial occupancy and anisotropy of the plasma and employed a large range of different initial conditions to study the subsequent evolution within classical-statistical lattice simulations.\\

We found that -- after a short transient regime -- the dynamics at late times becomes independent of the initial conditions. Similarly to the non-expanding case, the system then exhibits the universal self-similar dynamics characteristic of ``free'' wave turbulence. We obtained the universal scaling exponents and scaling functions and compared our results to different thermalization scenarios proposed in the literature. We found that, while the physics of plasma instabilities and free streaming describe the approach to the universal attractor, the self-similar dynamics of the turbulent regime is, within the accuracy of our simulations, consistent with the bottom up thermalization scenario~\cite{Baier:2000sb} and can be described entirely in terms of elastic scattering processes.\\

Most remarkably, the extraction of the scaling exponents in the kinetic theory framework only relies on conservation laws and the dominance of small angle scattering processes. The bottom up scenario only provides one particular realization of the associated scaling relations. Similarly to the non-expanding case, one can therefore expect to observe similar scaling phenomena also across very different energy scales. A straightforward test of these ideas is to apply the framework to the study of a longitudinally expanding scalar theory. This is work in progress and will be reported elsewhere.\\

The universal attractor also has the remarkable property that it conserves both energy and particle number in a single turbulent cascade. This can only be achieved in highly anisotropic systems and is a manifestation of the diverse nature of turbulent solutions in anisotropic systems~\cite{Falkovich}.\\

The competition between elastic scattering and the longitudinal expansion leads to an increase of the momentum space anisotropy of the system $\propto 1/\tau^{1/3}$ in the classical regime. Extrapolating our lattice results to later times, we concluded that this behavior persists up to the time scale $\tau_{\text{Quantum}}\sim\OneOverQ\alpha_s^{-3/2}$, when the system becomes dilute and quantum corrections can no longer be neglected. Since the classical-statistical framework can no longer be applied in this quantum regime, further progress at weak coupling relies on the ability to include quantum corrections dynamically in the non-equilibrium evolution. This remains an outstanding problem.\\ 

Interestingly, turbulent scaling phenomena have also been discussed for the quantum regime. In the context of jet propagation in a QCD medium, it was recently argued that the energy loss of the jet through multiple branchings can be viewed as a turbulent energy cascade towards lower momenta~\cite{Blaizot:2013hx,Iancu:2013ura}. Most remarkably, this process is completely analogous to the final stage of the bottom up thermalization scenario, where a small fraction of hard excitations decays into a soft thermal bath~\cite{Baier:2000sb}. This intriguing connection between in medium jet propagation and the thermalization process points to the fact that universal turbulent phenomena may play a larger role in the dynamics of the quark gluon plasma than earlier conceived.\\ 

While the universal attractor solution is robust in the weak coupling limit, one expects transient phenomena to become increasingly important at larger values of the coupling. The details of the initial conditions may then be essential to establish a quantitative understanding of heavy ion collisions at RHIC and LHC energies. However the straightforward application of the weak approach to realistic values of the coupling constant is not unambiguous given the conceptual and technical limitations of the classical-statistical framework. The extrapolation of the weak coupling results thus requires additional studies and will be addressed in the near future.

\begin{acknowledgments}
We would like to thank R.~Baier, J.~P.~Blaizot, D.~Boedeker, K.~Dusling, T.~Epelbaum, D.~Gelfand, F.~Gelis, A.~Kurkela, J.~Liao, L.~D.~McLerran, G.~D.~Moore, A.~Mueller and D.~Sexty for very valuable discussions. We would also like to express our gratitude to T.~Epelbaum for his contributions to app.~\ref{app:ModeVectors}. This work was supported in part by the German Research Foundation (DFG). S.S and R.V. are supported by US Department of Energy under DOE Contract No.~DE-AC02-98CH10886. The numerical results presented in this work were obtained on the bwGRiD (\url{http://www.bw-grid.de}), member of the German D-Grid initiative, funded by the Ministry for Education and Research (BMBF) and the Ministry for Science, Research and Arts Baden-Wuerttemberg (MWK-BW) and we gratefully acknowledge their support.
\end{acknowledgments}

\begin{appendix}
\begin{widetext}

\section{Free field solutions in co-moving coordinates}
\label{app:ModeVectors}
In this appendix we determine the basis of solutions to the linearized Yang-Mills evolution equations in $\tau,\eta$ coordinates. Similar computations can also be found in \cite{Makhlin:1996dr,Dusling:2011rz,Epelbaum:2013waa}.\footnote{The main difference to Makhlin's work \cite{Makhlin:1996dr} is the different choice of the residual gauge freedom in Fock-Schwinger $(A_{\tau}=0)$ gauge.} Since the solutions are labeled in terms of transverse momentum $\pvect$ and conjugate rapidity momentum $\nu$ we will denote the set of linearly independent solutions by
\begin{eqnarray}
a_{\mu}^{(\lambda)\pvect\nu\pm}(x)=\xi_{\mu}^{(\lambda)\pvect\nu\pm}(\tau)~e^{i(\pvect\xvect+\nu\eta)}\;,
\end{eqnarray}
where the index $\lambda=1,2,3$ labels the different polarizations, the index $\pm$ denotes the positive and negative frequency solutions and $\xi_{\mu}^{(\lambda)\pvect\nu\pm}(\tau)$ denote the time-dependent polarization vectors. Here we suppressed the color indices, since the free solutions display a diagonal structure in color space. The starting point for our discussion are the evolution equations, which in Fock-Schwinger ($a_\tau=0$) gauge take the form \cite{Makhlin:1996dr,Dusling:2011rz},
\begin{eqnarray}
\label{Modes:eom}
\partial_{\tau} \tau \partial_{\tau} a_i+\tau( \pt^2+\tau^{-2}\nu^2)a_i-\tau p_i p_j a_j-\tau^{-1}\nu p_i a_{\eta}&=&0\;, \nonumber \\
\partial_{\tau} \tau^{-1} \partial_{\tau} a_\eta +\tau^{-1} \pt^2 a_\eta -\tau^{-1}\nu p_i a_i&=&0\;,
\end{eqnarray}
where $\pt^2=p_x^2+p_y^2$ and summation over the transverse Lorenz index $i=x,y$ is implied. In addition to the above evolution equations, the solutions are required to satisfy the (Abelian) Gauss law constraint, which takes the form
\begin{eqnarray}
\label{Modes:Gauss}
p_i \tau \partial_{\tau} a_i+\nu \tau^{-1}\partial_{\tau} a_\eta=0\;.
\end{eqnarray}

In general there are five linearly independent solutions to the set of equations (\ref{Modes:eom}), taking into account the Gauss law constraint (\ref{Modes:Gauss}). However the remaining gauge freedom to perform time independent gauge transformations allows us to eliminate one of the solutions, such that we are left with four linearly independent solutions, which correspond to the negative and positive frequency solutions of the two transverse polarizations. We will exploit this fact and fix the remaining gauge freedom by implementing the Coulomb type gauge condition 
\begin{align}
\label{Modes:cg}
\left[p_ia_i+\tau^{-2}\nu a_\eta\right]_{\tau=\tau_0}=0\;,
\end{align}
at arbitrary time $\tau_0$. To see how this gauge fixing reduces the number of solutions, we first consider a solution of the type
\begin{align}
\xi^{(3)\pvect\nu}_{\mu}(\tau)=
\left(
\begin{array}{c} p_x \\
 p_y \\
\nu \end{array} \right)
\xi^{(3)\pvect\nu}(\tau)\;. 
\end{align}
The Gauss law constraint and the evolution equations then imply that $\partial_{\tau} \xi^{(3)\pvect\nu}(\tau)=0$, and hence $\xi^{(3)\pvect\nu}(\tau)=\xi^{(3)\pvect\nu}$ is a constant in time. However the gauge fixing condition (\ref{Modes:cg}) implies that $\xi^{(3)\pvect\nu}(\tau_0)=0$ vanishes such that this solution is eliminated by the choice of the gauge. We are therefore left with four physical solutions, which correspond to the positive and negative frequency solutions for the two transverse polarizations.

\subsubsection*{A.1 First set of physical solutions}
In order to construct the first set of physical solutions, we chose the ansatz
\begin{eqnarray}
\xi^{(1)\pvect\nu}_{\mu}(\tau)=
\left( \begin{array}{c} - p_y \\
 ~ p_x \\
0\end{array} \right)
\xi^{(1)\pvect\nu}(\tau)\;,
\end{eqnarray}
which complies with the Gauss law constraint by construction. The evolution equation for $a^{(1)\pvect\nu}(\tau)$ follows from eq. (\ref{Modes:eom}) and takes the form
\begin{eqnarray}
\label{Modes:BesselEq}
\left[\tau^{-1}\partial_{\tau} \tau \partial_{\tau}+\pt^2+\frac{\nu^2}{\tau^2}\right]\xi^{(1)\pvect\nu}(\tau)=0\;. 
\end{eqnarray}
This is the Bessel equation and its general solution can be expressed in terms of Hankel functions as
\begin{eqnarray}
\label{Modes:sol0}
\xi^{(1)\pvect\nu}(\tau)=c_1H^{(1)}_{i\nu}(\pt\tau)+c_2H^{(2)}_{i\nu}(\pt\tau)\;,
\end{eqnarray}
such that there are two linearly independent solutions. For each solution, the constants $c_1$ and $c_2$ can be fixed to yield the correct normalization and asymptotic behavior of the solution in terms of positive and negative frequency components. This is discussed in more detail below, where we obtain the final and correctly normalized result.

\subsubsection*{A.2 Second set of physical solutions}
The second set of solutions can be written in the general form
\begin{eqnarray}
\xi^{(2)\pvect\nu}_{\mu}(\tau)=
\left(
\begin{array}{c}
\nu  p_x/(\pt^2\tau_0^2)~R_\bot^{\pvect\nu}(\tau) \\
\nu p_y/(\pt^2\tau_0^2)~R_\bot^{\pvect\nu}(\tau) \\
-R_\eta^{\pvect\nu}(\tau) \end{array}
\right)\;,
\end{eqnarray}
where the residual gauge freedom is fixed at the time $\tau_0$ by the gauge condition (\ref{Modes:cg}), which implies $R_\eta^{\pvect\nu}(\tau_0)=R_\bot^{\pvect\nu}(\tau_0)$. Moreover the Gauss law constraint yields the relation
\begin{eqnarray}
\nu\left[\tau \tau_0^{-2} \partial_{\tau} R_\bot^{\pvect\nu}(\tau)-\tau^{-1}\partial_{\tau} R_{\eta}^{\pvect\nu}(\tau)\right]
=0\;, 
\end{eqnarray}
such that 
\begin{eqnarray}
\label{Modes:Gauss2}
\partial_{\tau} R_\eta^{\pvect\nu}(\tau)=\frac{\tau^2}{\tau_0^2}\partial_{\tau} R_\bot^{\pvect\nu}(\tau)\;, 
\end{eqnarray}
which can be used to eliminate $R_{\eta}^{\pvect\nu}(\tau)$ in favor of $R_\bot^{\pvect\nu}(\tau)$. The dynamic equations for $R_{\eta}^{\pvect\nu}(\tau)$ and $R_\bot^{\pvect\nu}(\tau)$ follow from the evolution equations (\ref{Modes:eom}) and take the form
\begin{align}
\label{Modes:eom2}
\partial_{\tau} \tau \partial_{\tau} R_\bot^{\pvect\nu}(\tau) + \frac{\nu^2}{\tau}R_\bot^{\pvect\nu}(\tau) +
\frac{(\pt^2\tau_0^2)}{\tau}R_\eta^{\pvect\nu}(\tau)=\null&0\;, \nonumber \\
\partial_{\tau} \tau^{-1} \partial_{\tau} R_\eta^{\pvect\nu}(\tau)+\tau^{-1}\pt^2 R_\eta^{\pvect\nu}(\tau)
+\tau^{-1}\frac{\nu^2}{\tau_0^2} R_\bot^{\pvect\nu}(\tau)=\null&0\;. 
\end{align}
In order to decouple the evolution equations, it is convenient to multiply the equations by appropriate factors of $\tau$ and subsequently differentiate with respect to $\tau$. By use of the relation (\ref{Modes:Gauss2}), this yields the set of equations
\begin{align}
\label{Modes:eom22}
\left[\tau^{-1}\partial_{\tau}\tau\partial_{\tau}+\pt^2+\frac{\nu^2}{\tau^2}\right]\tau\partial_{\tau} R_\bot^{\pvect\nu}(\tau)
=\null&0\;, \nonumber \\
\left[\tau^{-1}\partial_{\tau}\tau\partial_{\tau}+\pt^2+\frac{\nu^2}{\tau^2}\right]\tau^{-1}\partial_{\tau} R_\eta^{\pvect\nu}(\tau)
=\null&0\;, 
\end{align}
which are equivalent after the Gauss constraint (\ref{Modes:Gauss2}) is taken into account. It is important to note that by transforming the set of second order differential equations (\ref{Modes:eom2}) into the third order differential equation (\ref{Modes:eom22}), we have introduced an additional free parameter to the general solution. We will fix this parameter by requiring the solutions of (\ref{Modes:eom22}) to satisfy the original evolution equations (\ref{Modes:eom2}). In analogy to eq.~(\ref{Modes:sol0}), the general solution of the above equations takes the form
\begin{align}
\label{Modes:sol1}
R_\bot^{\pvect\nu}(\tau)=\null&
R_{\tau_0}^{\pvect\nu}
+\int_{\tau_0}^{\tau}~\dInt \tau'\,\tau'^{-1}\left[c_1H^{(1)}_{i\nu}(\pt\tau')+c_2H^{(2)}_{i\nu}(\pt\tau')\right]
\;, \\
\label{Modes:sol2}
R_\eta^{\pvect\nu}(\tau)=\null&
R_{\tau_0}^{\pvect\nu}+
\int_{\tau_0}^{\tau}~\dInt \tau'\,\frac{\tau'}{\tau_0^2}\left[c_1H^{(1)}_{i\nu}(\pt\tau')
+c_2H^{(2)}_{i\nu}(\pt\tau')\right]\;, 
\end{align}
where we have taken into account the Gauss law constraint to ensure that the constants $c_1$ and $c_2$ are the same for $R_\bot^{\pvect\nu}(\tau)$ and $R_\eta^{\pvect\nu}(\tau)$ and the gauge condition which ensures that the integration constant $R_{\tau_0}^{\pvect\nu}$ takes the same value. The value of the constant $R_{\tau_0}^{\pvect\nu}$ is fixed by requiring the solution to satisfy the original coupled set of second order differential equations. By inserting the solutions (\ref{Modes:sol1},\ref{Modes:sol2}) in the evolution equations (\ref{Modes:eom2}) we obtain the condition
\begin{align}
\label{Modes:fix}
&\partial_{\tau}\left[c_1H^{(1)}_{i\nu}(\pt\tau)+c_2H^{(2)}_{i\nu}(\pt\tau)\right] 
+\tau^{-1}\left[\nu^2+\pt^2\tau_0^2\right]R_{\tau_0}^{\pvect\nu} \notag\\
&+\tau^{-1}\int_{\tau_0}^{\tau}~\dInt \tau'\,\tau'\left[\frac{\nu^2}{\tau'^2}
+\pt^2\right]\left[c_1H^{(1)}_{i\nu}(\pt\tau')+c_2H^{(2)}_{i\nu}(\pt\tau')\right]=\null0\;,
\end{align}
and by use of the Bessel equation (\ref{Modes:BesselEq}), we can rewrite the integrand as
\begin{align}
\left[\frac{\nu^2}{\tau^2}+\pt^2\right]H^{(1/2)}_{i\nu}(\pt\tau)=\null&
-\tau^{-1}\partial_{\tau} \tau \partial_{\tau} H^{(1/2)}_{i\nu}(\pt\tau)\;, 
\end{align}
such that the powers of time under the integral cancel and we are left with the integration of a total derivative. We can then perform the time integration and find that the term from the upper bound of the integral cancels the first derivative term in eq. (\ref{Modes:fix}). In this way eq. (\ref{Modes:fix}) reduces to
\begin{align}
\left[\nu^2+\pt^2\tau_0^2\right]R_{\tau_0}^{\pvect\nu}
+\left.\tau_0~\partial_{\tau}\left[c_1H^{(1)}_{i\nu}(\pt\tau)+c_2H^{(2)}_{i\nu}(\pt\tau)\right]
\right|_{\tau=\tau_0}=\null&0\;,
\end{align}
which fixes the constant to be
\begin{align}
R_{\tau_0}^{\pvect\nu}=\null&
-\frac{\pt\tau_0~
\left[c_1{H'}^{(1)}_{i\nu}(\pt\tau_0)
+c_2{H'}^{(2)}_{i\nu}(\pt\tau_0)\right]}{\nu^2+\pt^2\tau_0^2}\;, 
\end{align}
where ${H'}^{(1/2)}_{i\nu}(x)=\partial_xH^{(1/2)}_{i\nu}(x)$ denotes the first derivative of the function evaluated at the respective argument. The first and second set of solutions are then both characterized by two free parameters, which we will fix in the following to obtain the correctly normalized positive and negative frequency solutions.

\subsubsection*{A.3 Normalization and asymptotic behavior}
In the previous section, we determined the orthogonal solutions to the free evolution equations in $\tau,\eta$ coordinates. In order to determine the normalization of the solutions it is important to realize that all solutions to the evolution equations (\ref{Modes:eom}) conserve the scalar product \cite{Dusling:2011rz}
\begin{align}
\label{Modes:sp}
(a|b)=-i\int ~\dInt^2\xvect~\dInt\eta~\tau~g^{\mu\nu}
\left[a^*_{\mu}(x)\partial_{\tau} b_{\nu}(x)-b_{\nu}(x) \partial_{\tau} a^*_{\mu}(x)\right]\;.
\end{align}
This can be checked explicitly by use of the equations of motion (\ref{Modes:eom}) and the relation can be used to properly normalize the solutions \cite{Dusling:2011rz}. 
\subsubsection{Normalization of the first set of solutions}
The first set of physical solutions takes the general form
\begin{align}
\xi^{(1)\pvect\nu}_{\mu}(x)=\null&
\left(
\begin{array}{c} - p_y \\
~ p_x \\
0\end{array} \right)
\left[c_1H^{(1)}_{i\nu}(\pt\tau)+c_2H^{(2)}_{i\nu}(\pt\tau)\right]\;,
\end{align}
and according to Ref.~\cite{Dusling:2011rz} the positive and negative frequency solution corresponds to the parts involving only the Hankel functions of the second and first kind respectively. If we focus on the positive and negative frequency parts $a^{(1)\pvect\nu\pm}$, with $c_1^+=0$ and $c_2^-=0$, the scalar products between the solutions take the form
\begin{align}
(a^{(1)\pvect\nu+}|a^{(1)\pvect'\nu'+})=\null&i\tau~(2\pi)^3\delta^2(\pvect-\pvect')\delta(\nu-\nu') 
~\pt^2~|c_2^+|^2~H^{(2)*}_{i\nu}(\pt\tau)\overleftrightarrow{\partial_{\tau}}H^{(2)}_{i\nu}(\pt\tau)\;, \\
(a^{(1)\pvect\nu-}|a^{(1)\pvect'\nu'-})=\null&i\tau~(2\pi)^3\delta^2(\pvect-\pvect')\delta(\nu-\nu')
~\pt^2~|c_1^-|^2~H^{(1)*}_{i\nu}(\pt\tau)\overleftrightarrow{\partial_{\tau}}H^{(1)}_{i\nu}(\pt\tau)\;, \\
(a^{(1)\pvect\nu+}|a^{(1)\pvect'\nu'-})=\null&i\tau~(2\pi)^3\delta^2(\pvect-\pvect')\delta(\nu-\nu') 
~\pt^2~ (c_2^+)^*c_1^-~H^{(2)*}_{i\nu}(\pt\tau)\overleftrightarrow{\partial_{\tau}}H^{(1)}_{i\nu}
(\pt\tau)\;.
\end{align}
and by use of the identities 
\begin{eqnarray}
\label{Modes:HankelID1}
H^{(2)*}_{i\nu}(x)\overleftrightarrow{\partial_x}H^{(2)}_{i\nu}(x)&=&-\frac{4i~e^{-\pi\nu}}{\pi x}\;, \nonumber \\
H^{(1)*}_{i\nu}(x)\overleftrightarrow{\partial_x}H^{(1)}_{i\nu}(x)&=&~\frac{4i~e^{+\pi\nu}}{\pi x}\;,
\end{eqnarray}
and
\begin{eqnarray}
\label{Modes:HankelID2}
H^{(2)*}_{i\nu}(x)\overleftrightarrow{\partial_x}H^{(1)}_{i\nu}(x)&=&0\;, \nonumber \\
H^{(1)*}_{i\nu}(x)\overleftrightarrow{\partial_x}H^{(2)}_{i\nu}(x)&=&0\;,
\end{eqnarray}
for the Hankel functions and their derivatives, we can evaluate the above expressions explicitly. With the choice of parameters
\begin{align}
c_2^{+}=\null&\frac{\sqrt{\pi}~e^{\frac{\pi\nu}{2}}}{2\pt}\;,&
\qquad c_1^{-}=\null&\frac{\sqrt{\pi}e^{-\frac{\pi\nu}{2}}}{2\pt} \;,
\end{align}
we obtain the usual normalization properties
\begin{align}
(a^{(1)\pvect\nu+}|a^{(1)\pvect'\nu'+})=\null&(2\pi)^3\delta^2(\pvect-\pvect')\delta(\nu-\nu')\;,\\
(a^{(1)\pvect\nu-}|a^{(1)\pvect'\nu'-})=\null&-(2\pi)^3\delta^2(\pvect-\pvect')\delta(\nu-\nu')\;,\\
(a^{(1)\pvect\nu+}|a^{(1)\pvect'\nu'-})=\null&0\;,
\end{align}
and the first set of final solutions takes the final form
\begin{align}
\xi^{(1)\pvect\nu\pm}_{\mu}(x)=\null&\frac{\sqrt{\pi}~e^{\pm\pi\nu/2}}{2\pt}
\left(
\begin{array}{c} - p_y \\
~ p_x \\ 
0\end{array} \right)
H^{(2/1)}_{i\nu}(\pt\tau)\;.
\end{align}
\subsubsection{Normalization of the second set of solutions}
The second set of solutions can be normalized in a similar way, by considering again the scalar product (\ref{Modes:sp}), between the different solutions. In Ref.~\cite{Dusling:2011rz} it is argued that the positive and negative frequency solutions are again the ones which involve only Hankel functions of the second and first kind respectively. The positive and negative frequency solutions then take the form
\begin{align}
\xi^{(2)\pvect\nu\pm}_{\mu}(x)=\null&
\left(
\begin{array}{c} \nu p_x/(\pt^2\tau_0^2)~R_\bot^{\pvect\nu\pm}(\tau) \\
\nu p_y/(\pt^2\tau_0^2)~R_\bot^{\pvect\nu\pm}(\tau) \\
-R_\eta^{\pvect\nu\pm}(\tau) \end{array}
\right)\;, 
\end{align}
where the time dependence is given by
\begin{align}
R_\bot^{\pvect\nu\pm}(\tau)=\null&-\frac{\pt\tau_0}{\nu^2+\pt^2\tau_0^2}~c_{2/1}^{\pm}
~H'^{(2/1)}_{i\nu}(\pt\tau_0)+\int_{\tau_0}^{\tau}~\dInt \tau'\,\tau^{'-1}c_{2/1}^{\pm}
H^{(2/1)}_{i\nu}(\pt\tau')\;, \\
R_\eta^{\pvect\nu\pm}(\tau)=\null&-\frac{\pt\tau_0}{\nu^2+\pt^2\tau_0^2}~c_{2/1}^{\pm}~
H'^{(2/1)}_{i\nu}(\pt\tau_0)+\int_{\tau_0}^{\tau}~\dInt \tau'\,\frac{\tau'}{\tau_0^2}
c_{2/1}^{\pm}H^{(2/1)}_{i\nu}(\pt\tau') \;.
\end{align}
Since the scalar product is constant in time, we can without loss of generality evaluate it at the time $\tau=\tau_0$, when the gauge condition (\ref{Modes:cg}) applies. The scalar products between the different solutions then take the form
\begin{align}
(a^{(2)\pvect\nu+}|a^{(2)\pvect'\nu'+})=\null&i\tau_0~(2\pi)^3\delta^2(\pvect-\pvect')\delta(\nu-\nu')
~\frac{|c_2^{+}|^2}{\tau_0^3(\pt\tau_0)}~
\left.\left[H^{(2)*}_{i\nu}(x)\overleftrightarrow{\partial_x}H^{(2)}_{i\nu}(x)\right]
\right|_{x=\pt\tau_0}\;, \\
\nonumber \\
(a^{(2)\pvect\nu-}|a^{(2)\pvect'\nu'-})=\null&i\tau_0~(2\pi)^3\delta^2(\pvect-\pvect')\delta(\nu-\nu')
~\frac{|c_1^{-}|^2}{\tau_0^3(\pt\tau_0)}~
\left.\left[H^{(1)*}_{i\nu}(x)\overleftrightarrow{\partial_x}H^{(1)}_{i\nu}(x)\right]
\right|_{x=\pt\tau_0}\;, \\
\nonumber \\
(a^{(2)\pvect\nu+}|a^{(2)\pvect'\nu'-})=\null&
i\tau_0~(2\pi)^3\delta^2(\pvect-\pvect')\delta(\nu-\nu')
~\frac{(c_2^+)^{*}c_1^{-}}{\tau_0^3(\pt\tau_0)}~ 
\left.\left[H^{(2)*}_{i\nu}(x)\overleftrightarrow{\partial_x}H^{(1)}_{i\nu}(x)\right]
\right|_{x=\pt\tau_0}\;,
\end{align}
where again we can evaluate the terms involving Hankel functions and their derivatives by use of the relations (\ref{Modes:HankelID1},~\ref{Modes:HankelID2}). With the choice of parameters
\begin{align}
c_2^{+}=\null&\tau_0~\frac{\sqrt{\pi}e^{\frac{\pi\nu}{2}}}{2}~\pt\tau_0\;, \\
c_2^{-}=\null&\tau_0~\frac{\sqrt{\pi}e^{-\frac{\pi\nu}{2}}}{2}~\pt\tau_0\;,
\end{align}
the solutions then satisfy the usual relations for the scalar product
\begin{align}
(a^{(2)\pvect\nu+}|a^{(2)\pvect'\nu'+})=\null&(2\pi)^3\delta^2(\pvect-\pvect')\delta(\nu-\nu')\;,\\
(a^{(2)\pvect\nu-}|a^{(2)\pvect'\nu'-})=\null&-(2\pi)^3\delta^2(\pvect-\pvect')\delta(\nu-\nu')\;,\\
(a^{(2)\pvect\nu+}|a^{(2)\pvect'\nu'-})=\null&0\;. 
\end{align}
The orthogonality of the solutions $a^{(2)\pvect\nu\pm}$ and $a^{(1)\pvect\nu\pm}$, in the sense that the scalar product of any combination of the two vanishes, follows directly from the structure of the polarization vectors. To complete the construction of the orthonormal basis of free modes in generalized Coulomb gauge, we have to consider also the cases where either $\nu$ or $\pt$ vanishes. We find that the previous set of solutions is well behaved in the limit $\nu\to 0$, whereas in the case $\pt\to 0$ it is more convenient to consider a different parametrization.

\subsubsection*{A.4 Special case: Zero transverse momentum}
In the case of vanishing transverse momentum, the Coulomb gauge condition (\ref{Modes:cg}) along with the Gauss constraint (\ref{Modes:Gauss}) imply that the physical solutions may be taken of the form
\begin{align}
\xi^{(1)\nu,\pvect=0}_{\mu}(\tau)=\null&
\left(\begin{array}{c} 1 \\
0 \\
0 \end{array}\right)
\xi^{\nu}(\tau)\;, &
\xi^{(2)\nu,\pvect=0}_{\mu}(\tau)=
\null&\left(
\begin{array}{c} 0 \\
1 \\
0 \end{array}\right)
\xi^{\nu}(\tau)\;.
\end{align}
The time dependence of the function $\xi^{\nu}(\tau)$ is governed by the evolution equation (\ref{Modes:eom}), which in this case takes the form
\begin{align}
\left[\tau^{-1}\partial_{\tau} \tau \partial_{\tau}+\frac{\nu^2}{\tau^2}\right]\xi^{\nu}(\tau)=\null&0 \;.
\end{align}
This equation has the general solution
\begin{align}
\xi^{\nu}(\tau)=\null&c_1\left(\frac{\tau}{\tau_0}\right)^{i\nu}
+c_2\left(\frac{\tau}{\tau_0}\right)^{-i\nu}\;,
\end{align}
and the positive and negative frequency solutions are given by $c_1^{+}=c_2^{-}=0$. With the appropriate normalization the solutions take the final form
\begin{align}
\xi^{\nu+}(\tau)=\null&\frac{1}{\sqrt{2\nu}}\left(\frac{\tau}{\tau_0}\right)^{-i\nu}\;,
&\xi^{\nu-}(\tau)=\null&\frac{1}{\sqrt{2\nu}}\left(\frac{\tau}{\tau_0}\right)^{i\nu}\;, 
\end{align}
and the scalar product between the solutions satisfies the usual relations
\begin{align}
(a^{(1/2)\nu,\pvect=0+}|a^{(1/2)\nu,\pvect'+})=\null&(2\pi)^3\delta^{2}(\pvect')\delta(\nu-\nu')\;, \\
(a^{(1/2)\nu,\pvect=0-}|a^{(1/2)\nu,\pvect'-})=\null&-(2\pi)^3\delta^{2}(\pvect')\delta(\nu-\nu')\;, \\
(a^{(1/2)\nu,\pvect=0+}|a^{(1/2)\nu,\pvect'-})=\null&0\;,
\end{align}
while the two sets of solutions $a^{(1)\nu,\pvect=0\pm}$ and $a^{(2)\nu,\pvect=0\pm}$ are orthogonal as can directly be observed from the polarization structure.

\section{Perturbative calculation of hard scales}
\label{app:HardScale}
In this appendix we compute the perturbative expressions for the gauge-invariant hard scale observables $\Lambda_T^2$ and $\Lambda_L^2$, which we introduced in sec.~\ref{sec:lattice}. We will frequently encounter the expectation values of equal-time correlation functions of the gauge fields at intermediate steps of the calculation. Since ultimately, we are interested only in gauge invariant quantities, we can evaluate all expressions in Fock-Schwinger gauge with the residual gauge freedom fixed by the generalized Coulomb gauge condition (c.f. app.~\ref{app:ModeVectors}). In order to evaluate equal-time correlation functions in this gauge, we first expand the gauge fields in terms of creation and annihilation operators according to
\begin{eqnarray}
\label{HardScale:ModeExp}
\langle A_{\mu}^a(\tau,\xvect,\eta)A_{\nu}^a(\tau,\xvect',\eta')\rangle= \int \frac{d^2\pvect}{(2\pi)^2}\frac{d\nu}{2\pi} \int&&\frac{\dInt^2\qvect}{(2\pi)^2}\frac{\dInt\nu'}{2\pi} 
\sum_{\lambda,\lambda'}
\left < 
\left[\xi_{\mu}^{(\lambda)\pvect\nu+}~{\bf a}^{\pvect\nu}_{\lambda,a}~e^{i(\pvect\xvect+\nu\eta)}+h.c.\right]
\right.\nonumber \\
&&\left.
\times\left[\xi_{\nu}^{(\lambda')\qvect\nu'+}~{\bf a}^{\qvect\nu}_{\lambda',a}~e^{i(\qvect\xvect'+\nu'\eta')}+h.c.\right]
\right>\;,
\end{eqnarray}
where $h.c.$ denotes the hermitian conjugate. In order to evaluate the above expectation values, we make use of the relations
\begin{eqnarray}
\langle {\bf a}^{\pvect\nu}_{\lambda,a} {\bf a}^{\dagger~\qvect\nu'}_{\lambda',b} \rangle&=&\delta_{ab}\delta_{\lambda\lambda'}~\Big(f(\pvect,\nu,\tau)+1\Big)~(2\pi)^3~\delta^{(2)}(\pvect-\qvect)~\delta(\nu-\nu')\;, \nonumber \\
\langle {\bf a}^{\dagger~\pvect\nu}_{\lambda,a} {\bf a}^{\qvect\nu'}_{\lambda',b} \rangle&=&\delta_{ab}\delta_{\lambda\lambda'}~f(\pvect,\nu,\tau)~(2\pi)^3~\delta^{(2)}(\pvect-\qvect)~\delta(\nu-\nu')\;, \nonumber \\
\end{eqnarray}
whereas all other terms appearing in eq.~(\ref{HardScale:ModeExp}) vanish identically. The expression in eq.~(\ref{HardScale:ModeExp}) can then be expressed as
\begin{eqnarray}
\langle A_{\mu}^a(\tau,\xvect,\eta)A_{\nu}^a(\tau,\xvect',\eta')\rangle= (N_c^2-1)\int \frac{\dInt^2\pvect}{(2\pi)^2}\frac{\dInt\nu}{2\pi}
\left[\Big(f(\pvect,\nu,\tau)+1\Big)~e^{i(\pvect(\xvect-\xvect')+\nu(\eta-\eta'))}\right. \nonumber \\
\left.
+f(\pvect,\nu)~e^{-i(\pvect(\xvect-\xvect')+\nu(\eta-\eta'))}\right]~\Pi^{\pvect\nu}_{\mu\nu}(\tau)\;,
\end{eqnarray}
where we defined the Lorentz tensor $\Pi^{\pvect\nu}_{\mu\nu}(\tau)$ according to
\begin{eqnarray}
\Pi^{\pvect\nu}_{\mu\nu}(\tau)=\sum_{\lambda}\xi_{\mu}^{(\lambda) \pvect\nu \pm}(\tau)\xi_{\nu}^{(\lambda) \pvect\nu \mp}(\tau)\;.
\end{eqnarray}
In order to evaluate this tensor, we consider the Coulomb gauge condition to be fixed at the time $\tau$ when the observables are calculated. We recall that the polarization vectors in this gauge take the form (c.f. Appendix~\ref{app:ModeVectors})
\begin{eqnarray}
\xi_{\mu}^{(1)\pvect\nu,\pm}(\tau)&=&\frac{\sqrt{\pi}e^{\pm \pi\nu/2}}{2\pt} \left (\begin{array}{c} -p_y \\ p_x \\ 0 \end{array} \right) H^{(2/1)}_{i\nu}(\pt \tau)\;, \nonumber \\
\xi_{\mu}^{(2)\pvect\nu,\pm}(\tau)&=&\frac{\sqrt{\pi}e^{\pm \pi\nu/2}}{2 \tau \psqr} \left (\begin{array}{c} \nu p_x \\ \nu p_y \\ -(\pt\tau)^2 \end{array} \right) H^{'(2/1)}_{i\nu}(\pt \tau)\;,
\end{eqnarray}
where $\pt=|\pvect|$ is the transverse momentum and $\psqr=\pt^2+\nu^2/\tau^2$ denotes the spatial momentum squared. The Lorentz tensor $\Pi^{\pvect\nu}_{\mu\nu}(\tau)$ can then be evaluated explicitly. In order to simplify the resulting tensor structure, we will approximate the behavior of the Hankel functions and their derivatives by the expansion for large time arguments
\begin{eqnarray}
\label{pert:HankelApprox}
\left[1+\frac{\nu^2}{x^2}\right]^{-1} H^{'(2)}_{i\nu}(x) H^{'(1)}_{i\nu}(x) \simeq  H^{(2)}_{i\nu}(x) H^{(1)}_{i\nu}(x) \simeq \frac{2}{\pi x}\;,
\end{eqnarray}
which effectively amounts to considering highly anisotropic systems, where the characteristic transverse momenta are much larger than the longitudinal momenta $(\pt\gg\nu/\tau)$. Within this approximation the Lorentz tensor can then be expressed as
\begin{eqnarray}
\Pi^{\pvect\nu}_{\mu\nu}(\tau)=\frac{1}{2 \pt\tau} \left[-g_{\mu\nu}-\frac{p_\mu p_\nu}{\psqr}\right]\;, 
\end{eqnarray}
where $g_{\mu\nu}=\text{diag}(-1,-1,-\tau^2)$ denotes the spatial components of the metric tensor and the spatial momentum vector is denoted as $p_{\mu}=(\pvect,\nu)$. In summary the equal-time correlation functions in eq. (\ref{HardScale:ModeExp}) can then be expressed as
\begin{eqnarray}
\label{pert:AA}
\langle A_{\mu}^a(\tau,\xvect,\eta)A_{\nu}^a(\tau,\xvect',\eta')\rangle&=&(N_c^2-1)\int \frac{\dInt^2\pvect}{(2\pi)^2}\frac{\dInt\nu}{2\pi}\left(\frac{1}{2\pt\tau}\right)
\left[\Big(f(\pvect,\nu,\tau)+1\Big)~e^{i(\pvect(\xvect-\xvect')+\nu(\eta-\eta'))}\right. \nonumber \\
&&+\left.f(\pvect,\nu,\tau)~e^{-i(\pvect(\xvect-\xvect')+\nu(\eta-\eta'))}\right]~\left[-g_{\mu\nu}-\frac{p_\mu p_\nu}{\psqr}\right]\;,
\end{eqnarray}
which we will use in the following in order to evaluate the perturbative expressions for gauge invariant quantities. We will also need the expectation value of equal-time correlation functions of time derivatives of the gauge fields, which can be calculated in a similar fashion. Here we will only present the result of this computation, which is given by
\begin{eqnarray}
\label{pert:EE}
&&\langle \partial_{\tau}A_{\mu}^a(\tau,\xvect,\eta)~\partial_{\tau}A_{\nu}^a(\tau,\xvect',\eta')\rangle= (N_c^2-1)\int \frac{\dInt^2\pvect}{(2\pi)^2}\frac{\dInt\nu}{2\pi}\left(\frac{1}{2\pt\tau}\right)
\Big[\Big(f(\pvect,\nu,\tau)+1\Big)\nonumber \\ 
&&\times\left.~e^{i(\pvect(\xvect-\xvect')+\nu(\eta-\eta'))}
+f(\pvect,\nu,\tau)~e^{-i(\pvect(\xvect-\xvect')+\nu(\eta-\eta'))}\right]~\left[-\psqr g_{\mu\nu}-p_\mu p_\nu\right]\;. 
\end{eqnarray}
In order to evaluate the perturbative expressions for the hard scale observables, we will first evaluate the perturbative expression for the energy density. Here we will consider separately the electric and magnetic components of the energy density, which can be calculated from eqns.~(\ref{pert:AA},\ref{pert:EE}) in a straightforward way.  The individual components of the magnetic energy density are given by
\begin{eqnarray}
\label{pert:BSqrDef}
\langle\mathcal{B}_x^2(\tau)\rangle=\frac{1}{V_\bot L_{\eta}} \int \dInt^2\xvect \dInt\eta~\langle \mathcal{F}_{y\eta}^a(x)\mathcal{F}^{y\eta}_a(x) \rangle\;, \nonumber \\
\langle\mathcal{B}_y^2(\tau)\rangle=\frac{1}{V_\bot L_{\eta}} \int \dInt^2\xvect \dInt\eta~\langle \mathcal{F}_{x\eta}^a(x)\mathcal{F}^{x\eta}_a(x) \rangle\;, \nonumber \\
\langle\mathcal{B}_\eta^2(\tau)\rangle=\frac{1}{V_\bot L_{\eta}} \int \dInt^2\xvect \dInt\eta~\langle \mathcal{F}_{xy}^a(x)\mathcal{F}^{xy}_a(x) \rangle \;,
\end{eqnarray}
where in the following we will consider only the Abelian part of the field strength tensor. The expressions in eq.~(\ref{pert:BSqrDef}) then reduce to 
\begin{eqnarray}
\langle\mathcal{B}_x^2(\tau)\rangle&=&\frac{1}{V_\bot L_{\eta}} \int \dInt^2\xvect \dInt\eta~\Delta^{\mu\nu}_{y\eta}(x_1,x_2)~\left.\langle A_{\mu}^a(x_1)A_{\nu}^a(x_2)\rangle\right|_{x_1=x_2=x}\;,  \nonumber \\
\langle\mathcal{B}_y^2(\tau)\rangle&=&\frac{1}{V_\bot L_{\eta}} \int \dInt^2\xvect \dInt\eta~\Delta^{\mu\nu}_{x\eta}(x_1,x_2)~\left.\langle A_{\mu}^a(x_1)A_{\nu}^a(x_2)\rangle\right|_{x_1=x_2=x}\;,  \nonumber \\
\langle\mathcal{B}_\eta^2(\tau)\rangle&=&\frac{1}{V_\bot L_{\eta}} \int \dInt^2\xvect \dInt\eta~\Delta^{\mu\nu}_{xy}(x_1,x_2)~\left.\langle A_{\mu}^a(x_1)A_{\nu}^a(x_2)\rangle\right|_{x_1=x_2=x}\;, 
\end{eqnarray}
where the differential operator $\Delta^{\mu\nu}_{\alpha\beta}(x_1,x_2)$ is given by
\begin{eqnarray}
\Delta^{\mu\nu}_{\alpha\beta}(x_1,x_2)=\partial^{x_1}_\alpha\partial_{x_2}^\alpha g^{\mu \beta}\delta^{\nu}_ {\beta}+\partial^{x_1}_\beta\partial_{x_2}^\beta g^{\mu \alpha}\delta^{\nu}_{ \alpha}-2\partial^{x_1}_{\alpha}\partial^{x_2}_{\beta} g^{\mu \beta}g^{\nu \alpha}\;,
\end{eqnarray}
(no summation over $\alpha,\beta$) where $\delta^{\alpha}_{\beta}$ denote the Kronecker symbol. By use of eq. (\ref{pert:AA}) for the equal time correlation functions in Coulomb gauge, one then obtains the final result
\begin{eqnarray}
\langle\mathcal{B}_x^2(\tau)\rangle&=&N_g\int \frac{d^2\pvect}{(2\pi)^2}\frac{d\nu}{2\pi}
~\frac{p_y^2+\nu^2/\tau^2}{2 \pt \tau}~\Big[f(\pvect,\nu,\tau)+1/2\Big]\;, \nonumber \\
\langle\mathcal{B}_y^2(\tau)\rangle&=&N_g\int \frac{d^2\pvect}{(2\pi)^2}\frac{d\nu}{2\pi}
~\frac{p_x^2+\nu^2/\tau^2}{2 \pt \tau}~\Big[f(\pvect,\nu,\tau)+1/2\Big]\;, \nonumber \\
\langle\mathcal{B}_\eta^2(\tau)\rangle&=&N_g\int \frac{d^2\pvect}{(2\pi)^2}\frac{d\nu}{2\pi}
~\frac{\pt^2}{2 \pt \tau}~\Big[f(\pvect,\nu,\tau)+1/2\Big]\;, \nonumber \\
\end{eqnarray}
where $N_g=2 (N_c^2-1)$ is the number of transverse gluons. Similarly, one can evaluate the electric components of the energy density according to
\begin{eqnarray}
\langle\mathcal{E}_x^2(\tau)\rangle&=&\frac{1}{V_\bot L_{\eta}} \int \dInt^2\xvect \dInt\eta ~\frac{[E^{x}_a(x)]^2}{\tau^2}=N_g\int \frac{d^2\pvect}{(2\pi)^2}\frac{d\nu}{2\pi}
~\frac{p_y^2+\nu^2/\tau^2}{2 \pt \tau}~\Big[f(\pvect,\nu,\tau)+1/2\Big]\;,\nonumber \\
\langle\mathcal{E}_y^2(\tau)\rangle&=&\frac{1}{V_\bot L_{\eta}} \int \dInt^2\xvect \dInt\eta ~\frac{[E^{y}_a(x)]^2}{\tau^2}=N_g\int \frac{d^2\pvect}{(2\pi)^2}\frac{d\nu}{2\pi}
~\frac{p_x^2+\nu^2/\tau^2}{2 \pt \tau}~\Big[f(\pvect,\nu,\tau)+1/2\Big]\;, \nonumber \\
\langle\mathcal{E}_\eta^2(\tau)\rangle&=&\frac{1}{V_\bot L_{\eta}} \int \dInt^2\xvect \dInt\eta ~[E^{\eta}_a(x)]^2=N_g\int \frac{d^2\pvect}{(2\pi)^2}\frac{d\nu}{2\pi}
~\frac{\pt^2}{2 \pt \tau}~\Big[f(\pvect,\nu,\tau)+1/2\Big]\;, 
\end{eqnarray}
such that the overall energy density $\epsilon(\tau)$ is given by (c.f. eqns.~(\ref{lat:EnergyPressureEB},\ref{lat:TotalEnergy}))
\begin{eqnarray}
\label{pert:Energy}
\epsilon(\tau)=2N_g\int \frac{d^2\pvect}{(2\pi)^2}\frac{d\nu}{2\pi}
~\frac{\psqr}{2 \pt \tau}~\Big[f(\pvect,\nu,\tau)+1/2\Big]\;.
\end{eqnarray}

We note that the factor in the denominator has the interpretation of the mode energy $\omega_p\simeq \pt$ in the limit of highly anisotropic systems, where the characteristic transverse momenta are much larger than the characteristic longitudinal momenta. Since this limit enters the approximation in eq.~(\ref{pert:HankelApprox}), we can also replace this factor to obtain the usual relativistic normalization. By absorbing the additional factor of $\tau$ into the integration over the longitudinal momentum $\pz=\nu/\tau$, one then recovers the standard textbook relations.\\

In order to evaluate the perturbative expressions for the gauge invariant hard scale observables $\Lambda_T^2$ and $\Lambda_L^2$, we also have to consider covariant derivatives of the field strength tensor according to (c.f. eq.~(\ref{lat:DFDef}))
\begin{eqnarray}
\label{pert:DFDef}
\langle\mathcal{H}^{\mu}_{~\mu}(\tau)\rangle=\frac{4}{V_\bot L_\eta}\int d^2\vec{x}_{\bot}~\dInt\eta~ \langle D_{\alpha}^{ab}(x)\mathcal{F}^{\alpha\mu}_{b}(x)~D^{\beta}_{ac}(x)\mathcal{F}_{\beta\mu}^{c}(x)\rangle\;,
\end{eqnarray}
(no summation over $\mu$) where summation over spatial Lorentz indices $\alpha,\beta=x,y,\eta$ is implied. We proceed as previously and consider only the Abelian part of the covariant derivative and the field strength tensor, such that the expression in eq.~(\ref{pert:DFDef}) reduces to
\begin{eqnarray}
\langle\mathcal{H}^{\mu}_{~\mu}(\tau)\rangle= \frac{4}{V_\bot L_\eta} \int \dInt^2\xvect \dInt\eta~\Gamma^{\mu,\gamma\delta}_{~\mu}(x_1,x_2)~\left.\langle A_{\gamma}^a(x_1)A_{\delta}^a(x_2) \rangle \right|_{x_1=x_2=x}
\end{eqnarray}
where we introduced the derivative operator
\begin{eqnarray}
\Gamma^{\mu,\gamma\delta}_{~\mu}(x_1,x_2)=\left[\partial_{\alpha}\partial^{\alpha}g^{\mu\gamma}-\partial_{\alpha}g^{\mu\nu}\partial_{\nu}g^{\alpha\gamma}\right]_{x_1}\left[\partial_{\beta}\partial^{\beta}\delta_{\mu}^{\delta}-g^{\beta\delta}\partial_{\beta}\partial_{\mu}\right]_{x_2}\;.
\end{eqnarray}
Since the second parts of the derivatives are of the form $\partial^{\gamma}A_{\gamma}(x)$, they do not contribute in the generalized Coulomb gauge. With this simplification, the above expression can then be evaluated explicitly according to
\begin{eqnarray}
\label{pert:DFResult}
\langle\mathcal{H}^{x}_{~x}(\tau)\rangle&=&4N_g\int \frac{d^2\pvect}{(2\pi)^2}\frac{d\nu}{2\pi}
~\frac{\psqr\left[p_y^2+\nu^2/\tau^2\right]}{2\pt \tau}~\Big[f(\pvect,\nu,\tau)+1/2\Big]\;, \nonumber \\
\langle\mathcal{H}^{y}_{~y}(\tau)\rangle&=&4N_g\int \frac{d^2\pvect}{(2\pi)^2}\frac{d\nu}{2\pi}
~\frac{\psqr\left[p_x^2+\nu^2/\tau^2\right]}{2\pt \tau}~\Big[f(\pvect,\nu,\tau)+1/2\Big]\;, \nonumber \\
\langle\mathcal{H}^{\eta}_{~\eta}(\tau)\rangle&=&4N_g\int \frac{d^2\pvect}{(2\pi)^2}\frac{d\nu}{2\pi}
~\frac{\psqr~\pt^2}{2\pt \tau}~\Big[f(\pvect,\nu,\tau)+1/2\Big]\;.
\end{eqnarray}
Combining the results in eqns.~(\ref{pert:Energy}) and (\ref{pert:DFResult}), we obtain the final result (c.f. eqns.~(\ref{lat:LambdaDef},\ref{lat:LambdaApprox}))
\begin{eqnarray}
\Lambda_{T}^2(\tau)&\simeq&\frac{\int d^2\pvect~\dInt\pz~2\pt^2~\omega_p~f(\pvect,\pz,\tau)}{\int d^2\pvect~\dInt\pz~\omega_p~f(\pvect,\pz,\tau)}\;, \quad
\Lambda_{L}^2(\tau)\simeq\frac{\int d^2\pvect~\dInt\pz~4\pz^2~\omega_p~f(\pvect,\pz,\tau)}{\int d^2\pvect~\dInt\pz~\omega_p~f(\pvect,\pz,\tau)}\;, \nonumber \\
\end{eqnarray}
where in the last step, we explicitly used $\omega_p\simeq p_T$ as the relativistic quasi-particle energy in the limit $p_T\gg\nu/\tau$ and identified $\pz=\nu/\tau$ as the longitudinal momentum.

\section{Occupation numbers and generalized Coulomb gauge on the lattice}
\label{app:GaugeFix}
In this appendix, we will discuss the procedure to compute gauge dependent quantities within the framework of classical-statistical lattice simulations. Since for practical purposes the Fock-Schwinger gauge $(A_\tau=0)$ needs to be employed throughout the entire time evolution, we focus here on fixing the residual gauge freedom to perform time independent gauge transformations. In this context it is of great advantage to choose the residual gauge freedom such that there is a clear interpretation of the physical degrees of freedom. We employ the generalized Coulomb gauge condition, which in the continuum takes the form
\begin{eqnarray}
\label{Gauge:CG}
\tau^{-2}\partial_{\eta}A_{\eta}(x)+\sum_i \partial_i A_i(x)=0\;. 
\end{eqnarray}
We emphasize that the gauge condition in eq.~(\ref{Gauge:CG}) can only be fixed once at an arbitrary time $\tau=\tau_0$ and does in general not hold for times $\tau\neq\tau_0$.\footnote{Due to the explicit time dependence of eq.~(\ref{Gauge:CG}) this is already not the case for the free theory. However, even in Minkowski spacetime, where the corresponding Coulomb type gauge condition is time independent, eq.~(\ref{Gauge:CG}) will in general not be satisfied at times $t\neq t_0$ due to the interactions of different momentum modes.} This is different in actual Coulomb gauge, where the gauge condition in eq.~(\ref{Gauge:CG}) is employed at all times at the expense of a non-vanishing temporal component $A_\tau$ of the gauge field. However, it can easily be shown that equal-time correlation functions of the spatial components of the gauge fields are the same as in actual Coulomb gauge. In particular, since this gauge provides a clear interpretation of the physical degrees of freedom of the system (c.f. Appendix \ref{app:ModeVectors}), we can safely use this prescription to develop a quasi-particle picture. The advantage of this procedure is that the gauge condition in eq.~(\ref{Gauge:CG}) can be employed at any time $\tau_0$, using only time independent gauge transformations. In practice, this implies that gauge dependent quantities at different times $\tau_0,\tau_1,...$ are effectively calculated in different gauges; however the physical interpretation of the quantities manifestly remains the same.\\

After these preliminary remarks, we will now discuss how the gauge condition can be employed in classical-statistical lattice simulations. The general procedure turns out to be very similar to Landau gauge fixing in standard vacuum or thermal equilibrium lattice QCD and can be formulated as a minimization procedure of the gauge fixing potential \cite{Cucchieri:1995pn}
\begin{align}
\label{Gauge:CGPotential}
\mathcal{E}_U[G]=\frac{1}{6 N_\bot^2 N_\eta} \sum_{\vec{x}_T,\eta}
\left\{
\frac{a_\bot^2}{\tau^2a_\eta^2}\text{tr}[\dblone-G(x)U_{\eta}(x)G^{\dagger}(x+\hat{\eta})]
+\sum_{i}\text{tr}[\dblone-G(x)U_{i}(x)G^{\dagger}(x+\ihat)]  
\right\}\;, \nonumber \\
\end{align}
with respect to time independent gauge transformations $G(x)\in SU(2)$. By variation of the gauge fixing potential in eq.~(\ref{Gauge:CGPotential}) with respect to infinitesimal gauge transformations around a local minimum $G(x)$ according to $G(x)\rightarrow [\dblone+i\alpha_a(x)\Gamma^a]G(x)$, it is straightforward to verify that the local minima\footnote{We will only consider local minima of the gauge fixing potential. The issue of identifying the global minimum of the gauge fixing potential in eq.~(\ref{Gauge:CGPotential}), usually referred to as 'Minimal Coulomb gauge', is related to the presence of Gribov copies and primarily affects the infrared sector (see e.g. Ref. \cite{Cucchieri:1997dx}). Since we expect the naive interpretation in terms of quasi-particle excitations to break down in the infrared, this does not affect our discussion.}  
\begin{eqnarray}
\left.\frac{\delta  \mathcal{E}_U[G]}{\delta \alpha_a(x)} \right|_{\alpha=0}=0\;,
\end{eqnarray}
satisfy the relation
\begin{eqnarray}
\label{Gauge:LatticeCGcriterion}
\frac{a_\bot^2}{\tau^2a_\eta^2}\text{tr}[i\Gamma^a (U_{\eta}^{(G)}(x)-U_{\eta}^{(G)}(x-\hat{\eta}))]+\sum_i \text{tr}[i\Gamma^a (U_{i}^{(G)}(x)-U_{i}^{(G)}(x-\ihat))]=0\;,
\end{eqnarray}
where $U_{\mu}^{(G)}(x)$ denotes the gauge transformed link variables $U_{\mu}^{(G)}(x)=G(x)U_{\mu}(x)G^{\dagger}(x+\hat{\mu})$ as discussed in sec. \ref{sec:lattice}. Since the expressions of the form $\text{tr}[i\Gamma^aU_{\mu}(x)]$ can be related to the gauge fields to leading order in lattice spacing, eq.~(\ref{Gauge:LatticeCGcriterion}) is the lattice analogue of the Coulomb gauge condition in eq.~(\ref{Gauge:CG}) in the continuum theory.\\

In practice, the minimization of the gauge fixing potential in eq.~(\ref{Gauge:CGPotential}), with respect to time independent gauge transformations $G(x)~\in~SU(2)$ can be achieved by a variety of different algorithms (see e.g. Ref.~\cite{Cucchieri:1995pn} for a review). Here we use the Fourier acceleration technique, where the gauge transformations $G(x)$ are iteratively updated according to 
\begin{eqnarray}
G_{\text{New}}(x)=\exp[i\mathcal{R}^a(x)\Gamma^a]~G_{\text{Old}}(x)\;,
\end{eqnarray}
with
\begin{eqnarray}
\mathcal{R}^a(x)&=&\alpha\left[\mathcal{F}^{-1}~\left(\frac{p_{\text{Max}}^2}{p^2}\right)~\mathcal{F}\right] ~\left\{\frac{a_\bot^2}{\tau^2a_\eta^2}\text{tr}[i\Gamma^a (U_{\eta}^{(G)}(x)-U_{\eta}^{(G)}(x-\hat{\eta}))] \sum_i \text{tr}[i\Gamma^a (U_{i}^{(G)}(x)-U_{i}^{(G)}(x-\ihat))]\right\}
\end{eqnarray}
until the gauge condition in eq.~(\ref{Gauge:LatticeCGcriterion}) is globally satisfied to a given accuracy. Here $\mathcal{F}$ and $\mathcal{F}^{-1}$ denote the fast Fourier transform and the inverse fast Fourier transform respectively and the factor $p_{\text{Max}}^2/p^2$ corresponds to the ratio of the maximal lattice momentum to the lattice momentum. The parameter $\alpha$ can be tuned to optimize the convergence of the algorithm and we typically use on the order of thousand iterations with $\alpha=0.005$ -- $0.025$ depending on the spatial lattice size.\\

We then perform the gauge transformation of the link variables to extract the lattice gauge fields according to
\begin{eqnarray}
\label{Gauge:FinalGaugeFields}
(g a_{\bot})~A_{i}^a(x+\ihat/2)&=&\log_{SU(2)}^{a}\left[G(x)U_{i}(x)G^{\dagger}(x+\ihat)\right]\;, \nonumber \\
(g a_{\eta})~A_{\eta}^a(x+\hat{\eta}/2)&=&\log_{SU(2)}^{a}\left[G(x)U_{\eta}(x)G^{\dagger}(x+\hat{\eta})\right]\;.
\end{eqnarray}
Similarly, one can extract the electric fields from the time-like plaquette variables according to
\begin{eqnarray}
\label{Gauge:FinalElectricFields}
(g a_\bot)~E^{i}_a(x+\ihat/2+\hat{\tau}/2)&=&~~\left(\frac{\tau}{a_\tau}\right)~~~\log_{SU(2)}^{a}\left[G(x)U_{\tau i}(x)G^{\dagger}(x)\right]\;, \nonumber \\
(g a_\bot^2)~E^{\eta}_a(x+\hat{\eta}/2+\hat{\tau}/2)&=&\left(\frac{a_\bot^2}{a_\tau\tau a_\eta}\right)~\log_{SU(2)}^{a}\left[G(x)U_{\tau \eta}(x)G^{\dagger}(x)\right]\;.
\end{eqnarray}
Finally, we perform a fast Fourier transformation of the fields in eqns.~(\ref{Gauge:FinalGaugeFields},\ref{Gauge:FinalElectricFields}) and evaluate the gluon distribution function $f(\pvect,p_z,\tau)$ according to eq.~(\ref{lat:ParticleNumber}), where we identify the longitudinal momentum at mid-rapidity as $p_z=\nu/\tau$ in the final step.

\section{Detailed analysis of self-similarity}
\label{app:LikelihoodDist}
In this appendix, we will provide a more detailed explanation of our method used in sec.~\ref{sec:VScalAna} to extract the scaling exponents  $(\alpha,\beta,\gamma)$ from the self-similar evolution of the single-particle distribution. Our strategy is to compare the rescaled moments of the distribution at different times $(Q\tau_{\text{Test}}=1250,1500,1750,2000)$
\begin{eqnarray}
f_{\text{Test}}^{(n,m)}(\pt,\pz)=\pt^n \pz^m~s^{-\alpha} f(s^{-\beta}\pt,s^{-\gamma}\pz,\tau_{\text{Test}})\;, \nonumber \\
\end{eqnarray}
where $s=(\tau_{\text{Test}}/\tau_{\text{Ref}})$ denotes the scale factor, with the reference values
\begin{eqnarray}
f_{\text{Ref}}^{(n,m)}(\pt,\pz)=\pt^n \pz^m~f(\pt,\pz,\tau_{\text{Ref}})\;,
\end{eqnarray}
at the reference time $Q\tau_{\text{Ref}} = 1000$. By use of the self-similarity relation (\ref{C3eq:scalingf}), one finds that $f_{\text{Test}}^{(n,m)}=f^{(n,m)}_{\text{Ref}}$ holds for the correct set of scaling exponents $(\alpha,\beta,\gamma)$. Since this equality is in general violated for different values of the scaling exponents, one can then attempt to minimize the deviation in order to determine the correct scaling exponents. However, the above equality will be violated even for the correct set of scaling exponents due to statistical uncertainties of the data as well as systematic deviations from the scaling behavior in eq.~(\ref{C3eq:scalingf}). We quantify these deviations in terms of 
\begin{eqnarray}
\label{eq:ChiNMDef}
\chi^2_{(n,m)}(\alpha,\beta,\gamma)=\frac{1}{N_{\text{Test}}}\sum_{\tau_{\text{Test}}} \frac{\int \dInt \pt \int \dInt \pz~ \Big(f_{\text{Test}}^{(n,m)}-f_{\text{Ref}}^{(n,m)}\Big)^2}{\int \dInt \pt \int \dInt \pz~ \Big(f_{\text{Ref}}^{(n,m)}\Big)^2}\;. \nonumber \\ 
\end{eqnarray}
In practice, we first divide all data into equidistant bins of size $\pt^{\text{Bin}}/Q=0.02$ and $\pz^{\text{Bin}}/Q=12.0/(Q\tau)$ respectively for transverse and longitudinal momenta to reduce statistical uncertainties. The integral is then evaluated as the sum over momentum bins. Since in general the rescaled momenta of the (binned) test data do not coincide with the center of any reference momentum bin, we use bi-cubic Bezier patches to perform additional interpolations and smoothing of the reference data.\\

As we are primarily interested in the behavior for hard excitations, we employ the moments $n=1,2,3$ and $m=0,1,2$ and impose a lower transverse momentum cut-off $0.7~Q \leq \pt$ on the integration in eq.~(\ref{eq:ChiNMDef}). In addition, we also impose a higher momentum cutoff $\pt \leq 1.7~Q$ and $\pz < 0.8~Q$ to restrict the comparison to the regime of high occupancies.\\

The combined deviation for all moments is then evaluated as the sum
\begin{eqnarray}
\chi^2(\alpha,\beta,\gamma)=\frac{1}{9} \sum_{n=1}^{3} \sum_{m=0}^{2} \chi^{2}_{(n,m)}(\alpha,\beta,\gamma)\;, 
\end{eqnarray}
which we calculate for different values of the scaling exponents. Clearly, a smaller overall deviation $\chi^{2}(\alpha,\beta,\gamma)$ for a given set of exponents shows a better realization of the scaling relation in eq.~(\ref{C3eq:scalingf}) and thus points to higher likelihood for that set of exponents. To quantify this behavior, we define the likelihood for a given set of scaling exponents $(\alpha,\beta,\gamma)$ as
\begin{eqnarray}
\label{eq:LHDistDef}
W(\alpha,\beta,\gamma)=\frac{1}{\mathcal{N}} \exp\left[-\frac{\chi^{2}(\alpha,\beta,\gamma)}{2 \chi^{2}_{\text{min}}}\right]\;.
\end{eqnarray}
Here $\chi^{2}_{\text{min}}$ denotes the smallest value of $\chi^2(\alpha,\beta,\gamma)$ obtained in the analysis and quantifies statistical fluctuations of the data set as well as systematic deviations from the scaling behavior in eq.~(\ref{C3eq:scalingf}). The smallest deviation $\chi^{2}_{\text{min}}=0.00033$ is obtained for the set of scaling exponents $\alpha-3\beta-\gamma=-1.05$, $\beta=-0.02$, $\gamma=0.285$.

\end{widetext}
\end{appendix}


\end{document}